\documentclass[12pt]{article}

\usepackage[T2A]{fontenc}
\usepackage[utf8]{inputenc}
\usepackage[english]{babel}
\usepackage{amsmath}
\usepackage{amsfonts}
\usepackage{mathrsfs}
\usepackage{amssymb}
\usepackage{graphicx}

\usepackage{epsfig,epsf}
\usepackage{amsmath}
\usepackage{amsthm}
\usepackage{latexsym}
\usepackage{amsfonts}
\usepackage{amssymb}
\usepackage{slashed}
\usepackage{amscd}

\newcommand{\be}{\begin{equation}}
\newcommand{\ee}{\end{equation}}
\newcommand{\bea}{\begin{eqnarray}}
\newcommand{\eea}{\end{eqnarray}}

\newcommand{\lb}{\label}
\newcommand{\p}[1]{(\ref{#1})}

\def\p{\hat{p}}
\def\q{\hat{q}}

\def\qed{\rule{5pt}{5pt}}

\newtheorem{proposition}{Proposition}

\begin{document}

\vspace*{0.7cm}

\begin{center}
{\LARGE\bf
 %Mellin-Barnes integral presentation
 %for ladder and zig-zag diagrams and operator formalism
 Ladder and zig-zag Feynman diagrams, \\  [0.2cm]
 operator formalism and conformal triangles}

\end{center}

\vspace{1cm}

\begin{center}
{\Large \bf  S.E. Derkachov${}^{a}$, A.P. Isaev${}^{b,c}$
and L.A. Shumilov${}^{a}$}
\end{center}

\vspace{0.2cm}

\begin{center}
{${}^a$ \it St.Petersburg Department of the Steklov Mathematical
Institute \\
\it of Russian Academy of Sciences,Fontanka 27, 191023, St.Petersburg,
Russia.}\vspace{0.1cm}

{${}^b$ \it
Bogoliubov  Laboratory of Theoretical Physics,\\
Joint Institute for Nuclear Research,
141980 Dubna, Russia}\vspace{0.1cm}

{${}^c$ \it Lomonosov Moscow State University, \\
 Physics Faculty, Russia}\vspace{0.3cm}

{\tt derkach@pdmi.ras.ru, isaevap@theor.jinr.ru, \\[0.2cm]
la\_shum@mail.ru}
\end{center}

\vspace{1cm}

\setcounter{footnote}{0}
\setcounter{equation}{0}

\begin{center}
{\bf Abstract}
\end{center}
 We develop an operator approach to the evaluation of
 multiple integrals for multiloop Feynman massless diagrams.
 A commutative family of graph building operators $H_\alpha$
 for ladder diagrams is constructed and
 investigated. The complete set of eigenfunctions and
  the corresponding
 eigenvalues for the operators $H_\alpha$ are found.
 This enables us to explicitly express a wide
 class of four-point ladder diagrams and a general
 two-loop propagator-type master diagram (with
 arbitrary indices on the lines) as
 Mellin-Barnes-type integrals. Special cases  of
 these integrals are explicitly evaluated.
 A certain class of zig-zag four-point and
two-point planar Feynman diagrams (relevant to
 the bi-scalar $D$-dimensional
 ''fishnet'' field theory and to the calculation of the
 $\beta$-function
 in $\phi^4$-theory) is considered.
The graph building operators and convenient integral representations
for these Feynman diagrams are obtained.
The explicit form of the
 eigenfunctions for the graph building operators
of the zig-zag diagrams is fixed by
 conformal symmetry and these eigenfunctions coincide
with the 3-point correlation functions in $D$-dimensional
conformal field theories.
By means of this approach, we exactly evaluate
the diagrams of the zig-zag series in special cases.
In particular, we find a fairly simple derivation of the values for
the zig-zag multi-loop two-point diagrams for $D=4$.
The role of conformal symmetry in this approach,
especially a connection of the considered graph
 building operators with conformal
invariant solutions of the Yang-Baxter equation
is investigated in detail.

\newpage

 \tableofcontents

 %\newpage

\vspace{1cm}

\section{Introduction}
\setcounter{equation}0

In multiloop calculations, the number of diagrams
in quantum field theories grows
faster than $(2L+1)!!$ for the order of
perturbations $L \to \infty$
(see \cite{RJR} for estimates of this number growth in various
theories).
Since numerical calculations give errors that
increase drastically
with the number of diagrams, analytical evaluation become important.
Up to now, there are no universal methods
for analytical evaluation of higher loop
diagrams (see however the method of differential equations
for master integrals, \cite{Kot3,Remi,Lee}, which is based on the
''integration by parts'' reduction method, Refs. \cite{Tkach, TChT} ;
see also an approach to Feynman period computations \cite{BO}).
Thus, it becomes important to propose methods to compute integrals
for a certain infinite class of special
 diagrams with an arbitrary number of loops.
The well known examples of such classes of diagrams
are ladder diagrams
in $\phi^3_D$-theory \cite{DU1,DBro,Isa},
their generalizations \cite{Drum},
zig-zag diagrams for $\phi^4_D$-theory
\cite{BrKr,Schnetz,Schnetz2,BS,DISh} and Basso-Dixon
fishnet diagrams \cite{BD,BD1,DKO,DO,DO1,DFO}. \\

In this paper, we
present an effective method for the analytical
evaluation of multi-loop
massless Feynman diagrams, which in
particular gives us an easy way to evaluate the ladder
and zig-zag series of diagrams.
 The impressive fact is that our approach to the analytical
evaluation of perturbative
 multiple Feynman integrals uses the full power
of the methods of $D$-dimensional conformal field theories
\cite{Polyakov1,Polyakov2,FradPal,DobMac,ToMiPe,F1,F2,F3,F4,DolOsb1,DolOsb2,Osb3}.

Let us itemize the main ideas.
\begin{itemize}
\item In special cases, the expression for the Feynman diagram
can be interpreted as integral kernel of some integral operator.
In this case, the most natural representation is the spectral
representation for the corresponding operator instead of
the initial coordinate or momentum representation.

\item Special examples are diagrams of iterative type containing some
    repeated
elementary building block (a similar effect occurs in the study of
the Bethe-Salpeter equations \cite{ItZub}).
This elementary building block deserves a
special name and is usually called the graph building operator
(see e.g. \cite{GKK}).
The Feynman diagram containing convolution of $n$ building blocks
corresponds to
the $n$-th power of eigenvalues
of the graph building operator. This means that the spectral
representation for the graph building
 operator allows one to write the general
expression for an infinite number of Feynman diagrams obtained by
iteration of the single graph building operator.
We should note that the so-called fishnet conformal field theories
\cite{KG}, \cite{GGKK}, \cite{KO}, \cite{GKK}, \cite{GKKNS},
\cite{CK}, \cite{BCF}, \cite{KO23}
produce many examples of nontrivial graph building operators.

\item Of course, the possibility of a constructive use
of the spectral representation for graph building operator heavily
depends on the effective construction of the
 eigenfunction and the statements
of orthogonality and completeness.
This can be done in interesting cases of ladder diagrams and more
sophisticated zig-zag diagrams and Basso-Dixon diagrams.

\item In interesting cases, one obtains a family of
commuting graph building operators with a nontrivial
symmetry group.

\item All this is very close in spirit to the operator approach
to the evaluation of Feynman diagrams initiated by one
of the authors \cite{Isa}, \cite{Isa2}, \cite{IsaB}.
\end{itemize}

The plan of the paper is the following.
In Section \ref{opfor} we recall the basic facts about
the operator method \cite{Isa}, \cite{Isa2}  and \cite{GI}
of multiloop evaluations,
that will be used in this paper.
Section \ref{2ladder} is devoted to the simplest case,
where a special family of commuting graph building operators
$H_\alpha$: $H_\alpha H_\beta = H_\beta H_\alpha$ is
considered. The operators $H_\alpha$ are constructed
 from the generators of one copy of the
D-dimensional Heisenberg algebra and should be
equivalently  understood as integral
operators acting in the space of functions of one vector
$x \in \mathbb{R}^D$
with coordinates $x^\mu$ $(\mu=1, \dots, D)$.
The main examples of the diagrams in this situation are
 $2$-point and $4$-point
 ladder diagrams and in particular the general two-loop master
 propagator type diagram with arbitrary indices on the lines.

In Section \ref{zig-zag}, we consider a special class of
2-point and 4-point zig-zag diagrams in
D-dimensional $\phi^4$ quantum field theory.
The 2-point zig-zag diagrams give considerable contributions to
the renormalization group $\beta$-function  in
$\phi^4_{D=4}$  theory. An intriguing history of the
 evaluation of renormalization group functions (up to seven loops)
 in $\phi^4_{D=4}$ theory  and
 the 2-point zig-zag diagrams (for $D=4$)
 is outlined in \cite{BrKr} and in \cite{Schnetz2},
 \cite{DISh} (see also references therein). Here, in Section
\ref{zig-zag}, the graph building operator is constructed
 for zig-zag diagrams as an element of the product of two copies
of the D-dimensional Heisenberg algebras,
 or equivalently it is an integral
operator acting on the space of functions of two vector arguments
 $x_1,x_2 \in \mathbb{R}^D$.
It is interesting that a repeating block in the case of
zig-zag diagrams can be represented as a square of
a simpler operator $\hat{Q}_{12}$.
It is a particular example of a more general phenomenon.
As we discuss in Section \ref{YB}, the
initial repeating block is associated with
the $R$-operator, which is a conformally
invariant solution of the famous Yang-Baxter
equation and its factorization in the product
of two operators $\hat{Q}_{12}$ is a particular example
of the factorization of the $R$-operator \cite{Der,ChDeIs}.

We consider integrals for massless Feynman diagrams,
which possess invariance under conformal transformations.
In fact, due to conformal invariance of the graph building operators
the orthogonality and completeness of their
eigenfunctions are well known
from $D$-dimensional conformal field theory.
In Section \ref{eigpro}, we discuss
the needed properties of the
corresponding conformal triangles in detail and
then demonstrate bydirect calculation
that conformal triangles are eigenfunctions
of graph building operators and calculate the corresponding
eigenvalues.
In this section, we also try to present all the
needed proofs and
perform all calculations directly step by step to demonstrate how it
works.

In Section \ref{BK24}, we use the worked technique
to calculate the general 2-point and 4-point
 zig-zag diagrams and
to prove (outlined in \cite{DISh}) of the
Broadhurst and Kreimer conjecture \cite{BrKr}
for the zig-zag diagrams in four-dimensional $\phi^4$ quantum field
theory. Another proof was given in \cite{BS}.

The last Section \ref{YB} of the main part
of the paper is devoted to a discussion
of the role of conformal symmetry and the
connection of the considered graph
building operators with the $R$-operator, the solution of the
Yang-Baxter equation.
In the previous sections, we tried to carry
 out all the calculations explicitly
and combined in a reasonable way the integral identities and
their translation to the
corresponding operator forms. In the last section, we use mainly the
compact operator formalism, which is set out in the second section.

In Sections \ref{project} -- \ref{FF} (Appendixes),
we give proofs of some statements and provide
the technical details of the derivation of useful
formulas that we use in the main text of the paper.

\section{Operator formalism and diagram technique\label{opfor}}
\setcounter{equation}0

In this section, we
briefly recall the result of
\cite{Isa}, \cite{DerShum} and \cite{GI} which will be used
below.

Consider a $D$-dimensional Euclidean space
with coordinates $x^\mu$ where $\mu=1, \dots, D$ and
denote $x^{2\alpha} = (\sum_\mu \, x^\mu \, x^\mu)^{\alpha} $.
Let $\{ \q^\mu , \; \p^\nu \}$
be hermitian generators of the $D$-dimensional
Heisenberg algebra ${\cal H}$:
\be
\lb{gr001}
[\q^\mu , \, \p^\nu ] = i \, \delta^{\mu \nu} \; .
\ee
 The algebra ${\cal H}$ acts in the space $V$ and
we introduce two sets
of basis states $|x \rangle$ and $|k \rangle$ in $V$
which respectively diagonalize the
operators $\q^\mu$ and $\p^\nu$
\be
\lb{gr002}
\q^\mu |x \rangle = x^\mu |x \rangle \; , \;\;\;\;\;
\p^\mu |k \rangle = k^\mu |k \rangle \; .
\ee
We also introduce the dual states $\langle x |$
and $\langle k |$ such that the orthogonality and
completeness conditions are valid
\be
\lb{ortcom}
\langle x | x' \rangle = \delta^D (x-x')\, , \;\;\;\;
\langle k | k' \rangle = \delta^D (k-k') \, , \;\;\;\;
\int d^D x \,|x \rangle \langle x | = I =
\int d^D k \,|k \rangle \langle k | \, .
\ee
Relations (\ref{gr001}), (\ref{gr002}),
(\ref{ortcom}) are consistent if we have
$$
k^\mu \langle x | k \rangle =
\langle x | \p^\mu |k \rangle = -i \frac{\partial}{\partial x^\mu}
\langle x | k \rangle \;\;\; \Rightarrow \;\;\;
\langle x | k \rangle = \frac{1}{(2\pi)^{D/2}}
e^{i k^\mu x^\mu} \; ,
$$
where the normalization constant $(2\pi)^{-D/2}$ is fixed by
(\ref{ortcom}).

Define the inversion operator ${\cal I}$ such that (see \cite{Isa})
\be
\lb{inver}
\begin{array}{c}
{\cal I}^2 = 1 \; , \;\;\;\; {\cal I}^\dagger = {\cal I} \cdot \q^{2D}
\; , \;\;\;\;  \langle x | {\cal I} = \langle \frac{1}{x} |
\; , \;\;\;\;  {\cal I} | x \rangle = x^{-2D} | \frac{1}{x} \rangle
\;\;\;\;\;\;\;\;\; (\frac{1}{x} : = x^\mu/x^2) \; ,\\ [0.2cm]
{\cal I} \cdot \q^\mu \cdot {\cal I} = \frac{\q^\mu}{\q^2} \; , \;\;\;
{\cal I} \cdot \p^\mu \cdot {\cal I}  =
\q^2 \, \p^\mu - 2 \, \q^\mu \, (\q \, \p) \equiv K^\mu \;\;\;\;
  \Rightarrow \\ [0.2cm]
{\cal I} \cdot  (\q \, \p) \cdot {\cal I}  = - (\q \, \p) \; ,
\;\;\;\;
{\cal I} \cdot  \q^{2\alpha} \cdot {\cal I} = \q^{-2\alpha} \; ,
\;\;\;\;
{\cal I} \cdot  \p^{2\alpha} \cdot {\cal I} =
\q^{2(\alpha+\frac{D}{2})} \cdot \p^{2\alpha} \cdot
\q^{2(\alpha-\frac{D}{2})} \; ,
\end{array}
\ee
where $(\q \, \p) := \q^\mu \, \p^\mu$.
Below we also use the shifted inversion
operator
\be
\lb{shin}
{\cal I}_\Delta := {\cal I} \; \q^{2\Delta} \; ,
\ee
 for which
formulas in (\ref{inver})
are modified (we will need these formulas in Subsect.
{\bf \ref{roper}})
\be
\lb{inver2}
\begin{array}{c}
{\cal I}_\Delta^{\;2} = 1 \, , \;\;\;\;\;
({\cal I}_\Delta)^\dagger = {\cal I}_{D-\Delta} \, , \;\;\;\;\;
{\cal I}_\Delta \cdot \q^\mu \cdot {\cal I}_\Delta =
\frac{\q^\mu}{\q^2} \, , \\ [0.2cm]
{\cal I}_\Delta | x \rangle =
x^{2(\Delta-D)} | \frac{1}{x} \rangle  \, , \;\;\;\;
 \langle x | {\cal I}_\Delta =
x^{-2\Delta} \langle \frac{1}{x}| \;\;\;\;\; \Rightarrow
 \;\;\;\;\; \langle x | {\cal I}_\Delta | \Phi \rangle =
x^{-2\Delta} \; \langle \frac{1}{x}|\Phi \rangle\, , \\ [0.2cm]
{\cal I}_\Delta \cdot  \q^{2\alpha} \cdot {\cal I}_\Delta
 = \q^{-2\alpha} \; , \;\;\;\;
{\cal I}_\Delta \cdot  \p^{2\alpha} \cdot {\cal I}_\Delta =
\q^{2(\alpha+\frac{D}{2}-\Delta)} \cdot \p^{2\alpha} \cdot
\q^{2(\alpha-\frac{D}{2}+\Delta)} \; ,
\end{array}
\ee
and for generators of special conformal transformations
we deduce
\be
\lb{shK}
 K^{(\Delta)}_\mu : = \;
{\cal I}_\Delta \cdot \p^\mu \cdot {\cal I}_\Delta  =
\q^2 \, \p^\mu - 2 \, \q^\mu \, (\q \, \p) +
 2 i \Delta \, \q^\mu \; .
 \ee
 Here we also need a special case of
 ${\cal I}_\Delta$, i.e.
${\cal I}' := {\cal I}_{\Delta}|_{\Delta = \frac{D}{2}}
= {\cal I}\, \q^{2(\frac{D}{2})}$
that is Hermitian ${\cal I}^{\, \prime \dagger} =
{\cal I}^{\, \prime}$
with respect to (\ref{ortcom})
and
 \be
 \lb{inver1}
 ({\cal I}^{\prime})^2 = 1 , \;\;\;\;
 \bigl\langle x \bigr| \; {\cal I}' =
x^{-2({D \over 2})} \; \bigl\langle \frac{1}{x} \bigr|
, \;\;\;\;   {\cal I}' \; \bigr| x \bigl\rangle   =
x^{-2({D \over 2})} \; \bigr| \frac{1}{x}  \bigl\rangle
, \;\;\;\;\; {\cal I}' \cdot  \p^{2\alpha} \cdot {\cal I}' =
\q^{2\alpha} \cdot \p^{2\alpha} \cdot
\q^{2\alpha}   \; ,
 \ee
 where the last formula
in (\ref{inver}), (\ref{inver2}) is simplified.
Then, from the obvious identity
${\cal I}' \cdot \p^{2(\alpha + \beta)} \cdot {\cal I}' =
{\cal I}' \cdot \p^{2\alpha} \cdot
({\cal I}')^2 \cdot \p^{2\beta} \cdot {\cal I}'$,
we deduce the operator version \cite{Isa} of
the star-triangle relation
\be
\lb{startr}
\p^{2\alpha} \, \q^{2(\alpha + \beta)} \, p^{2\beta} =
\q^{2\beta} \, \p^{2(\alpha + \beta)} \, q^{2\alpha}
 \;\;\;\;\;\;
(\forall \;\; \alpha,\beta)  \; .
\ee

 We stress here that the operators $\q^{2\alpha}$ and
 $\p^{2\beta}$ with noninteger $\alpha$ and $\beta$ are
 understood as integral operators defined
 via their integral kernels. Namely,
 the operators $\q^{2\alpha}$ and
 $\p^{2\beta}$ act on the states $|\psi \rangle$ in
 the coordinate representation as
 $\langle x | \q^{2\alpha} |\psi \rangle =
 \int d^D y \langle x | \q^{2\alpha} | y \rangle
 \langle y |\psi \rangle = x ^{2\alpha}
\langle x | \psi \rangle$ and
 $$
\langle x | \p^{-2\alpha} |\psi \rangle =
\int d^D y \; \langle x | \p^{-2\alpha} |y \rangle
 \langle y |\psi \rangle =
 a(\alpha) \; \int d^D y
 \frac{\langle y | \psi \rangle }{(x-y)^{2(D/2-\alpha)}} \; ,
 $$
 where we applied the well known formula for Fourier transformation
 \be
\lb{gr3}
\begin{array}{c}
\displaystyle \langle x| \p^{-2\alpha} |y\rangle =
\int d^D k \langle x| \p^{-2\alpha} |k \rangle
\langle k  |y\rangle =
 \int  \, \frac{d^D k}{(2 \pi)^{D} } \,
 \frac{e^{i k ( x - y)}}{k^{2 \alpha} }=
\frac{a(\alpha)}{(x-y)^{2 \alpha'} } \; , \\ [0.5cm]
\displaystyle
a(\alpha):=
 \frac{1}{2^{2 \alpha} \, \pi^{D/2}} \,
 \frac{\Gamma(\alpha') }{ \Gamma(\alpha) } \; , \;\;\;\;\;\;\;
 a(\alpha) \; a(\alpha') = (2\pi)^{-D}
 \; , \;\;\;\;\;\;\; \alpha' := D/2 - \alpha \; .
 \end{array}
\ee
Then, we write the operator
identity (\ref{startr}) as
 $$
 \begin{array}{c}
 \int d^D z \;
\langle y| \p^{-2\alpha'} \, | z \rangle\,
\langle z| \,  \q^{-2(\alpha' + \beta')}
\, p^{-2\beta'} | x \rangle =
\langle y| \q^{-2\beta'} \, \p^{-2(\alpha' + \beta')}
\, \q^{-2\alpha'} | x \rangle \; ,
\end{array}
$$
and, taking into account eq. (\ref{gr3}), we represent it
 in the familiar form of the star-triangle relation
 \cite{Parisi,FradPal,Zam,VPH,Kaz1,AN}
 \be
 \lb{startr2}
\int \frac{d^D z}{(y-z)^{2\alpha} \;
z^{2\gamma} \; (z-x)^{2\beta}} \; = \;
\frac{a(\gamma)}{a(\alpha^{\, \prime})\;a(\beta^{\, \prime})} \;\cdot
\;
\frac{1}{y^{2\beta'}}\; \frac{1}{(x-y)^{2\gamma^{\, \prime}}}
\; \frac{1}{x^{2\alpha^{\, \prime}}} \;\; ,
 \ee
 where $(\alpha + \beta + \gamma) = D$ and $\frac{a(\gamma)}{
a(\alpha^{\, \prime})a(\beta^{\, \prime})} = \pi^{D/2}
\frac{\Gamma(\alpha')\Gamma(\beta')\Gamma(\gamma')}{
\Gamma(\alpha)\Gamma(\beta)\Gamma(\gamma)}$. Analogously from
the identity
$$
\p^{-2\alpha'} \p^{-2\beta'} = \p^{-2(\alpha' + \beta')}
 \;\; \Rightarrow
\;\;
 \int d^D z \;
\langle y| \p^{-2\alpha'} \, | z \rangle\,
\langle z| \,
 p^{-2\beta'} | x \rangle =
 \langle y| \p^{-2(\alpha' + \beta')} | x \rangle \; ,
 $$
 we deduce the chain relation
 \be
 \lb{chain}
  \int d^D z \; \frac{1}{(x-z)^{2\alpha}\, (z-y)^{2\beta}} =
 \frac{a(\alpha'+\beta')}{
 a(\alpha')\; a(\beta')} \cdot
 \frac{1}{(x-y)^{2(\alpha +\beta -D/2)}} \; ,
 \ee
 and the coefficients in the right-hand sides of (\ref{startr2})
 and (\ref{chain}) are the same.

\noindent
{\bf Remark 1.}  Of course, the coincidence of the coefficients
has a simple explanation -- the point is that
relations (\ref{chain}) and (\ref{startr2}) are equivalent \cite{AN}.
Indeed, after inversion of all variables $x\to 1/x,y\to 1/y$ and $z\to
1/z$
in (\ref{chain}) one obtains relation (\ref{startr2}).
 On the other hand, after shifting all the variables
$x\to x+w,y\to y+w$ and $z\to z+w$ in (\ref{startr2})
and then sending $w \to \infty$, one obtains (\ref{chain}).

\begin{figure}[h]
\begin{minipage}[h]{0.35\linewidth}
\center{\includegraphics[width=1\linewidth]{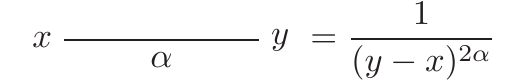}} (a) \\
\end{minipage}
\hfill
\begin{minipage}[h]{0.35\linewidth}
\center{\includegraphics[width=1\linewidth]{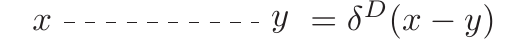}} \\(b)
\end{minipage}
\hfill
\begin{minipage}[h]{0.15\linewidth}
\center{\includegraphics[width=1\linewidth]{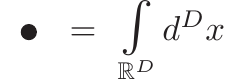}} \\(c)
\end{minipage}
\caption{\small\it (a) - Propagator; (b) - Delta function; (c) - Bold
vertex depict integration over the whole space $\mathbb{R}^D$.}
\label{f-rules1}
\end{figure}

Integral kernels and relations (\ref{startr2}), (\ref{chain})
can be depicted as Feynman graphs.
The Feynman rules that will be used in this paper
 are shown in Fig. \ref{f-rules1}.

Using the language of Feynman graphs, we can easily represent different
calculations. For example, the chain relation \eqref{chain} is shown on
Fig. \ref{f-chain}.

\begin{figure}[h]
\unitlength=8mm
\begin{picture}(25,1.6)(-2,0)

\put(0.6,0.8){\line(1,0){4.1}}
\put(2.5,0.7){$\bullet$}
\put(1.55,1){\footnotesize $\alpha$}
\put(3.6,1){\footnotesize $\beta$}

\put(0.5,0.4){$x$}
\put(4.5,0.4){$y$}
\put(2.5,0.3){$z$}

\put(5.8,0.65){$=$}

\put(6.8,0.65){$\displaystyle
\pi^{D/2} \frac{\Gamma(\alpha')\Gamma(\beta')\Gamma(\gamma')}{
\Gamma(\alpha)\Gamma(\beta)\Gamma(\gamma)}\;\;\; \cdot$}

\put(12.7,0.8){\line(1,0){3.2}}
\put(13.5,1.1){\scriptsize $\alpha \! + \! \beta \! - \! \frac{D}{2}$}

\put(12.7,0.4){$x$}
\put(15.7,0.4){$y$}

%\put(2.6,0.7){\circle{1}}
%\put(2.6,1.25){\line(0,1){0.4}}
%\put(2.6,0.15){\line(0,-1){0.4}}
\end{picture}
\caption{\small\it Scalar chain rule}
\label{f-chain}
\end{figure}

 \noindent
 {\bf Remark 2.}
 In the framework of the dimensional regularization scheme
we have the following identity \cite{GI}:
\be
\lb{gori}
\int \, d^D x \, \frac{1}{x^{2(D/2 + i \alpha)}} = \pi \, \Omega_D \,
\delta(\alpha) \; , \;\;\;\;\;
\Omega_D = \frac{2 \pi^{D/2}}{\Gamma(D/2)} \; ,
\ee
where $\alpha$
is a real number, $\delta(\alpha)$ is the standard delta-function
and $\Omega_D$ is the volume
of the unit sphere $\mathbb{S}^{D-1}$. For applications,
it turns out to be convenient to use formula
(\ref{gori}) in the form
\be
\lb{gori1}
\left. \int \, d^D x \, \frac{f(z)}{x^{\,2z}}
\right|_{{\rm Re}(z) = \frac{D}{2}}= \pi \, \Omega_D
 \,f(D/2) \,  \delta\bigl({\rm Im}(z)\bigr) \; ,
 %\int \, d^D x \, \frac{1}{x^{2(D/2 + \alpha)}}
\ee
where $z$ is a complex number and $f(z)$ is
 a function analytic in $z=0$. This form permits one \cite{GI}
 to bring the evaluation of propagator-type
perturbative integrals
 to the evaluation
of vacuum perturbative integrals
(it also permits to search their symmetries).
Further, breaking any of the lines (the propagators)
 in the corresponding vacuum diagram,
one can obtain another propagator type integral and thus
deduce many remarkable nontrivial
relations between the propagator type integrals
 in $D$ dimensions.
Sometimes these relations are called
 the ''glue-and-cut'' symmetry.
For details see \cite{GI}, \cite{Chet})
(for $D=2$ such relations were used in \cite{DerSpir}).
 We apply relation (\ref{gori1}) in the next section
(see e.g. (\ref{twoloop})).

\section{Operators $H_\alpha$, two-loop master
diagram and ladder diagrams\label{2ladder}}
\setcounter{equation}0

\subsection{Definition of the operators $H_\alpha$
and two-loop master diagram\label{master}}

We note that the star-triangle relation (\ref{startr})
can be written as the commutativity condition
 \be
\lb{hhhh1}
[\p^{2\beta} \; \q^{2\beta}, \;
 \p^{2(\alpha+\beta)} \; \q^{2(\alpha+\beta)}]=0
 \;\;\;\; \Longrightarrow \;\;\;\;
 [H_\alpha, \; H_{\alpha +\beta}] = 0 \; ,
   \;\;\;\; \forall \alpha,\beta \; ,
 \ee
\be
\lb{hhhh}
H_\alpha := \p^{2\alpha} \, \q^{2\alpha} \; .
\ee
This section will focus on applying the
operators $H_\alpha$, which will be used
 for analytical evaluation of the
two loop master-diagram and the $L$-loop ladder diagrams.
The two loop master-diagram and related diagram
for the 3-point function are respectively depicted
in Fig. \ref{fig3} and Fig. \ref{fig4}.
\begin{figure}[h!!]
%picture 1
\unitlength=7mm
\begin{picture}(25,4)

 %\put(0,2){\line(1,0){1.8}}
 %\put(2.2,2){\line(1,0){1.8}}
 %\put(1.1,2.2){\footnotesize $\alpha_6$}
\put(0,2){\line(2,1){2}}
\put(0,2){\line(2,-1){2}}
\put(0.7,2.7){\footnotesize $\alpha_3$}
\put(0.7,1.1){\footnotesize $\alpha_2$}
\put(2.2,2){\footnotesize $\alpha_5$}
\put(-0.5,2){\footnotesize $0$}

\put(2,1){\line(0,1){2}}
\put(1.85,0.9){$\bullet$}
\put(1.85,2.9){$\bullet$}
 %\put(3.85,1.9){$\bullet$}

\put(2,3){\line(2,-1){2}}
\put(3,2.7){\footnotesize $\alpha_4$}
\put(2,1){\line(2,1){2}}
\put(3,1.1){\footnotesize $\alpha_1$}
\put(2,0.6){$y$}
\put(2,3.2){$z$}

\put(4.3,2){$x$}
\put(5.5,2.2){$=$}
\put(6.5,2.2){\footnotesize ${\displaystyle \int \!
\frac{d^D y \;\; d^D z}{(x-y)^{2\alpha_1} \, y^{2\alpha_2} \,
 (y-z)^{2\alpha_5} \, z^{2\alpha_3} \, (z-x)^{2\alpha_4}}
 \, \equiv \, \frac{C(\alpha_1,\alpha_2,\alpha_6;
 \alpha_3,\alpha_4,\alpha_5)}{x^{2(D/2-\alpha_6)}}}$ ,}

 \put(14,0.5){$\alpha_6 := (3D/2 - \alpha_{1...5})$.}

\end{picture}
\caption{\small\it Two-loop master diagram.\label{fig3}}
\end{figure}
\begin{figure}[!h]
%picture 2
\unitlength=7mm
\begin{picture}(25,3)(0,1)

\put(1,1.5){\line(1,0){6}}
\put(3.7,1.7){\footnotesize $\alpha_5$}
\put(4,3.8){$0$}

\put(3,1.5){\line(1,2){1}}
\put(2.9,1.4){$\bullet$}
\put(1.8,1.7){\footnotesize $\alpha_1$}
\put(2.7,2.5){\footnotesize $\alpha_2$}

\put(5,1.5){\line(-1,2){1}}
\put(4.9,1.4){$\bullet$}
\put(5.8,1.7){\footnotesize $\alpha_4$}
\put(4.8,2.5){\footnotesize $\alpha_3$}

\put(0.5,1.5){$x$}
\put(7.4,1.5){$y$}
\put(8.7,2.5){$=$}
\put(10,2.5){$\displaystyle
\frac{1}{a(\alpha'_1)\,  a(\alpha'_4)
\, a(\alpha'_5)} \;
 \langle x| \p^{-2\alpha_1'} \q^{-2\alpha_2}
\p^{-2\alpha_5'} \q^{-2\alpha_3} \p^{-2\alpha_4'} |y \rangle$ ,}

%\put(9,0.5){\bf Fig.4}

\end{picture}
\caption{\small\it The 3-point diagram.\label{fig4}}
\end{figure}
The functions $a(\beta)$
in Fig. \ref{fig4} are defined in (\ref{gr3}).
We note that
the function for the master-diagram in Fig. \ref{fig3} is obtained
from the 3-point function
in Fig. \ref{fig4} by fixing $y=x$.
The tetrahedron vacuum diagram (which is related to the master
two-loop diagram in Fig. \ref{fig3}) is presented in Fig.\ref{fig5}.
All diagrams in Fig.\ref{fig3}--Fig.\ref{fig5}
are understood as diagrams in the configuration space and boldface
vertices (see Fig.\ref{f-rules1})
 denote integration over $\mathbb{R}^D$.
\begin{figure}[h!]
\unitlength=7mm
\begin{picture}(25,3)(-2,1)

\put(0,2){\line(1,0){1.8}}
\put(2.2,2){\line(1,0){1.8}}
\put(1.1,1.7){\footnotesize $\alpha_6$}
\put(0,2){\line(2,1){2}}
\put(0,2){\line(2,-1){2}}
\put(0.8,2.8){\footnotesize $\alpha_3$}
\put(0.8,1){\footnotesize  $\alpha_2$}
\put(2.1,2.3){\footnotesize $\alpha_5$}
\put(-0.5,2){$0$}

\put(2,1){\line(0,1){2}}
\put(1.85,0.9){$\bullet$}
\put(1.85,2.9){$\bullet$}
\put(3.85,1.9){$\bullet$}

\put(2,3){\line(2,-1){2}}
\put(3,2.8){\footnotesize $\alpha_4$}
\put(2,1){\line(2,1){2}}
\put(3,1){\footnotesize $\alpha_1$}

\put(4.2,2){$x$}
\put(5.5,2){$=$}
\put(6.5,2){$\displaystyle
\int \,
\frac{d^D x}{a(\alpha'_1) a(\alpha'_4)
a(\alpha'_5)} \,
 \langle x| \p^{-2\alpha_1'} \q^{-2\alpha_2}
\p^{-2\alpha_5'} \q^{-2\alpha_3} \p^{-2\alpha_4'}
\q^{-2\alpha_6} |x \rangle$ .}
 %\put(5,1.0){$(\alpha'_1 + \alpha'_4 + \alpha'_5 =
 %\alpha_2 + \alpha_3 + \alpha_6$ $\;\; \Rightarrow  \;\;$
 %$\alpha_1+...+\alpha_6 = \frac{3D}{2})$}

 %\put(8,0.3){\bf Fig5}

\end{picture}
\caption{\small\it The tetrahedron vacuum diagram.\label{fig5}}
\end{figure}

 %\noindent
According toidentity (\ref{gori}) the integral in
the right-hand side of equality in Fig.\ref{fig5}
(after the substitution of the relation in Fig.\ref{fig3})
 is not equal to zero only in the case when
 \be
 \lb{alfix}
 \alpha_6 \, = \, \alpha^0_6 \; := \; \frac{3D}{2} -\alpha_{1...5}
  \;\;\;\;\; \Longleftrightarrow \;\;\;\;\;
  \alpha_1'+\alpha_5'+\alpha_4'=\alpha_2 +
\alpha_3+\alpha_6 \; .
\ee
 Here we use the convenient notation
$\alpha_{b_1... b_k} := \sum_{i=1}^k \alpha_{b_i}$. In particular,
the $D$-dimensional ''glue-and-cut'' symmetry \cite{GI}
(mentioned in Remark in previous Section
{\bf \ref{opfor}}), which follows
from the vacuum diagram in Fig.\ref{fig5},
gives the tetrahedral symmetry
 \cite{GI} of the
two-loop master diagram in Fig.\ref{fig3}:
 \be
 \lb{tesym}
\begin{array}{rl}
C(\alpha_1,\alpha_2,\alpha_6; \alpha_3,\alpha_4,\alpha_5) & = \;
 C(\alpha_6,\alpha_1,\alpha_2; \alpha_5,\alpha_3,\alpha_4) \, , \\
 [0.2cm]
 & = \; C(\alpha_6,\alpha_3,\alpha_4; \alpha_5,\alpha_1,\alpha_2)
 \, , \\ [0.2cm]
 & = \; C(\alpha_1,\alpha_6,\alpha_2; \alpha_3,\alpha_5,\alpha_4)  \;
 .
 \end{array}
 \ee
Here the first two symmetries
 are rotations of the tetrahedron in Fig.\ref{fig5}
with respect of the vertices $[\alpha_3,\alpha_4,\alpha_5]$ and
$[\alpha_1,\alpha_4,\alpha_6]$,
and the last symmetry is a
 reflection of the tetrahedron in Fig.\ref{fig5} with
 respect to the plane perpendicular to the
 edge $[\alpha_1]$ (which is reduced to
the exchange of  vertices $[\alpha_1,\alpha_4,\alpha_6]$
and $[\alpha_1,\alpha_2,\alpha_5]$)  in Fig.\ref{fig5}.
All symmetries (\ref{tesym}) are elements of
the tetrahedral group $S_4$ which is
 the permutation group of four vertices
 of the tetrahedron in Fig. \ref{fig5}.
 The whole $S_6 \times \mathbb{Z}_2$ symmetry of
 the coefficient function $C(\alpha_1,\alpha_2,\alpha_6;
 \alpha_3,\alpha_4,\alpha_5)$, which is discovered in \cite{Broad},
 \cite{Broad1} (see also \cite{Isa}), is generated by
 transformations (\ref{tesym}) and by additional symmetry
 \be
 \lb{sttr2}
 \begin{array}{c}
 C(\alpha_1,\alpha_2,\alpha_6; \alpha_3,\alpha_4,\alpha_5) = \\
 [0.3cm]
  %\displaystyle
 = \frac{a\bigl(\alpha_1'-\alpha_2+\alpha_5' \bigr)
 a(\alpha_2)}{
 a(\alpha_1')a(\alpha_5')}\;
 C\bigl(\alpha_1+\alpha_2-\alpha_5',\alpha_5',\alpha_6;
 \alpha_2+\alpha_3- \alpha_5',\alpha_4,\alpha_2'\bigr) \; ,
 \end{array}
 \ee
 which is produced by the identity
 $$
  \p^{-2\alpha_1'} \q^{-2\alpha_2}
\p^{-2\alpha_5'} \q^{-2\alpha_3} \p^{-2\alpha_4'}  \q^{-2\alpha_6} =
 \p^{-2(\alpha_1'-\alpha_2+\alpha_5')} \q^{-2\alpha_5'}
\p^{-2\alpha_2} \q^{-2(\alpha_2 -\alpha_5' +\alpha_3)}
\p^{-2\alpha_4'}
 \q^{-2\alpha_6} \; ,
 $$
following from the star-triangle relation (\ref{startr}).

Now we note that the operators $H_\alpha $ (\ref{hhhh})
 satisfy
\be
\lb{setcom}
 %H_\alpha = \p^{2\alpha} \; \q^{2\alpha} \; , \;\;\;\;
H_\alpha^\dagger = \q^{2\alpha} \; \p^{2\alpha}
\equiv {\cal I}' \cdot H_\alpha \cdot {\cal I}'
\; , \;\;\;\;\;\;\;\;
 H_\alpha^\dagger \equiv (H_{-\alpha})^{-1} \; ,
 \ee
 and expressions, shown in
 Fig.\ref{fig5}, could be expressed
 in terms of $H_\alpha$.
Indeed,
the product of the operators $\p^{2\alpha}$ and $\q^{2\beta}$
 in the right-hand side of the equality in Fig.\ref{fig5}
 is written  as
 $$
 \begin{array}{c}
 \p^{-2\alpha_1'} \, \q^{-2\alpha_2} \,
\p^{-2\alpha_5'} \, \q^{-2\alpha_3} \, \p^{-2\alpha_4'} \,
\q^{-2\alpha^{0}_6}  \,
\q^{-2(\alpha_6-\alpha^{0}_6)} =  \\ [0.2cm]
= \bigl( H_{\alpha_1-\frac{D}{2}}
H^\dagger_{\frac{D}{2}-\alpha_{12}}
H_{\alpha_{125}-D} H^\dagger_{D-\alpha_{1235}}
H_{\alpha_{12345}-\frac{3D}{2}} \bigr) \,
\q^{-2(\alpha_6-\alpha^{0}_6)} \; ,
\end{array}
 $$
  where $\alpha^{0}_6$ is defined in (\ref{alfix}).
 So we get the following
 expression for the diagram in Fig.\ref{fig5}
\begin{equation}\label{twoloop}
    \int \,
\frac{d^D x}{a(\alpha'_1) a(\alpha'_4)
a(\alpha'_5)} \,
 \langle x|H_{\alpha_1-\frac{D}{2}}
H^\dagger_{\frac{D}{2}-\alpha_{12}}
H_{\alpha_{125}-D} H^\dagger_{D-\alpha_{1235}}
H_{\alpha_{12345}-\frac{3D}{2}} \;|x \rangle
\, x^{-2(\alpha_6-\alpha^{0}_6)} \; .
\end{equation}

\subsection{The L-loop ladder diagrams\label{ladder}}
Moreover, the L-loop ladder diagrams
can also be expressed as an integral kernel of the products
of operators $H_\alpha$.
We start with dimensionaly
and analytically regularized massless integrals
\be
\lb{br1}
D_L (p_0, p_{L+1}, p) =
\left[ \prod_{k=1}^{L} \int
\frac{d^D p_k}{p^{2\alpha_k}_k \, (p_k - p)^{2\beta_k}} \right]
\prod_{m=0}^{L} \frac{1}{(p_{m+1} - p_m)^{2\gamma_m}}
\ee
which correspond to the diagram
depicted in Fig.\ref{fig6} ($x = p_0$, $y = p_{L+1}$, $z =p$).
The dual to this diagram for the case
$\alpha_k = \beta_k = \gamma_k =1$
is the $L$-loop ladder diagram for massless $\phi^3$ theory
presented in Fig.\ref{fig7}.

\begin{figure}[h]
%picture 4
\unitlength=6mm
\begin{picture}(25,5)(-6,0)

\put(1,3){\line(1,0){12}}
\put(6.8,5.5){$0$}

\put(2.9,2.8){$\bullet$}
\put(5,2.8){$\bullet$}
\put(3.9,3.2){\footnotesize $\gamma_1$}
\put(1.8,3.2){\footnotesize $\gamma_0$}
\put(4,4.2){\footnotesize $\alpha_1$}
\put(4.5,1.4){\footnotesize $\beta_1$}

\put(5.8,3.2){\footnotesize $\gamma_2$}
\put(6.3,4.1){\footnotesize $\alpha_2$}
\put(6.1,2){\footnotesize $\beta_2$}
\put(7,4.1){$\cdots$}
\put(6.5,3.2){$\cdots$}
\put(7,2){$\cdots$}

\put(5.2,3){\line(3,4){1.8}}
\put(3,3){\line(5,3){4}}
\put(8.8,3){\line(-3,4){1.8}}
\put(11,3){\line(-5,3){4}}

\put(5.2,3){\line(3,-4){1.8}}
\put(3,3){\line(5,-3){4}}
\put(8.8,3){\line(-3,-4){1.8}}
\put(11,3){\line(-5,-3){4}}

\put(9.1,1.4){\footnotesize $\beta_L$}
\put(9.3,4.1){\footnotesize $\alpha_L$}
\put(9.1,3.2){\footnotesize $\gamma_{L-1}$}
\put(11.5,3.2){\footnotesize $\gamma_L$}

\put(8.6,2.8){$\bullet$}
\put(10.8,2.8){$\bullet$}

\put(-0.2,2.5){$x=p_0$}
\put(12.7,2.5){$y=p_{_{L+1}}$}
\put(6.5,0.2){$z=p$}

\put(3,2.4){\footnotesize $p_1$}
\put(4.9,2.4){\footnotesize $p_2$}
\put(8.6,2.5){\footnotesize $p_{L-1}$}
\put(11,2.5){\footnotesize $p_L$}

 %\put(9,0){\bf Fig. 6}
\end{picture}
\caption{\small\it Graphical representation of the integral
(\ref{br1}).\label{fig6}}
\end{figure}

\begin{figure}[h]
%picture 5
\unitlength=11mm
\begin{picture}(25,2.5)(-3,0.5)

\put(1,2.5){\vector(1,0){5}}
%\put(-0.2,2.9){($\beta_i$)}
\put(6,1){\vector(-1,0){5}}
%\put(-0.2,0.9){($\alpha_i$)}

\put(2,1){\vector(0,1){1.5}}
\put(3.2,1){\vector(0,1){1.5}}
\put(5,1){\vector(0,1){1.5}}
\put(3.5,1.8){$........$}

\put(0.5,2.8){\footnotesize $p_0$ - $p$}
\put(1,0.7){\footnotesize $p_0$}
\put(2.2,2.8){\footnotesize $p_1$ - $p$}
\put(2.3,0.7){\footnotesize $p_1$}
\put(5.7,2.8){\footnotesize $p_{_{L+1}}$ - $p$}
\put(6,0.7){\footnotesize $p_{_{L+1}}$}

\put(1.5,1.8){\footnotesize $p_{10}$}
\put(2.7,1.8){\footnotesize $p_{21}$}
\put(5.1,1.8){\footnotesize $p_{_{L+1\,L}}$}

\put(2.4,2.3){\scriptsize $\beta_1$}
\put(3.5,2.3){\scriptsize $\beta_2$}
\put(4,2.3){\scriptsize $.....$}
\put(4.5,2.3){\scriptsize $\beta_L$}
\put(2.4,1.1){\scriptsize $\alpha_1$}
\put(3.5,1.1){\scriptsize $\alpha_2$}
\put(4,1.1){\scriptsize $.....$}
\put(4.5,1.1){\scriptsize $\alpha_L$}
\put(2,1.4){\scriptsize $\gamma_0$}
\put(3.2,1.4){\scriptsize $\gamma_1$}
\put(4.7,1.4){\scriptsize $\gamma_L$}

\put(8.5,1.8){$(p_{mk} := p_m - p_k)$}

 %\put(5,-0.0){\bf Fig. 6}

\end{picture}
\caption{\small\it The $L$-loop ladder diagram in momentum
space for massless $\phi^3$ theory.\label{fig7}}
\end{figure}

\noindent
The integral (\ref{br1}) is written
in the following operator form \cite{Isa}
\be
\lb{br1a}
D_L (x, y, z; \alpha, \beta, \gamma) =
\left( \prod_{k=0}^L \, \frac{1}{a(\gamma_k')} \right) \,
\langle x| \frac{1}{\p^{2\gamma_0'}} \left( \prod_{k=1}^L \,
\frac{1}{\q^{2\alpha_k}} \frac{1}{(\q-z)^{2\beta_k}}
\frac{1}{\p^{2\gamma_k'}} \right)  |y\rangle \; .
\ee
Let the indices on the lines in the diagram in Fig.6 satisfy
the conformal condition
$$
\gamma_{k-1} + \alpha_k + \beta_k + \gamma_k = D \; , \;\;\;\;\;
k=1,..., L \; .
$$
Then by using (\ref{inver1}) and relation ${\cal I}'(\q-z)^2 =
\frac{z^2}{\q^2}(\q-\frac{1}{z})^2 {\cal I}'$ we obtain for
 the matrix element in the right-hand side of (\ref{br1a}):
$$
\langle x| ({\cal I}')^2 \frac{1}{\p^{2\gamma_0'}} \left( \prod_{k=1}^L
\,
\frac{1}{\q^{2\alpha_k}} \frac{1}{(\q-z)^{2\beta_k}}
\frac{1}{\p^{2\gamma_k'}} \right)  |y\rangle =
$$
$$
= \langle x| {\cal I}' \; \q^{-2\gamma_0'} \,
 \frac{1}{\p^{2\gamma_0'}} \left( \prod_{k=1}^L \,
\frac{z^{-2\beta_k}}{(\q-\frac{1}{z})^{2\beta_k}}
\frac{1}{\p^{2\gamma_k'}} \right) \; \q^{-2\gamma_L'}
\; {\cal I}' |y\rangle =
  $$
  \be
\lb{bra2}
\begin{array}{c}
\displaystyle
= x^{-2\gamma_0} y^{-2\gamma_L} z^{-2\beta_{1...L}}
\bigl\langle \frac{1}{x} \bigr| \,
 \frac{1}{\p^{2\gamma_0'}} \left( \prod_{k=1}^L \,
\frac{1}{(\q-\frac{1}{z})^{2\beta_k}}
\frac{1}{\p^{2\gamma_k'}} \right) \,\bigl| \frac{1}{y}
\bigr\rangle = \\ [0.2cm]
\displaystyle  = x^{-2\gamma_0} y^{-2\gamma_L} z^{-2\beta_{1...L}} \;
\bigl\langle u\bigr| \,
 \frac{1}{\p^{2\gamma_0'}} \left( \prod_{k=1}^L \,
\frac{1}{\q^{2\beta_k}}
\frac{1}{\p^{2\gamma_k'}} \right) \,
\bigl| w\bigr\rangle\; .
\end{array}
\ee
where $u = \frac{1}{x} -\frac{1}{z}$, $w = \frac{1}{y} -\frac{1}{z}$
and $(\frac{1}{x})_i = \frac{x_i}{x^2}$ etc. Thus, to evaluate
$L$-loop ladder diagrams (\ref{br1a}), we need to calculate the matrix
element
 \be
 \lb{bra2i}
\bigl\langle u\bigr| \,
 \frac{1}{\p^{2\gamma_0'}} \left( \prod_{k=1}^L \,
\frac{1}{\q^{2\beta_k}}
\frac{1}{\p^{2\gamma_k'}} \right) \,
\bigl| w\bigr\rangle =
\bigl\langle u\bigr| \, H_{-\gamma_0'} H^\dagger_{\gamma_0' - \beta_1}
H_{\beta_1 - \gamma'_{01}}
%H^\dagger_{\gamma_{01}' - \beta_{12}}
\cdots H_{\beta_{1...L} - \gamma_{01...L}'}
 \q^{2 \delta_{1...L}} \,
\bigl| w\bigr\rangle \; ,
\ee
where $\delta_{1...L} := \gamma_{01...L}' - \beta_{1...L}
\equiv \gamma_{0}' +\gamma_1' + ...+\gamma_L' - \beta_{1}-
\beta_{1}-...-\beta_{L}$.
Now, in view of relations (\ref{twoloop})
and (\ref{bra2i}),
the problem of evaluating
 the two-loop master-diagram and L-loop ladder
 diagrams is reduced to the spectral problem of
 the operators $H_\alpha$.

\subsection{Diagonalization of operator $H_\alpha$}

 As we mentioned above,
 the star-triangle relation (\ref{startr}) is written as
 a commutativity condition (\ref{hhhh1}),
 and this means that the operators $H_\alpha$ form a commutative set
 $[H_\alpha, \; H_\beta] = [H_\alpha, \; H^\dagger_\beta] =
 [H^\dagger_\alpha, \; H^\dagger_\beta]=0$ for all parameters
 $\alpha,\beta$.
 To investigate the spectrum of operators
 (\ref{hhhh}), (\ref{setcom}),
 we firstly note that the elements $\q^2$, $\p^2$ and dilatation
 operator
 $$
 H := \frac{i}{2}\bigl(  (\q \, \p) + (\p \, \q) \bigr)
 =i \, (\q \, \p) + D/2 \; ,
 $$
 generate the $s\ell(2)$ Lie algebra:
 \be
 \lb{sl2}
 [\q^2, \; \p^2] = 4 H \; , \;\;\;
 H\, \q^2 = \q^2 \, (H +2)\; , \;\;\;
 H\, \p^2 = \p^2 \, (H -2) \; .
 \ee
 For the dilatation operator we have
 $H \cdot f_{k,\ell}(\p,\q) = f_{k,\ell}(\p,\q) \cdot (H +\ell -k)$,
 where $f_{k,\ell}(\p,\q)$ is a monomial of order $k$ in $\p_\mu$
 and of order $\ell$ in $\q_\nu$. Thus,
 the dilatation operator $H=-H^\dagger$ (the Cartan element for
 the Lie algebra $s\ell(2)$) commutes with
 all operators (\ref{hhhh}), (\ref{setcom}).
 The quadratic Casimir operator
 for the $s\ell(2)$ algebra (\ref{sl2}):
 \be
 \lb{cas2}
 \begin{array}{c}
C_{(2)} : = \frac{1}{2} (\p^2 \, \q^2 + \q^2 \, \p^2) +  H^2
= (\p^2 \q^2 + 2H + H^2)  = (H_1 + (H + 1)^2 - 1) \; ,
\end{array}
\ee
 is related to the
 operator $H_1$ from the set (\ref{hhhh}).
Further we need the following useful relations
which generalize (\ref{sl2}):
\be
 \lb{sl2b}
 H\, \q^{2\alpha} = \q^{2\alpha}  \, (H +2\alpha)\; ,
 \;\;\;\;\;\;\;
 H\, \p^{2\alpha} = \p^{2\alpha} \, (H -2\alpha) \; ,
 \ee
\be
\lb{is1}
 \p^{2 \beta} \, \q^2 =  \q^2 \, \p^{2\beta} -
4  \beta \, ( H + \beta - 1)\, \p^{2(\beta -1)} \; ,
 %\equiv \q^2 \, \p^{2\beta} -
 %4 \, \beta \, \p^{2(\beta -1)}  ( H - \beta + 1) \; ,
\ee
\be
\lb{is2}
 \q^{2 \beta} \, \p^{2} = \p^{2} \, \q^{2 \beta} +
4  \beta \, ( H - \beta + 1) \, \q^{2(\beta -1)} \; .
\ee

 %\vspace{0.2cm}
 %\noindent
 %{\bf Proposition 1.}

\begin{proposition}\label{prop1}{\it Let
$|\psi_{j,\nu} \rangle$ be common eigenvectors
 of the operators $H_\beta$ (\ref{hhhh}):
 \be
 \lb{spect01}
 H_\beta \; |\psi_{j,\nu} \rangle =
\tau_{j,\nu}(\beta) \; |\psi_{j,\nu} \rangle \; ,
\ee
where $\tau_{j,\nu}(\beta)$ are the corresponding eigenvalues.
We numerate  $|\psi_{j,\nu} \rangle$
by two real numbers $\nu$ and $j$ which are respectively
fixed by the eigenvalues of the Cartan element $H=-H^\dagger$
 and quadratic Casimir operator $C_{(2)}$:
\be
 \lb{is5d}
H \, |\psi_{j,\nu} \rangle = -2i \nu |\psi_{\nu,j} \rangle \; ,
\;\;\;\;
C_{(2)} \, |\psi_{j,\nu} \rangle = 4j(j - 1)\, |\psi_{j,\nu} \rangle \;
.
 \ee
Then, the eigenvalues $\tau_{j,\nu}(\beta)$ in
(\ref{spect01}) are
\be
 \lb{is5}
\tau_{j,\nu}(\beta) = 4^\beta \;
\frac{\Gamma(j+\beta - i\nu)\, \Gamma(j +i\nu)}{
\Gamma(j-\beta + i\nu)\, \Gamma(j - i\nu)} \; .
 \ee
 }
\end{proposition}
{\bf Proof.} Relations (\ref{is1}), (\ref{is2}) are
 written as identities
 \be
 \lb{is3}
 \begin{array}{c}
H_\beta = \Bigl(H_{1}  - 4 (\beta - 1) (H +\beta) \Bigr)
 H_{\beta-1} \;\;\; \Rightarrow  \\ [0.3cm]
 H_\beta = \Bigl(C_{2} -H^2-2H - 4 (\beta - 1) (H +\beta) \Bigr)
 H_{\beta-1} \; ,
 \end{array}
 \ee
and the last identity gives the functional equation
for the eigenvalues $\tau_{j,\nu}(\beta)$ of the operators $H_\beta$.
Indeed, from (\ref{spect01}), (\ref{is5d}) and identity
 (\ref{is3}) we deduce the equation
 \be
 \lb{is4}
\tau_{j,\nu}(\beta) = 4 \bigl(j +  i \nu - \beta \bigr)
\bigl(j-1  + \beta - i \nu \bigr) \tau_{j,\nu}(\beta-1) \; ,
 \ee
which is solved as
$$
\tau_{j,\nu}(\beta) = 4^\beta \frac{\Gamma(j+\beta - i\nu)}{
\Gamma(j-\beta +i\nu)} \; c(j,\nu) \; ,
$$
where $c(j,\nu)$ is a function independent of $\beta$.
This function is fixed by obvious condition $\tau_{j,\nu}(0)=1$
and we finally obtain (\ref{is5}). \hfill \qed

\vspace{0.1cm}

\noindent
{\bf Consequence.}
Equation (\ref{is4}) gives for $\beta=1$ the equality
 \be
 \lb{is6}
 \tau_j(1) = 4 \bigl(j - i \nu \bigr) \bigl(j +  i \nu - 1 \bigr) \; ,
 \ee
which is used below.

As it is seen from the examples of subsections
{\bf \ref{master}, \ref{ladder}},
we deal with the
algebra ${\cal U}$ of polynomials which are generated by the operators
$H$, $\q^{2\alpha}$ and $\p^{2\beta}$, i.e. the basis elements
in ${\cal U}$ are
\be
 \lb{eee}
E_{\alpha_1,...,\alpha_k;\beta_1,...,\beta_k} =
\p^{2\alpha_1}\q^{2\beta_1}\p^{2\alpha_2}\q^{2\beta_2}
\cdots \p^{2\alpha_k}\q^{2\beta_k} \; , \;\;\;\;\; k=0,1,2,3,...
 \ee
and any element of ${\cal U}$ is written as a formal
series
 \be
 \lb{aaa}
a = \sum_{k=0}^\infty \phi_{k}(H) \;
E_{\alpha_1,...,\alpha_k;\beta_1,...,\beta_k} \; ,
 \ee
where $\phi_{k}(H)$ are functions in $H$. The
 operator $C_2$ defined in (\ref{cas2})
commutes with any element (\ref{aaa}) in ${\cal U}$.
Indeed, the Casimir operator $C_2$ commutes
with an arbitrary function $\phi_{k}(H)$ of the Cartan element $H$
as well as with the operators $\q^{2\alpha}$ and $\p^{2\beta}$
 $(\forall \alpha,\beta)$.
Note that the basis elements (\ref{eee})
 are written in the form
\be
\lb{hhh}
\begin{array}{c}
 E_{\alpha_1,...,\alpha_k;\beta_1,...,\beta_k} =
H_{\alpha_1} \, H_{\beta_1-\alpha_1}^\dagger
\, H_{\alpha_{12}-\beta_1} \, H_{\beta_{12}-\alpha_{12}}^\dagger
\cdots H_{\alpha_{1...k}-\beta_{1...(k-1)}} \,
\q^{2(\beta_{1...k}-\alpha_{1...k})}  = \\ [0.2cm]
 = H_{\alpha_1} \, H_{\alpha_1-\beta_1}^{-1}
\, H_{\alpha_{12}-\beta_1} \, H_{\alpha_{12}-\beta_{12}}^{-1}
\cdots H_{\alpha_{1...k}-\beta_{1...(k-1)}} \,
\q^{2(\beta_{1...k}-\alpha_{1...k})} \; ,
\end{array}
\ee
where the commutative operators $H_\alpha$ and
 $H_\beta^\dagger$ are defined in (\ref{hhhh}),
(\ref{setcom}), and we use concise notation
$\alpha_{1...\ell}= \alpha_1 + ... +\alpha_\ell$.
Equation (\ref{hhh}) means that for $\beta_{1...k}=\alpha_{1...k}$
 an alternating product of
$\q^{2\alpha}$ and $\p^{2\beta}$ in (\ref{eee})
commutes with any element of the set (\ref{hhhh}),
(\ref{setcom}).

In the paper \cite{DerShum}, a complete set of orthogonal common
eigenvectors $| \psi^{\mu_1 ... \mu_n}_\nu \rangle$ for
 the operators (\ref{hhhh}),
(\ref{setcom}) was explicitly constructed:
\be
\lb{eigfun}
\langle x | \psi^{\mu_1 ... \mu_n}_\nu \rangle =
\frac{x^{\mu_1 ... \mu_n}}{x^{2(D/4+n/2+i \nu)}}  \; .
 \;\;\;\;\;\;\;
\ee
These eigenfunctions satisfy the following relations
 \be
\lb{eigfun1}
\sum_{n=0}^\infty \; \mu(n)
\int\limits_{-\infty}^{+\infty} d\nu \;
| \psi^{\mu_1 ... \mu_n}_\nu \rangle
\langle \psi^{\mu_1 ... \mu_n}_\nu | \; = \; I \; ,
\;\;\;\;\; \mu(n):=
  \frac{2^{n-1}\Gamma(D/2+n)}{\pi^{D/2+1} \, n!}\; ,
\ee
 %\marginpar{\bf \tiny Is.03.01.23 Вставил новую формулу}
\be
\lb{spect}
 \begin{array}{c}
 \displaystyle
 \p^{2\alpha} \, \q^{2\alpha} \cdot
 \frac{x^{\mu_1 ... \mu_n}}{x^{2\beta}} = 4^\alpha \,
 \frac{\Gamma(D/2-\beta+\alpha+n)\, \Gamma(\beta)}{
 \Gamma(\beta-\alpha)\,  \Gamma(D/2-\beta+n)} \;
 \frac{x^{\mu_1 ... \mu_n}}{x^{2\beta}} \;\;\;
  %\;\;\;\;\;
  %a_n(w) := \frac{\Gamma(D/2-w+n)}{\Gamma(w)} \;
 \Rightarrow  \\ [0.4cm]
H_\alpha \, | \psi^{\mu_1 ... \mu_n}_\nu \rangle =
\tau_{n,\nu}(\alpha) \,  | \psi^{\mu_1 ... \mu_n}_\nu \rangle  \; ,
\;\;\;\;\;\;\;
H_\alpha^\dagger \, | \psi^{\mu_1 ... \mu_n}_\nu \rangle =
\tau_{n,-\nu}(\alpha) \,  | \psi^{\mu_1 ... \mu_n}_\nu \rangle  \; ,
\end{array}
\ee
\be
\lb{spect2}
 H \, | \psi^{\mu_1 ... \mu_n}_\nu \rangle = - 2i\nu \,
 | \psi^{\mu_1 ... \mu_n}_\nu \rangle \; , \;\;\;\;\;\;\;
 {\cal I}' \, | \psi^{\mu_1 ... \mu_n}_\nu \rangle  =
 | \psi^{\mu_1 ... \mu_n}_{-\nu} \rangle \; ,
 %\end{array}
\ee
 \be
 \lb{funct}
\tau_{n,\nu}(\alpha) = 4^\alpha \frac{a_n(D/4 + n/2-\alpha+i\nu)}{
a_n(D/4 + n/2 + i \nu)}   \, , \;\;\;\;\;\;
 a_n(w) := \frac{\Gamma(D/2-w+n)}{\Gamma(w)} \, ,
 \ee
 \be
\lb{orthog}
 \langle \psi^{\mu_1 ... \mu_n}_\nu |
 \psi^{\nu_1 ... \nu_m}_{\nu'} \rangle =
\frac{\pi^{D/2+1} n!}{2^{n-1} \Gamma(D/2+n)}
\delta(\nu - \nu') \delta_{nm}
P^{\mu_1 ... \mu_n}_{\nu_1 ... \nu_m}  \; ,
 \ee
where $x^{\mu_1 ... \mu_n} = P^{\mu_1 ... \mu_n}_{\nu_1 ... \nu_m}
 \; x^{\nu_1} \cdots x^{\nu_n}$ is the traceless
symmetric homogeneous polynomial in $x^{\mu}$ and
$P^{\mu_1 ... \mu_n}_{\nu_1 ... \nu_m}$ is
the projector on such polynomials. We stress
that the key spectral
formulas (\ref{spect}) are operator versions of
the well known relation~\cite{AN,KT}
\begin{equation}\label{dchain}
\int d^{D} x \dfrac{1}{\left(y - x\right)^{2\alpha}}
\dfrac{x^{\mu_1\ldots\mu_n}}{x^{2\beta}} =
\pi^{D/2}\dfrac{a_n\left(\beta\right)a_0\left(\alpha\right)}{
a_n{\left(\alpha + \beta - D/2\right)}}\dfrac{y^{\mu_1\ldots\mu_n}}{
y^{2\left(\alpha + \beta - D/2\right)}},
\end{equation}
where $a_n\left(\alpha\right)$ is defined in (\ref{funct}).
The detailed information
 about the projector
$P^{\mu_1 ... \mu_n}_{\nu_1 ... \nu_m}$ and the proof of
the orthogonality condition (\ref{orthog}) for
 the eigenvectors $| \psi^{\nu_1 ... \nu_m}_{\nu} \rangle$
 are given in Appendix {\bf A} (see Section {\bf \ref{project}}).
 The convolution product of
the symmetrized polynomials has the form
 (see \cite{DerShum} and Appendix {\bf A})
 \be
 \lb{conv}
x^{\mu_1 ...\mu_n}x^{\mu_1 ...\mu_n} =
\frac{\Gamma(n+D-2)\Gamma(D/2-1)}{
2^n \Gamma(n+D/2-1)\Gamma(D-2)} x^{2n} \; .
 \ee
 We stress that the orthogonality condition (\ref{orthog})
 is consistent with the resolution of unity in
 (\ref{eigfun1}).

 Note that
 for the quadratic Casimir operator (\ref{cas2}) we have
 (cf. (\ref{is5d}))
 $$
 C_{(2)} \; | \psi^{\mu_1 ... \mu_n}_\nu \rangle =
 \frac{1}{4} (D + 2n) (D + 2n-4)
  \; | \psi^{\mu_1 ... \mu_n}_\nu \rangle
 \;\;\;\;\;\; \Rightarrow \;\;\;\;\;\;
 j=\frac{D}{4} + \frac{n}{2} \; ,
 $$
 and, for these $j$, the eigenvalue $\tau_{n,\nu}(\beta)$ of $H_\beta$,
 given in (\ref{funct}),
 is equal to the function (\ref{is5}) from Proposition {\bf
 \ref{prop1}}.
 This fact indicates that vectors (\ref{eigfun})
 could form a complete set since the statement of
  Proposition {\bf \ref{prop1}}
   gives the general form for any
  possible eigenvalues for $H_\beta$. The proof of completeness is
  given in \cite{DerShum}.
 Then, we note that there are useful identities for
 functions (\ref{funct}):
 \be
 \lb{relfun}
 a_n(u +n/2) \; a_n(u' +n/2) = 1 \;\;\;\;
 \Rightarrow \;\;\;\; \tau_{n,\nu}(u) =  \frac{1}{\tau_{n,-\nu}(-u)}
 \;\; \Leftrightarrow \;\;
 H_u \cdot H^\dagger_{-u} = 1 \, ,
 \ee
 where $u' = D/2-u$, and for $u,\nu \in \mathbb{R}$ we have
 $\tau_{n,-\nu}(u) = \tau_{n,\nu}^*(u)$.
 %Equation (\ref{orthog}) is the orthogonality condition for
 %eigenvectors $| \psi^{\nu_1 ... \nu_m}_{\nu} \rangle$.

   \noindent
 {\bf Remark 1.} The second equation in (\ref{spect})
 follows from identities (\ref{relfun}).
The second relation in (\ref{spect2}) is
 proven as
 $$
 \begin{array}{c}
\langle x |\,  {\cal I}^{\prime} \,| \psi^{\mu_1 ... \mu_n}_\nu \rangle
=
x^{2(-\frac{D}{2})} \; \langle \frac{1}{x} |
 \psi^{\mu_1 ... \mu_n}_\nu \rangle = \displaystyle
 \frac{x^{\mu_1 ... \mu_n}}{x^{2(D/4+n/2-i \nu)}} =
 \langle x | \psi^{\mu_1 ... \mu_n}_{-\nu} \rangle \; .
 \end{array}
 $$
  \noindent
 {\bf Remark 2.} The Gegenbauer polynomial technique \cite{CKT}
 (see also \cite{Kot}) is based on the identity \cite{DerShum}
 \be
 \lb{gegenb}
 \begin{array}{c}
 \langle x | \p^{-2\alpha'} | y \rangle =
 \sum\limits_{n=0}^\infty \mu(n)
 \int\limits_{-\infty}^{+\infty} d\nu
\langle x | \q^{2\alpha'} \, H^\dagger_{-\alpha'}
| \psi^{\mu_1 ... \mu_n}_\nu \rangle
\langle \psi^{\mu_1 ... \mu_n}_\nu | y \rangle = \\ [0.2cm]
 = \sum\limits_{n=0}^\infty
 \mu(n) \int\limits_{-\infty}^{+\infty} d\nu \;
 \tau_{n,-\nu}(\alpha-D/2) \; x^{2\alpha'}
\langle x | \psi^{\mu_1 ... \mu_n}_\nu \rangle
\langle \psi^{\mu_1 ... \mu_n}_\nu | y \rangle = \\ [0.2cm]
 = 2^{2\alpha - D}
 \sum\limits_{n=0}^\infty \mu(n)
 \int\limits_{-\infty}^{+\infty} d\nu \;
 \displaystyle \frac{a_n(\frac{3D}{4}+\frac{n}{2}-\alpha-i\nu)}{
 a_n(\frac{D}{4}+\frac{n}{2}-i\nu)} \;
 \frac{x^{\mu_1 ...\mu_n} \; y^{\mu_1 ...\mu_n}}
 {x^{2(n/2+i\nu +\alpha -D/4)} y^{2(D/4 + n/2-i\nu)}} \; ,
\end{array}
 \ee
 where in the right-hand side of this chain of relations
 one can make a shift of the coordinates
 $x,y \to (x-z),(y-z)$
 with an arbitrary vector  $z \in \mathbb{R}^D$ since the
 left-hand side of (\ref{gegenb}) is invariant under this shift. \\

 \noindent
 {\bf Remark 3.} Below we also need the following matrix element
 \be
 \lb{gori2}
\begin{array}{c}
 \langle \psi^{\mu_1 ... \mu_n}_\nu |
\q^{2(\frac{3D}{2} - \alpha_{1...6})}
| \psi^{\mu_1 ... \mu_n}_\nu  \rangle =
 \int d^Dx \langle \psi^{\mu_1 ... \mu_n}_\nu |
\q^{2(\frac{3D}{2} - \alpha_{1...6})}
|x \rangle
\langle x | \psi^{\mu_1 ... \mu_n}_\nu  \rangle =
\\ [0.2cm]
= \int d^Dx
\frac{x^{\mu_1 ... \mu_n} x^{\mu_1 ... \mu_n}}{
x^{2(D/2+n)}}
x^{2(\frac{3D}{2} - \alpha_{1...6})} =
 \frac{\Gamma(n+D-2)\Gamma(D/2-1)}{
2^n \Gamma(n+D/2-1)\Gamma(D-2)}
\int d^Dx \frac{1}{x^{2(\alpha_{1...6} - D)}} = \\ [0.2cm]
= \frac{\Gamma(n+D-2)\Gamma(D/2-1)\pi \Omega_D}{
2^n \Gamma(n+D/2-1)\Gamma(D-2)}
\delta(\alpha_{1...6} - \frac{3D}{2})
= \frac{\Gamma(n+D-2)\pi^{D/2+1}}{
2^{n-2} \Gamma(n+D/2-1)\Gamma(D-1)}
\delta(\alpha_{1...6} - \frac{3D}{2}) \; .
\end{array}
 \ee
The evaluation of (\ref{gori2})
illustrates the application of the formula (\ref{gori}). Here
we also use (\ref{eigfun}) and
 identity (\ref{conv}).

\subsection{Answer for the two loop master-diagram}

Expression (\ref{twoloop})
for the vacuum diagram in Fig.\ref{fig5} is represented
in the form
$$
\frac{1}{a(\alpha'_1) a(\alpha'_4)
a(\alpha'_5)} \sum_{n=0}^\infty \mu(n)
 \int\limits_{-\infty}^{+\infty} d\nu \,
 \langle \psi^{\mu_1 ... \mu_n}_\nu |
 \p^{-2\alpha_1'} \q^{-2\alpha_2}
\p^{-2\alpha_5'} \q^{-2\alpha_3} \p^{-2\alpha_4'}
\q^{-2\alpha^{0}_6}
\q^{-2(\alpha_6 -\alpha^{0}_6)}
| \psi^{\mu_1 ... \mu_n}_\nu  \rangle =
$$
$$
\begin{array}{c}
= \frac{1}{a(\alpha'_1) a(\alpha'_4)
a(\alpha'_5)} \sum\limits_{n=0}^\infty
 \mu(n) \int\limits_{-\infty}^{+\infty} d\nu \,
\tau_{n,\nu}(\alpha_1-\frac{D}{2})
\tau_{n,\nu}^*(\frac{D}{2}-\alpha_{12})
\tau_{n,\nu}(\alpha_{125} - D)
\tau_{n,\nu}^*(D-\alpha_{1235})
\tau_{n,\nu}(-\alpha^0_{6}) \cdot \\ [0.2cm]
\cdot \; \langle \psi^{\mu_1 ... \mu_n}_\nu |
\q^{2(\frac{3D}{2} - \alpha_{1...6})}
| \psi^{\mu_1 ... \mu_n}_\nu  \rangle  =
\end{array}
$$
 \be
 \lb{2loop}
\begin{array}{rl}
 \displaystyle
= \frac{1}{a(\alpha'_1) a(\alpha'_4)
a(\alpha'_5)} & \displaystyle
\sum\limits_{n=0}^\infty \mu(n) \;
\frac{\Gamma(n+D-2)\Gamma(D/2-1)\pi \Omega_D}{
2^n \Gamma(n+D/2-1)\Gamma(D-2)} \;\; \times \\ [0.3cm]
 \times & \displaystyle
 \int\limits_{-\infty}^{+\infty} d\nu \,
\frac{\tau_{n,\nu}(\alpha_1-\frac{D}{2})
\tau_{n,\nu}(\alpha_{125} - D)\; \tau_{n,\nu}(-\alpha^0_{6})}{
\; \tau_{n,\nu}(\alpha_{12}-\frac{D}{2})
\; \tau_{n,\nu}(\alpha_{1235}-D)}
\; \delta(\alpha_{1...6} - \frac{3D}{2})  \; ,
\end{array}
 \ee
where $\mu(n)$ is defined in (\ref{eigfun1});
in the last equality, we use identities (\ref{gori2}),
(\ref{relfun}) and $\tau_{n,-\nu}(u) = \tau_{n,\nu}^*(u)$.
 Comparing the expression in Fig.\ref{fig3} with
eq. (\ref{2loop}), we deduce the formula
 \be
 \lb{2loop2}
 \begin{array}{c}
  C(\alpha_1,\alpha_2,\alpha_6;
 \alpha_3,\alpha_4,\alpha_5) =
\frac{\Gamma(D/2-1)/\Gamma(D-2)}{a(\alpha'_1) a(\alpha'_4)
a(\alpha'_5)} \sum\limits_{n=0}^\infty \; \mu(n) \,
\frac{\Gamma(n+D-2)}{
2^n \Gamma(n+D/2-1)} \; \times \\ [0.3cm]
 \times \; \int\limits_{-\infty}^{+\infty} d\nu \,
\frac{\tau_{n,\nu}(\alpha_1-\frac{D}{2})
\tau_{n,\nu}(\alpha_{125} - D)\; \tau_{n,\nu}(-\alpha^0_{6})}{
\; \tau_{n,\nu}(\alpha_{12}-\frac{D}{2})
\; \tau_{n,\nu}(\alpha_{1235}-D)}  \; .
 \end{array}
 \ee

\vspace{0.2cm}

We note that the infinite sum (\ref{2loop2})
of the Mellin-Barnes integrals is invariant under
 the $S_6 \times \mathbb{Z}_2$ symmetry discovered in \cite{Broad},
 \cite{Broad1} (see also \cite{Isa}). This $S_6 \times \mathbb{Z}_2$
 symmetry
 is generated by the tetrahedral symmetry (\ref{tesym})
  and by additional symmetry (\ref{sttr2}) which is produced by
 the star-triangle relation.

  \vspace{0.2cm}

  The example of calculation of the
  coefficient function $C(\alpha_1,\alpha_2,\alpha_6;
 \alpha_3,\alpha_4,\alpha_5)$ in (\ref{2loop2})
  for a special case of indices
  $$
  \alpha_1 = \alpha_4 = \alpha_5 = D/2 -1 \; , \;\;\;\;
  \alpha_2 = \beta \; , \;\;\;\;
  \alpha_3 = \alpha \; ,
  $$
  is given in Appendix {\bf B} (see Section {\bf \ref{TkaChe}}).
   Analytical answer for
   the coefficient function
   (this result was firstly obtained in \cite{ChTk2}
   by means of the integration-by-parts method)
   is related to the case of the dual master two-loop diagram,
   Fig.\ref{fig3}) when three indices on the
   lines are fixed
   $\alpha_1=\alpha_2=\alpha_5 =1$ and indices
   $\alpha_3,\alpha_4$ are arbitrary.

 \vspace{0.2cm}

\noindent
{\bf Remark 4.} One can obtain from eq. (\ref{2loop2})
the known results
for 2-loop master diagrams in Fig.\ref{fig3},
 when  $\alpha_1=\alpha_2=\alpha_3=\alpha_4=1$
 (see e.g. \cite{Kot}, \cite{KT} and references therein),
or say $\alpha_3=\alpha_4=\alpha_5=1$,
which is exactly evaluated by the
integration-by-parts method and represented as
a finite sum (see \cite{Tkach}, \cite{TChT}).
 A rather general result for the 2-loop master integral (\ref{2loop2})
was obtained in the paper \cite{Mikh}.
\vspace{0.2cm}

\noindent
 %{\bf Problem 2.}
For physical applications one has to know the
explicit expansion of
function (\ref{2loop2}) in the limit $\epsilon \to 0$,
when $D=4-2\epsilon$,
$\alpha_i = n_i + m_i \epsilon$ $(n_i,m_i \in \mathbb{Z})$.

\subsection{Answer for the L-loop ladder diagram}

To evaluate
$L$-loop ladder diagrams (\ref{br1a}), we need to calculate the matrix
element in (\ref{bra2}), (\ref{bra2i}):
$$
\bigl\langle u\bigr| \,
 \frac{1}{\p^{2\gamma_0'}} \left( \prod_{k=1}^L \,
\frac{1}{\q^{2\beta_k}}
\frac{1}{\p^{2\gamma_k'}} \right) \,
\bigl| w\bigr\rangle =
\bigl\langle u\bigr| \, H_{-\gamma_0'} H^\dagger_{\gamma_0' - \beta_1}
H_{\beta_1 - \gamma'_{01}} H^\dagger_{\gamma_{01}' - \beta_{12}}
\cdots H_{\beta_{1...L} - \gamma_{01...L}'}
 \q^{2(\gamma_{01...L}' - \beta_{1...L})} \,
\bigl| w\bigr\rangle =
$$
$$
= w^{2(\gamma_{01...L}' - \beta_{1...L})}
\sum_{n=0}^\infty \mu(n) \int\limits_{-\infty}^{+\infty} d\nu
\bigl\langle u\bigr| \, H_{-\gamma_0'} H^\dagger_{\gamma_0' - \beta_1}
H_{\beta_1 - \gamma'_{01}}
 %H^\dagger_{\gamma_{01}' - \beta_{12}}
\cdots H_{\beta_{1...L} - \gamma_{01...L}'}
| \psi^{\mu_1 ... \mu_n}_\nu \rangle
\langle \psi^{\mu_1 ... \mu_n}_\nu
\bigl| w\bigr\rangle =
$$
\be
\lb{Lloop}
= w^{2(\gamma_{01...L}' - \beta_{1...L})}
\sum_{n=0}^\infty \mu(n) \int\limits_{-\infty}^{+\infty} d\nu
\frac{\tau_{n,\nu}(-\gamma_0')\, \tau_{n,\nu}(\beta_1 -
\gamma'_{01})\cdots }{
\tau_{n,\nu}(\beta_1-\gamma_0')\, \tau_{n,\nu}(\beta_{12}-\gamma_{01}')
\cdots}
\left( \frac{u^{\mu_1 ... \mu_n} w^{\mu_1 ... \mu_n}}{
u^{2(D/4+n/2+i\nu)} w^{2(D/4+n/2-i\nu)}}\right)  \; ,
\ee
where $\gamma'_{01...k} = \gamma'_0 + \gamma'_1 + ... +\gamma'_k =
(k+1) \frac{D}{2} - \gamma_{01...k}$ and $\mu(n)$
is defined in (\ref{eigfun1}).
 For physical applications one has to know the
infinite sum of
Mellin-Barnes type integrals (\ref{Lloop}) for $D=4-2\epsilon$,
$\alpha_i = n_i + m_i \epsilon$ $(n_i,m_i \in \mathbb{Z})$
and an explicit expansion of (\ref{Lloop})
in the limit $\epsilon \to 0$.

\vspace{0.3cm}

In the case $\gamma_k' = \beta_k = \beta$,
we write the generating function for matrix elements in
the left-hand side of (\ref{Lloop}) as follows:
$$
\begin{array}{c}
\bigl\langle u\bigr| \, \frac{1}{\p^{2\beta} -
g \, \q^{-2\beta}} \,
\bigl| w\bigr\rangle = \bigl\langle u\bigr| \q^{2\beta} \,
\frac{1}{\p^{2\beta} \q^{2\beta}-
g \, } \, \bigl| w\bigr\rangle = u^{2\beta} \,
\bigl\langle u\bigr| \frac{1}{H_{\beta}- g \, } \,
\bigl| w\bigr\rangle = \\ [0.2cm]
\displaystyle
= u^{2\beta} \,
\sum\limits_{n=0}^\infty \mu(n)
\int\limits_{-\infty}^{+\infty} d\nu
\bigl\langle u\bigr| \frac{1}{(H_{\beta}- g)} \,
| \psi^{\mu_1 ... \mu_n}_\nu \rangle
\langle \psi^{\mu_1 ... \mu_n}_\nu
\bigl| w\bigr\rangle = \\ [0.2cm]
\displaystyle = u^{2\beta} \,
\sum\limits_{n=0}^\infty \mu(n)
  \int\limits_{-\infty}^{+\infty} d\nu
\frac{1}{(\tau_{n,\nu}(\beta)- g)} \, \frac{
u^{\mu_1 ... \mu_n}\, w^{\mu_1 ... \mu_n}}{
(u^2\, w^2)^{(D/4+n/2)}(u^2/w^2)^{i\nu}} =  \\ [0.2cm]
\displaystyle =
u^{2\beta} \,
\sum\limits_{n=0}^\infty \mu(n) \, \frac{
u^{\mu_1 ... \mu_n}\, w^{\mu_1 ... \mu_n}}{
(u^2\, w^2)^{(D/4+n/2)}}
\int\limits_{-\infty}^{+\infty} d\nu
 \, \frac{(w^2/u^2)^{i\nu}}{(\tau_{n,\nu}(\beta)- g)}
\end{array}
$$
Then for $\beta=1$ we have Green's
function for conformal quantum mechanics
$$
\bigl\langle u\bigr| \, \frac{1}{\p^{2} -
g \, \q^{-2}} \,\bigl| w\bigr\rangle =
u^{2} \,
\sum\limits_{n=0}^\infty \mu(n) \, \frac{
u^{\mu_1 ... \mu_n}\, w^{\mu_1 ... \mu_n}}{
(u^2\, w^2)^{(D/4+n/2)}}
\int\limits_{-\infty}^{+\infty} d\nu
 \, \frac{(w^2/u^2)^{i\nu}}{(\tau_{n,\nu}(1)- g)} =
$$
\be
\lb{grn0}
 = u^2 \, \sum_{L=0}^\infty \; g^L \sum\limits_{n=0}^\infty \mu(n)
 \, \frac{ u^{\mu_1 ... \mu_n}\, w^{\mu_1 ... \mu_n}}{
 (u^2\, w^2)^{(D/4+n/2)}}
 \int\limits_{-\infty}^{+\infty} d\nu \;
 \frac{(w^2/u^2)^{i\nu}}{\bigl(\tau_{n,\nu}(1)\bigr)^{L+1}} \; ,
\ee
where we expand the function $1/(\tau_{n,\nu}(1)- g)$ over $g$.
 Now we take into account relation (\ref{is6}):
 $\tau_{n,\nu}(1) =
 4(\frac{D}{4}+\frac{n}{2}-i \nu)(\frac{D}{4}+\frac{n}{2}+i \nu-1)$
and write
 expansion (\ref{grn0}) over $g$ in the form
 \be
\lb{grn2}
\langle u| \, \frac{1}{(\p^{2} - g/\q^{2})} \, |w \rangle =
 \sum^\infty_{L=0}
\frac{1}{L!} \left({g \over 4} \right)^L \, \Phi_L(u,w)
\ee
where for $\Phi_L(u,w)$ we have the representation
\be
\lb{grn5}
\Phi_L(u,w) = \frac{u^2 L!}{4}
\sum\limits_{n=0}^\infty \mu(n) \frac{
u^{\mu_1 ... \mu_n}\, w^{\mu_1 ... \mu_n}}{
(u^2\, w^2)^{(D/4+n/2)}}
\int\limits_{-\infty}^{+\infty}
  \frac{d\nu \;\; (w^2/u^2)^{i\nu}}{
 \Bigl((\frac{D}{4}+\frac{n}{2}-i \nu)
 (\frac{D}{4}+\frac{n}{2}+i \nu-1)\Bigr)^{L+1}} \, .
\ee

Note that the function $\Phi_L(u,w)$ is related to the
4-point $L$-loop ladder integral (\ref{br1}),
(\ref{br1a}) with fixed indices
$\alpha_k=\beta_k=\gamma_k'=1$ of the propagators.
This relation is given by the formula
(see (\ref{bra2}) and (\ref{grn2}))
\be
\lb{grn4}
D_L(x,y,z;1,1,D/2-1) \; = \; \frac{(x^2y^2)^{(1-D/2)}z^{-2L}}{
L!\, (4 \, a(1))^L}
\Phi_L\Bigl(\frac{1}{x}-\frac{1}{z},
\frac{1}{y}-\frac{1}{z} \Bigr) \; ,
 \ee
where $x=p_0$, $y=p_{L+1}$, $z=p$. We recall
that in \cite{Isa} the expansion
(\ref{grn2}) (for any $D$) was also considered and
another integral representation
for the coefficients $\Phi_L(u,w)$ was deduced:
  \be
\lb{grn3}
\Phi_L(u,w) = \frac{a(1)}{(L!)}
\int^\infty_0 \, dt \, t^L \,
\left[ \log \left( \frac{u^2}{w^2}\right) + t\right]^L \,
\partial_t \left( \frac{e^{-t}\;\;\;}{
 (u -  e^{-t}w)^{2}}\right)^{({D \over 2}-1)} \; ,
 \ee
 Then we reproduced \cite{Isa} by means of
 (\ref{grn3}) for $D=4$  the famous  result \cite{DU1}:
\begin{equation}
\label{d4}
    \Phi_{L}(u,w) = \dfrac{1}{4\pi^2}\dfrac{1}{u^2}
    \dfrac{1}{(z - \bar{z})}\sum_{k = 0}^{L}
    \dfrac{(-1)^k(2L - k)!}{k!(L -
    k)!}\operatorname{Log}^k(z\bar{z})\left[\operatorname{Li}_{2L -
    k}\left(z\right) - \operatorname{Li}_{2L -
    k}\left(\bar{z}\right)\right]  \; ,
\end{equation}
$$
z + \bar{z} = \dfrac{2\, u\cdot w}{u^2 } \; , \;\;\;\;\;
  z\bar{z} = \dfrac{w^2}{u^2} \; ,
$$
which is needed, in view of (\ref{grn4}),
for explicit evaluation of ladder diagrams.
Formula (\ref{d4}) was
intensively applied to the evaluation of planar
amplitudes in $D=4$ conformal field theories and,
in particular, in the $N=4$ supersymmetric Yang-Mills theory
(see, e.g. \cite{AlRo},\cite{DKS},\cite{DolOsb1}
and references therein).

 By making use (\ref{grn2}),
 we easily find the first coefficient
 $\Phi_0=a(1) (u-w)^{2(1-D/2)}$ (it is
 instructive for $D=4$ to deduce this coefficient
 from (\ref{d4}))
 and find the symmetry of $\Phi_L(u,w)$ for all $L$
$$
\Phi_L(u,w) = \Phi_L(w,u) =  (u^2 w^2)^{(1 - {D \over 2})} \,
\Phi_L \Bigl({1 \over u}, {1 \over w}\Bigr)  \; .
$$

 %\add{Der 17.01 Я вставил подправленные
 %вычисления интеграла по вычетам, которые присылал Леня.}

Here we show, as an example of the
effectiveness of the proposed methods, that the
integral in (\ref{grn5}) gives the same result (\ref{d4}).
The integral in (\ref{grn5})
is calculated by residues. By condition $u^2 > w^2$ the contour
can be closed in a lower half-plane and it remains to calculate
residue
at the $(L+1)$-order pole $i\nu = D/4 + n/2$
\begin{align*}
& Res =
\dfrac{1}{L!}\left(\dfrac{d}{d\nu}\right)^L\left[\dfrac{(u^2/w^2)^{-i\nu}}{(\nu
- i(D/4 + n/2 - 1))^{L +1}}\right]\bigg|_{\nu = -i(D/4 + n/2)} = \\
 %&\dfrac{1}{L!}\sum_{k = 0}^{L}\binom{L}{k}(-i)^k
 %\dfrac{\operatorname{Log}^{k}(u^2/w^2)}{
 %(u^2/w^2)^{D/4 + n/2}}(-1)^{L - k}
 %\dfrac{(2L - k)!}{L!}\dfrac{i^{2L - k + 1}}{
 %(D/2 + n - 1)^{2L - k + 1}} = \\
& = \dfrac{i}{L!}\sum_{k = 0}^{L}\binom{L}{k}\dfrac{(2L -
k)!}{L!}(u^2/w^2)^{-D/4 - n/2}
\dfrac{\operatorname{Log}^k(u^2/w^2)}{
(D/2 + n - 1)^{2L - k + 1}} \; .
\end{align*}
In this way we obtain the following representation for the function
$\Phi_L(u, w)$:
\begin{equation}
    \Phi_L(u, w) = 2\pi\dfrac{u^2}{4}\sum_{n =
    0}^{\infty}\mu(n)\dfrac{u^{\mu_1\ldots\mu_n}
    w^{\mu_1\ldots\mu_n}}{(u^2)^{D/2 + n}}\sum_{k = 0}^{L}\dfrac{(2L -
    k)!}{k!(L - k)!}\dfrac{\operatorname{Log}^k(u^2/w^2)}{
    (D/2 + n - 1)^{2L - k + 1}}
\end{equation}
which is transformed to
\begin{multline}
    \Phi_L(u, w) = \pi^{-D/2}\dfrac{u^2}{4(u^2)^{D/2}}
    \sum_{n = 0}^{\infty}\dfrac{\Gamma(D/2 + n)\Gamma(D/2 -
    1)}{\Gamma(D/2 - 1 +
    n)}\dfrac{C^{(D/2-1)}_n\left(\hat{u}\hat{w}\right)}{
    (u^2/w^2)^{n/2}}\\
    \sum_{k = 0}^{L}\dfrac{(2L - k)!}{k!(L -
    k)!}\dfrac{\operatorname{Log}^k(u^2/w^2)}{(D/2 + n - 1)^{2L - k +
    1}}.
\end{multline}
where we use the explicit formula (\ref{eigfun1}) for
 the measure $\mu(n)$ and
 the formula with the Gegenbauer polynomials
\begin{equation}
    u^{\mu_1\ldots\mu_n}w^{\mu_1\ldots\mu_n} = \dfrac{n!\Gamma\left(D/2
    - 1\right)}{2^n\Gamma\left(n + D/2 -
    1\right)}C^{(D/2-1)}_n\left(\hat{u}\hat{w}\right)
    \left(u^2w^2\right)^{n/2} .
\end{equation}
In the case $D = 4$, we have a
 simple expression for the Gegenbauer polynomials
\begin{equation}
    C_n^{(1)}(\hat{u}\hat{w}) = \dfrac{\sin((n +
    1)\theta)}{\sin\theta},
\end{equation}
where $\theta$ is the angle between the vectors $u$ and $w$.
Then, after a shift $n \to (n-1)$ in the summation we have
\begin{align}
\nonumber
&\Phi_L(u, w) =
\dfrac{1}{(u^2w^2)^{\frac{1}{2}}4 \pi^2}\sum_{n = 1}^{\infty}n
\dfrac{\sin(n\theta)}{(u^2/w^2)^{n/2}\sin\theta}\sum_{k =
0}^{L}\dfrac{(2L - k)!}{k!(L -
k)!}\dfrac{\operatorname{Log}^k(u^2/w^2)}{n^{2L - k + 1}}\; ,
\end{align}
where the sum over $n$ can be expressed in terms of polylogs
$$
{\rm Li}_k(z) = \sum\limits_{n=0}^\infty \frac{z^n}{n^k}
\;\;\; \Rightarrow \;\;\;
    \sum_{n = 1}^{\infty}\dfrac{e^{in\theta}}{
    (u^2/w^2)^{n/2}n^{2L - k}} = \operatorname{Li}_{2L -
    k}\left(\dfrac{e^{i\theta}}{(u^2/w^2)^{1/2}}\right) \; .
$$
As a result, we deduce
\begin{multline}
\label{d4p}
\Phi_{L}(u, w) = \dfrac{1}{(u^2w^2)^{1/2}}
\dfrac{1}{4 \pi^2}\dfrac{1}{(e^{i\theta} - e^{-i\theta})} \sum_{k =
0}^L\dfrac{(-1)^k(2L - k)!}{k!(L -
k)!}\operatorname{Log}^k\left(\dfrac{w^2}{u^2}\right)\times \\
\times\left[\operatorname{Li}_{2L -
k}\left(\dfrac{e^{i\theta}}{(u^2/w^2)^{1/2}}\right) -
\operatorname{Li}_{2L -
k}\left(\dfrac{e^{-i\theta}}{(u^2/w^2)^{1/2}}\right)\right]
\end{multline}
We use the parametrization
\begin{equation}
    z = \dfrac{e^{i\theta}}{(u^2/w^2)^{1/2}}, ~~~~~ \bar{z} =
    \dfrac{e^{-i\theta}}{(u^2/w^2)^{1/2}} \;\;\;\;~~
  \Rightarrow  \;\;\;\;~~
  z + \bar{z} = \dfrac{2\, u\cdot w}{u^2 } \; , \;\;\;\;
  z\bar{z} = \dfrac{w^2}{u^2} \; ,
\end{equation}
and finally obtain from (\ref{d4p}) the known result (\ref{d4}).

\vspace{0.2cm}

\noindent
{\bf Remark 5.} In view of eq. (\ref{grn4}),
after some renormalization and
change of variables, the Green's function (\ref{grn2})
is related to the sum of the 4-point ladder integrals
(\ref{grn5}).
 For $D=4$ this sum  was investigated
in \cite{DBro} and then was written
 in another form
  of a single integral over the Bessel function $J_0$ with
 a trigonometric measure (see eqs. (14), (15) in \cite{DBro}).

\section{Operator formalism and zig-zag diagrams\label{zig-zag}}
\setcounter{equation}0

Below we consider a special class of Feynman perturbative
4-point $G_4^{(M)} (x_1, x_{2}; y_1, y_{2})$
and 2-point $G_2^{(M)} (x_{2}, y_1)$
 massless integrals
\begin{align}
\label{zgzg44}
G_4^{(2N)} (x_1, x_{2}; y_1, y_{2}) &= (y_1 - y_2)^{2\beta}\,
\int \left[ \prod_{i=1}^{N} d^D z_i \,
d^D \widetilde{z}_i \right]\,
\prod\limits_{j=1}^{N}
\frac{1}{z_{j,j+1}^{2\beta'} \,
\widetilde{z}_{j,j+1}^{2\beta'} \,
z_{j,\,\widetilde{j+1}}^{2\beta} \,
z_{j+1,\widetilde{j+1}}^{2\beta}}\,, \\
\label{zg44}
G_4^{(2N+1)} (x_1, x_{2}; y_1, y_{2}) &=
\int d^Dz \,
\frac{G_4^{(2N)} (x_1, x_{2}; y_2, z)}{
(z - y_1)^{2\beta'} (z - y_2)^{2\beta}}\,,\\
\label{zgzg22}
G^{(M)}_2(x_2,y_1) &= \int d^D x_1 \; d^D y_2 \;
\dfrac{G^{(M)}_4(x_1,x_2;y_1,y_2)} {
(x_1-x_2)^{2\beta}(y_1-y_2)^{2\beta}}\,,
\end{align}
where $z_0=x_1$, $\widetilde{z}_0=x_2$,
$z_{N+1}=y_1$, $\widetilde{z}_{N+1}=y_2$ and we use compact notation
$z_{ik} = z_i - z_k\,, \widetilde{z}_{ik} =
\widetilde{z}_i - \widetilde{z}_k\,,
z_{i\widetilde{k}} = z_i - \widetilde{z}_k$ and $\beta' = D/2-\beta$.

The integrals (\ref{zgzg44}) -- (\ref{zgzg22})
are called zig-zag integrals and are
visualized below as Feynman diagrams (zig-zag diagrams)
in Figs. \ref{zgzg1},  \ref{zgzg2} and  \ref{zgzg3},
\ref{zgzg4}.

\subsection{Definition of the
graph building operator $\hat{Q}_{12}$ }

 The first step is to obtain
 operator expressions for the zig-zag integrals
 (\ref{zgzg44}) -- (\ref{zgzg22})
  using the operator formalism discussed in Sections
  {\bf \ref{opfor}} and {\bf \ref{2ladder}}.
  In these Sections we used only one $D$-dimensional
  Heisenberg algebra ${\cal H}$ with defining relations
  \eqref{gr001}. For the purposes of evaluation
  of zig-zag integrals, it would be convenient to consider a
 direct product of several Heisenberg algebras \eqref{gr001}.
 Consider the algebra $\mathcal{H}^{(n)}$
 consisting of $n$ copies of the $D$-dimensional Heisenberg algebras
 $\mathcal{H}_i \equiv \mathcal{H}$ $(i=1,...,n)$, so
 $\mathcal{H}^{(n)} = \otimes_{i=1}^n \mathcal{H}_i$,
 with the commutation relations of generators
\begin{equation}
\lb{deff}
 [\hat{q}^{\mu}_i, \hat{q}^{\nu}_j] = 0 =
  [\hat{p}^{\mu}_i, \hat{p}^{\nu}_j] \, , \;\;\;\;\;\;\;\;
    [\hat{q}^{\mu}_i, \hat{p}^{\nu}_j] =
    i\delta^{\mu\nu}\delta_{ij} \; , \;\;\;\; i,j = 1,...,n
    \; , \;\;\;\; \mu,\nu = 1,...,D \; .
\end{equation}
Further we introduce states $|x_i \rangle$ and $|k_i \rangle$
which respectively diagonalize $\q^\mu_i$
and $\p^\nu_i$ as in \eqref{gr002} for each copy of the algebras
$\mathcal{H}_i$.
 Such states form a basis in the space $V_i$, where the subalgebra
 $\mathcal{H}_i$ acts. The whole algebra $\mathcal{H}^{(n)}$ acts in
 the space
 $V_1 \otimes \cdots \otimes V_n$
  with the basis vectors
\begin{equation}
    |x_1,..., x_n \rangle : = |x_1\rangle \otimes
    \cdots \otimes |x_n\rangle.
\end{equation}

To obtain
 operator expressions for the zig-zag integrals
 (\ref{zgzg44}) -- (\ref{zgzg22}),
we need only the
algebra $\mathcal{H}^{(2)}= \mathcal{H}_1 \otimes \mathcal{H}_2$
consisting of two copies of the $D$-dimensional Heisenberg algebra.
The entire subsequent procedure of building the diagram technique that
uses the operator formalism in the algebra $\mathcal{H}^{(2)}$ repeats
the methods of Section {\bf \ref{opfor}}, except for the fact that all
operators carry an additional index $i=1,2$.

We want to obtain, in the case of zig-zag diagrams,
 expressions similar to \eqref{twoloop} and \eqref{bra2},
 so we introduce analogs of the operators $H_\alpha =
 \p^{2\alpha} q^{2\alpha}$, which
 we call graph building operators. The original problem of
 calculating zig-zag diagram was formulated in $D = 4$, so firstly we
 will consider the
 four-dimensional case and then its generalization to any $D$.

Let us define the four-dimensional
graph-building operator $\hat{Q}_{12}$ in the following form:
\begin{equation}
\label{q-4}
    \hat{Q}_{12} =
    (2\pi)^2\mathcal{P}_{12}\hat{p}_1^{-2}\hat{q}_{12}^{-2},
\end{equation}
where we use the convenient notation
 $\hat{q}^{\mu}_{ij} = \hat{q}^{\mu}_i - \hat{q}^{\mu}_j$ and
 $\mathcal{P}_{12}$ is a permutation operator
 \be
 \lb{perm}
 {\cal P}_{12} \, \q_1 = \q_2 \, {\cal P}_{12} \; , \;\;\;\;
 {\cal P}_{12} \, \p_1 = \p_2 \, {\cal P}_{12} \; , \;\;\;\; \langle
 x_1,x_2 |{\cal P}_{12}| \Psi \rangle  =
\langle  x_2,x_1 | \Psi \rangle \; , \;\;\;\; ({\cal P}_{12})^2 = I \;
.
 \ee

For the calculation we need to know the
integral kernel of the operator $\hat{Q}_{12}$:
\begin{equation}\label{q-kern}
\langle x_1,x_2 |
\hat{Q}_{12}  | y_1,y_2 \rangle \, =
\, (2\pi)^2 \, \cdot \, \langle x_1,x_2 |\;
{\cal P}_{12} \,\p_1^{-2} \,
\hat{q}_{12}^{-2} \, | y_1,y_2 \rangle \, =
\frac{\delta^4(x_1-y_2)}{(x_2-y_1)^2 \, (y_1-y_2)^2} \; .
\end{equation}
The visualization of the integral
kernel (\ref{q-kern}) is given in Fig. \ref{qkrn}

\begin{figure}[h]
  \unitlength=5mm
\begin{picture}(25,4)(-4,1)

\put(1,2.8){\footnotesize  $\langle x_1,x_2 |
\hat{Q}_{12}  | y_1,y_2 \rangle \; =
\; {\cal P}_{12} \; \cdot$}
%%%%%%%%%%%%%%%%%%%%%%%

 \put(10.1,4){\footnotesize $x_1$}
 %\put(8.1,1.9){\tiny $\beta$}
 %\put(6.7,0.5){\tiny $\beta'$}
\put(10.2,1.8){\footnotesize $x_2$}
\put(13.2,1.8){\footnotesize $y_2$}
\put(13.2,4){\footnotesize $y_1$}

\put(13,2){\line(0,1){2}}
 \put(13,4){\line(-1,0){2}}
 \multiput(13,2)(-0.24,0){10}{\circle*{0.1}}

 \put(15,2.8){\footnotesize $ =$}
%%%%%%%%%%%%%%%%%%%%

\put(16.2,4){\footnotesize $x_1$}
 %\put(3.1,1.9){\tiny $\beta$}
 %\put(1.3,1.6){\tiny $\beta'$}
\put(16.2,2){\footnotesize $x_2$}
\put(19.2,1.8){\footnotesize $y_2$}
\put(19.2,4){\footnotesize $y_1$}

\put(19,2){\line(0,1){2}}
 \put(19,4){\line(-1,-1){2}}
 \multiput(19,2)(-0.18,0.18){12}{\circle*{0.1}}

 %%%%%%%%%%%%%%%%%%%%%%%%%
 %\put(22,1.5){\bf (a)}

\end{picture}
\caption{\small
\it The diagram that represents the integral kernel $\langle x_1,x_2 |
\hat{Q}_{12}  | y_1,y_2 \rangle$ (we do not
indicate the index $1$ on the solid lines).
 \label{qkrn}}
\end{figure}

 To illustrate why
 $\hat{Q}_{12}$ is  the graph-building operator for the zig-zag
 diagrams,
 we consider the operator $\bigl(\hat{Q}_{12}\bigr)^2$ and its integral
 kernel
 \begin{equation}
    \bigl(\hat{Q}_{12}\bigr)^2 =
    (2\pi)^4\hat{p}_2^{-2}\hat{q}_{12}^{-2}
    \hat{p}_1^{-2}\hat{q}_{12}^{-2}
\end{equation}
\begin{equation}
    \langle x_1, x_2| \bigl(\hat{Q}_{12}\bigr)^2|y_1,y_2\rangle =
    \dfrac{1}{(x_1 - y_1)^2(x_2 - y_2)^2(x_1 - y_2)^2(y_1 - y_2)^2}
\end{equation}
 The visualization of the
 evaluation of the integral kernel for
 $\bigl(\hat{Q}_{12}\bigr)^2$ is given in
   Fig. \ref{q-krnl}. The integral operator
   (\ref{q-kern}) was
considered in \cite{GKK} and denoted there as $H_{1\!\!1}$.

\begin{figure}[h!]
  \unitlength=4.5mm

\begin{picture}(25,7)(-2,-0.5)

\put(0,5.8){\footnotesize $\langle x_1,x_2 |
\hat{Q}_{12}\; \hat{Q}_{12} | y_1,y_2 \rangle
\;\;\;\;\; =$}

\put(-1.2,4){\footnotesize
$\overbrace{I=\int d^D z_1 d^D z_2 | z_1,z_2 \rangle
\langle z_1,z_2 |}$}
%%%%%%%%%%%%%%%%%%%%%%%

\put(2.6,1.3){\footnotesize
$= \;\; \int d^D z_1 \; d^D z_2
\; {\cal P}_{12} \; \cdot$}

 \put(10.1,2.5){\footnotesize $x_1$}
 %\put(8.1,1.9){\tiny $\beta$}
 %\put(6.7,0.5){\tiny $\beta'$}
\put(10.2,0.3){\footnotesize $x_2$}
\put(13.2,0.3){\footnotesize $z_2$}
\put(13.2,2.5){\footnotesize $z_1$}

\put(13,0.5){\line(0,1){2}}
 \put(13,2.5){\line(-1,0){2}}
 \multiput(13,0.5)(-0.24,0){10}{\circle*{0.1}}

 \put(14,1.3){\footnotesize $\cdot \; {\cal P}_{12} \; \cdot $}
%%%%%%%%%%%%%%%%%%%%

\put(16.1,2.5){\footnotesize $z_1$}
 %\put(3.1,1.9){\tiny $\beta$}
 %\put(1.3,1.6){\tiny $\beta'$}
\put(16.1,0.3){\footnotesize $z_2$}
\put(19.2,0.3){\footnotesize $y_2$}
\put(19.2,2.5){\footnotesize $y_1$}

\put(19,0.5){\line(0,1){2}}
 \put(19,2.5){\line(-1,0){2}}
 \multiput(19,0.5)(-0.24,0){10}{\circle*{0.1}}

 \put(20,1.3){\footnotesize $=$}

 %%%%%%%%%%%%%%%%%%%%%%%%%
 %%%%%%%%%%%%%%%%%%%%%%%%%%%

  \put(21.1,2.7){\footnotesize $x_1$}
  \put(21.2,0){\footnotesize $x_2$}

 \put(23.8,0.3){\footnotesize $\bullet$}
  \put(23.8,2.3){\footnotesize $\bullet$}

\put(24,-0.1){\footnotesize $z_2$}
\put(24,2.8){\footnotesize $z_1$}

\put(24,0.5){\line(0,1){2}}
 \put(24,0.5){\line(-1,0){2}}
 \multiput(24,2.5)(-0.24,0){10}{\circle*{0.1}}

%%%%%%%%%%%%%%%%%%%%

\put(26.2,0.1){\footnotesize $y_2$}
\put(26.2,2.7){\footnotesize $y_1$}

\put(26.1,0.5){\line(0,1){2}}
 \put(26.1,2.5){\line(-1,0){2}}
 \multiput(26.1,0.5)(-0.24,0){10}{\circle*{0.1}}

  \put(28,1.3){\footnotesize $=$}

 %%%%%%%%%%%%%%%%%%%%%%%%%%%%%%%%%%%%%%
 \put(29.1,2.5){\footnotesize $x_1$}
\put(29.1,0.5){\footnotesize $x_2$}
\put(32,0.5){\line(0,1){2}}
\put(30,0.5){\line(1,0){2}}
\put(30,2.5){\line(1,0){2}}
 \put(32,0.5){\line(-1,1){2}}
  \put(32.2,2.5){\footnotesize $y_1$}
\put(32.2,0.5){\footnotesize $y_2$}
 %%%%%%%%%%%%%%%%%%
 %\put(34,0.5){\bf (b)}

\end{picture}
\caption{\small
\it The diagram that represents the integral kernel  $\langle x_1,x_2
|
\bigl(\hat{Q}_{12}\bigr)^2  | y_1,y_2 \rangle$.
Two operators ${\cal P}_{12}$
 turn over the first graph for
 $\langle x_1,x_2 |
 \hat{p}_1^{-2}\hat{q}_{12}^{-2} | z_1,z_2 \rangle$ and
the result of the calculation is
obtained by shrinking dashed lines
($\delta$-functions) into points.
 \label{q-krnl}}
\end{figure}

 %\noindent

The definition of the graph-building
operator \eqref{q-4} sufficient for the calculation
of the standard zig-zag series of diagrams,
 arised in the calculation of the $\beta$-function in four-dimensional
 $\varphi^4$ theory, but this approach can be extended for the
 calculation of a more general series of zig-zag diagrams with
 non-unit indices on the lines in the case
 of an arbitrary $D$. Consider the generalized version of
 the operator \eqref{q-4} (for  any dimension $D$)
\begin{equation}
\label{q-d}
 \hat{Q}_{12}^{(\beta)} : = \dfrac{1}{a(\beta)} \;
{\cal P}_{12} \; (\p_1)^{-2\beta} \; (\hat{q}_{12})^{-2\beta} \; ,
\end{equation}
where $\beta$ is an arbitrary parameter
and the function $a(\beta)$ is defined in
 (\ref{gr3}).
The integral kernel of this operator is
\begin{equation}
 \langle x_1,x_2 |
\hat{Q}_{12}^{(\beta)}  | y_1,y_2 \rangle \, = \;\;
\frac{ \delta^D(x_1-y_2)}{(x_2-y_1)^{2\beta'} \, (y_1-y_2)^{2\beta}},
\end{equation}
and we can build a zig-zag series
of diagrams with arbitrary parameters $\beta$
and $\beta'=D/2-\beta$
on the lines by using the kernel of the operator
$\left(\hat{Q}^{(\beta)}_{12}\right)$ as a building block. Indeed,
consider four point planar zig-zag Feynman graphs:
\\
for even loops\footnote{The number
of loops coincides
with the number of bold face vertices.}

\unitlength=5mm
\begin{picture}(25,3.5)(1,1)

 \put(4.6,1.9){\tiny $\beta$}
\put(3.5,1.8){\tiny $\beta$}
 %\put(8.6,1.9){\tiny $\beta$}
 %\put(7.5,1.8){\tiny $\beta$}

 \put(3.8,3.2){\tiny $\beta'$}
 \put(3.8,0.5){\tiny $\beta'$}
  \put(5.8,3.2){\tiny $\beta'$}
 \put(5.8,0.5){\tiny $\beta'$}
 %\put(7.8,0.5){\tiny $\beta'$}
 %\put(8,3.2){\tiny $\beta'$}

\put(2.1,3){\footnotesize $x_1$}
\put(2.1,1){\footnotesize $x_2$}
\put(5,1){\line(0,1){2}}
\put(3,1){\line(1,0){2}}
\put(3,3){\line(1,0){2}}
 \put(5,1){\line(-1,1){2}}

%%%%%%%%%%%%%%%%%%%%%%%%%%%%

 \put(6.6,1.9){\tiny $\beta$}
 \put(5.5,1.8){\tiny $\beta$}

\put(7,1){\line(0,1){2}}
\put(5,1){\line(1,0){2}}
\put(5,3){\line(1,0){2}}
 \put(7,1){\line(-1,1){2}}
 \put(4.85,0.8){$\bullet$}
 \put(4.85,2.8){$\bullet$}

 %%%%%%%%%%%%%%%%%%%%%%%%%%%%%%

 \put(6.85,0.8){$\bullet$}
 \put(6.85,2.8){$\bullet$}
 %\put(9,1){\line(0,1){2}}
 %\put(7,1){\line(1,0){2}}
 %\put(7,3){\line(1,0){2}}
 %\put(9,1){\line(-1,1){2}}

 %\put(8.85,0.8){$\bullet$}
 %\put(8.85,2.8){$\bullet$}
%%%%%%%%%%%%%%%%%%%%%%%%%%%%%%
 \put(8,2){$. \; .\; . \; . \; .\; . $}
%%%%%%%%%%%%%%%%%%%%%%%%%%%%%%

 \put(11.8,0.5){\tiny $\beta'$}
 \put(12,3.2){\tiny $\beta'$}

\put(13,1){\line(0,1){2}}
\put(11,1){\line(1,0){2}}
\put(11,3){\line(1,0){2}}
 \put(13,1){\line(-1,1){2}}

%%%%%%%%%%%%%%%%%%%%%%%%%%%%

\put(13.8,0.5){\tiny $\beta'$}
\put(12.1,1.9){\tiny $\beta$}
\put(14.1,1.9){\tiny $\beta$}
\put(14,3.2){\tiny $\beta'$}

\put(13,1){\line(1,0){2}}
\put(13,3){\line(1,0){2}}
 \put(15,1){\line(-1,1){2}}
 \put(12.85,0.8){$\bullet$}
\put(12.85,2.8){$\bullet$}

\put(15.2,0.8){\footnotesize $y_2$}
\put(15.2,3){\footnotesize $y_1$}

 %\put(18.5,2){\footnotesize $=$}
 %\put(4,-2){\footnotesize
\put(17,2){\footnotesize $= \;\;\;
\langle x_1,x_2 | (\hat{Q}_{12}^{(\beta)})^{2N}| y_1,y_2 \rangle
(y_1-y_2)^{2\beta} = $}

\end{picture}

\vspace{-0.5cm}

\be
\lb{zgzg1}
{}
\ee

%\vspace{0.5cm}

\unitlength=1.8mm
\begin{picture}(4,8)(-30,-1)

%%%%%%%%%%%%%%%%%%%%%%%%%%%%

\put(-4,2.7){$=$}

\put(0.3,6.2){\footnotesize $x_1$}
\put(0.3,0.2){\footnotesize $x_2$}

%%%%%%%%%%%%Pervaya petlja%%%%%%%%%%%%%%%%

\qbezier(4.2,2.3)(5.1,-1.6)(6,2.5)

 \qbezier(6,2.5)(6.2,6.1)(2.5,5.7)

 \qbezier(4.2,2.3)(3.9,3.5)(5.2,4.6)

 %%%%%%%%%%%%%%%%%Vtoraya petlja%%%%%%%%%%%%%%%%%%%

 \qbezier(8,2.3)(8.9,-1.6)(9.8,2.5)
 \qbezier(9.8,2.5)(9.8,7)(6.3,5)

 \qbezier(8,2.3)(7.7,3.5)(9,4.6)

%%%%%%%%%%%%tret'ja petlja%%%%%%%%%%%%%%%%

 \qbezier(11.5,2.3)(12.4,-1.6)(13.3,2.5)
 \qbezier(13.3,2.5)(13.3,7)(9.8,5)

 \qbezier(11.5,2.3)(11.2,3.5)(12.5,4.6)

 %%%%%%%%%%%%%%%%%%Poslednie petli%%%%%%%%%%%%%%%%%%%%%%%%
 %%%%%%%%%%%%%%%%%%%%%%%%%%%%%%%%%%%%%%%%

  \put(14.5,3){$. \; .\; . \; . \; .\; . $}

 %%%%%%%%%%%%%%%%%4 petlja%%%%%%%%%%%%%%%%%%%

 \qbezier(22,2.3)(22.9,-1.6)(23.8,2.5)
 \qbezier(23.8,2.5)(23.8,7)(20.3,5)

 \qbezier(22,2.3)(21.7,3.5)(23,4.6)

%%%%%%%%%%%%5 petlja%%%%%%%%%%%%%%%%

 \qbezier(25.5,2.3)(26.4,-1.6)(27.3,2.5)
 \qbezier(27.3,2.5)(27.3,7)(23.8,5)

 \qbezier(25.5,2.3)(25.2,3.5)(26.5,4.6)

 \qbezier(27.5,5.1)(29.1,6.4)(30.2,4.8)

 \qbezier(30.2,4.8)(31,3)(30.1,0.4)

 \put(2.7,0.4){\line(1,0){27.5}}
 \put(3.4,5.7){\line(1,0){29}}

 %%%%%%%%%%%%%%%%%%%%%%%%%%%%%%%%%%
 \put(4.5,-0.2){$\bullet$}
 \put(8.4,-0.2){$\bullet$}
  \put(11.8,-0.2){$\bullet$}
  \put(22.4,-0.2){$\bullet$}
  \put(25.8,-0.2){$\bullet$}

 \put(7.5,5.2){$\bullet$}
  \put(11,5.2){$\bullet$}
  \put(21.6,5.2){$\bullet$}
  \put(25,5.2){$\bullet$}
   \put(28.4,5.2){$\bullet$}

 \put(33,6.2){\footnotesize $y_1$}
\put(30.8,0){\footnotesize $y_2$}

\put(36,3){\bf ;}
 %%%%%%%%%%%%%%%%%Primery%%%%%%%%%%%%%%%%%%
 %\qbezier(42,2.5)(42,3.5)(42.9,4.4)
 %\qbezier(43.8,3.5)(45.1,-1.6)(46,2.5)
 %\qbezier(12.7,4.1)(15.1,-0.6)(15.6,2.5)
 %\qbezier(44.5,2.5)(45.5,5)(46,2.5)
 %%%%%%%%%%%%%%%%%%%%%%%%%%

\end{picture}

\noindent
for odd loops

\unitlength=4.5mm
\begin{picture}(25,4)(2,1)

\put(2.2,3.2){\footnotesize $x_1$}
\put(2.1,0.6){\footnotesize $x_2$}

 %\put(4.5,3.4){\footnotesize $z_1$}
 %\put(4.7,0.5){\footnotesize $z_2$}
 %\multiput(5,1)(-0.25,0){9}{\circle*{0.13}}

\put(5,1){\line(0,1){2}}
\put(3,3){\line(1,0){2}}
 \put(5,1){\line(-1,1){2}}
 \put(3,1){\line(1,0){2}}

\put(3.8,0.5){\tiny $\beta'$}
 \put(4,3.2){\tiny $\beta'$}
 \put(5.8,0.5){\tiny $\beta'$}
 \put(6,3.2){\tiny $\beta'$}
 \put(7.8,0.5){\tiny $\beta'$}
 \put(8,3.2){\tiny $\beta'$}

 \put(4.6,1.9){\tiny $\beta$}
\put(3.5,1.8){\tiny $\beta$}
 \put(6.6,1.9){\tiny $\beta$}
\put(5.5,1.8){\tiny $\beta$}
 \put(8.6,1.9){\tiny $\beta$}
\put(7.5,1.8){\tiny $\beta$}
%%%%%%%%%%%%%%%%%%%%%%%%%%%%

\put(7,1){\line(0,1){2}}
\put(5,1){\line(1,0){2}}
\put(5,3){\line(1,0){2}}
 \put(7,1){\line(-1,1){2}}
 \put(4.85,0.8){$\bullet$}
\put(4.85,2.8){$\bullet$}

 %%%%%%%%%%%%%%%%%%%%%%%%%%%%%%

 \put(6.85,0.8){$\bullet$}
\put(6.85,2.8){$\bullet$}
 \put(9,1){\line(0,1){2}}
\put(7,1){\line(1,0){2}}
\put(7,3){\line(1,0){2}}
 \put(9,1){\line(-1,1){2}}

\put(8.85,0.8){$\bullet$}
\put(8.85,2.8){$\bullet$}
%%%%%%%%%%%%%%%%%%%%%%%%%%%%%%
 \put(10,2){$. \; .\; . \; . $}
%%%%%%%%%%%%%%%%%%%%%%%%%%%%%%

\put(14,1){\line(0,1){2}}
\put(12,1){\line(1,0){2}}
\put(12,3){\line(1,0){2}}
 \put(14,1){\line(-1,1){2}}

%%%%%%%%%%%%%%%%%%%%%%%%%%%%

\put(14.8,0.5){\tiny $\beta'$}
\put(15,3.2){\tiny $\beta'$}
\put(12.9,3.2){\tiny $\beta'$}
\put(12.7,0.5){\tiny $\beta'$}
\put(16.9,0.5){\tiny $\beta'$}
\put(16.1,1.9){\tiny $\beta$}
\put(14.6,1.7){\tiny $\beta$}
\put(13.6,1.9){\tiny $\beta$}
\put(12.6,1.7){\tiny $\beta$}

\put(16,1){\line(0,1){2}}
\put(14,1){\line(1,0){2}}
\put(14,3){\line(1,0){2}}
 \put(16,1){\line(-1,1){2}}
  %\put(11.85,0.8){$\bullet$}
  %\put(11.85,2.8){$\bullet$}
 \put(13.85,0.8){$\bullet$}
\put(13.85,2.8){$\bullet$}

\put(16,1){\line(1,0){2}}
 \put(15.85,0.8){$\bullet$}
\put(18.2,0.8){\footnotesize $y_1$}
\put(16.2,3){\footnotesize $y_2$}

\put(20, 2){\footnotesize $= \;
 %\put(4,-2){\footnotesize $= \;
\langle x_1,x_2 |
(\hat{Q}_{12}^{(\beta)})^{2N+1}| y_1,y_2 \rangle
(y_1-y_2)^{2\beta} =$}

\end{picture}
\vspace{-0.5cm}

\be
\lb{zgzg2}
{}
\ee

%\vspace{0.5cm}

\unitlength=1.8mm
\begin{picture}(4,8)(-30,-1)

%%%%%%%%%%%%%%%%%%%%%%%%%%%%

\put(-4,2.7){$=$}

\put(0.3,6.2){\footnotesize $x_1$}
\put(0.3,0.2){\footnotesize $x_2$}

%%%%%%%%%%%%Pervaya petlja%%%%%%%%%%%%%%%%

\qbezier(4.2,2.3)(5.1,-1.6)(6,2.5)

 \qbezier(6,2.5)(6.2,6.1)(2.5,5.7)

 \qbezier(4.2,2.3)(3.9,3.5)(5.2,4.6)

 %%%%%%%%%%%%%%%%%Vtoraya petlja%%%%%%%%%%%%%%%%%%%

 \qbezier(8,2.3)(8.9,-1.6)(9.8,2.5)
 \qbezier(9.8,2.5)(9.8,7)(6.3,5)

 \qbezier(8,2.3)(7.7,3.5)(9,4.6)

%%%%%%%%%%%%tret'ja petlja%%%%%%%%%%%%%%%%

 \qbezier(11.5,2.3)(12.4,-1.6)(13.3,2.5)
 \qbezier(13.3,2.5)(13.3,7)(9.8,5)

 \qbezier(11.5,2.3)(11.2,3.5)(12.5,4.6)

 %%%%%%%%%%%%%%%%%%Poslednie petli%%%%%%%%%%%%%%%%%%%%%%%%
 %%%%%%%%%%%%%%%%%%%%%%%%%%%%%%%%%%%%%%%%

  \put(14.5,3){$. \; .\; . \; . \; .\; . $}

 %%%%%%%%%%%%%%%%%4 petlja%%%%%%%%%%%%%%%%%%%

 \qbezier(22,2.3)(22.9,-1.6)(23.8,2.5)
 \qbezier(23.8,2.5)(23.8,7)(20.3,5)

 \qbezier(22,2.3)(21.7,3.5)(23,4.6)

%%%%%%%%%%%%5 petlja%%%%%%%%%%%%%%%%

 \qbezier(25.5,2.3)(26.4,-1.6)(27.3,2.5)
 \qbezier(27.3,2.5)(27.3,7)(23.8,5)

 \qbezier(25.5,2.3)(25.2,3.5)(26.5,4.6)

 \qbezier(27.3,5)(28.5,5.8)(29.5,5.7)

 \put(2.7,0.4){\line(1,0){27.5}}
 \put(3.4,5.7){\line(1,0){26}}

 %%%%%%%%%%%%%%%%%%%%%%%%%%%%%%%%%%
 \put(4.5,-0.2){$\bullet$}
 \put(8.4,-0.2){$\bullet$}
  \put(11.8,-0.2){$\bullet$}
  \put(22.4,-0.2){$\bullet$}
  \put(25.8,-0.2){$\bullet$}

 \put(7.5,5.2){$\bullet$}
  \put(11,5.2){$\bullet$}
  \put(21.6,5.2){$\bullet$}
  \put(25,5.2){$\bullet$}
   %\put(28.4,5.2){$\bullet$}

 \put(30,6.2){\footnotesize $y_2$}
\put(30.8,0){\footnotesize $y_1$}

\put(36,3){\bf .}
 %%%%%%%%%%%%%%%%%Primery%%%%%%%%%%%%%%%%%%
 %\qbezier(42,2.5)(42,3.5)(42.9,4.4)
 %\qbezier(43.8,3.5)(45.1,-1.6)(46,2.5)
 %\qbezier(12.7,4.1)(15.1,-0.6)(15.6,2.5)
 %\qbezier(44.5,2.5)(45.5,5)(46,2.5)
 %%%%%%%%%%%%%%%%%%%%%%%%%%

\end{picture}
%\end{widetext}
\vspace{0.8cm}

\noindent
Thus, to find the corresponding four point
zig-zag correlation function in $(M-2)$-loop order,
 we need to calculate
\begin{equation}
\label{g-4}
    G^{(M)}_4(x_1,x_2;y_1,y_2) = \langle x_1,x_2 |
    \bigl(\hat{Q}_{12}^{(\beta)}\bigr)^{M}| y_1,y_2 \rangle
(y_1-y_2)^{2\beta}.
\end{equation}
In eqs. (\ref{zgzg1}), (\ref{zgzg2})
we present the spiral visualization of the
zig-zag diagrams which are specific for the
conformal biscalar fishnet
theories \cite{KG}, \cite{KO}.

The two-point zig-zag Feynman graphs
and the corresponding integrals are
 defined via the 4-point
 functions (\ref{g-4}) as follows:
\\
\noindent
for even loops

\unitlength=4mm
\begin{picture}(25,3.5)(2,1)

\put(2.6,1.9){\tiny $\beta$}
 \put(4.6,1.9){\tiny $\beta$}
\put(3.5,1.8){\tiny $\beta$}
\put(8.6,1.9){\tiny $\beta$}
\put(7.5,1.8){\tiny $\beta$}

 \put(3.8,3.2){\tiny $\beta'$}
 \put(3.8,0.5){\tiny $\beta'$}
  \put(5.8,3.2){\tiny $\beta'$}
 \put(5.8,0.5){\tiny $\beta'$}
 \put(7.8,0.5){\tiny $\beta'$}
 \put(8,3.2){\tiny $\beta'$}

\put(2.1,3){\footnotesize $x_1$}
\put(2.1,1){\footnotesize $x_2$}
\put(5,1){\line(0,1){2}}
\put(3,1){\line(1,0){2}}
\put(3,3){\line(1,0){2}}
 \put(5,1){\line(-1,1){2}}

 %\put(3,1){\line(0,1){2}}
 %\put(2.85,2.85){$\bullet$}
 %\put(16.85,0.85){$\bullet$}
%%%%%%%%%%%%%%%%%%%%%%%%%%%%

 \put(6.6,1.9){\tiny $\beta$}
 \put(5.5,1.8){\tiny $\beta$}

\put(17.4,0.8){\footnotesize $y_2$}
\put(17.3,3){\footnotesize $y_1$}
\put(7,1){\line(0,1){2}}
\put(5,1){\line(1,0){2}}
\put(5,3){\line(1,0){2}}
 \put(7,1){\line(-1,1){2}}
 \put(4.85,0.8){$\bullet$}
\put(4.85,2.8){$\bullet$}

 %%%%%%%%%%%%%%%%%%%%%%%%%%%%%%

 \put(6.85,0.8){$\bullet$}
\put(6.85,2.8){$\bullet$}
 \put(9,1){\line(0,1){2}}
\put(7,1){\line(1,0){2}}
\put(7,3){\line(1,0){2}}
 \put(9,1){\line(-1,1){2}}

\put(8.85,0.8){$\bullet$}
\put(8.85,2.8){$\bullet$}
%%%%%%%%%%%%%%%%%%%%%%%%%%%%%%
 \put(9.8,2){$. \; .\; . \; . \; .\; . $}
%%%%%%%%%%%%%%%%%%%%%%%%%%%%%%

 \put(13.8,0.5){\tiny $\beta'$}
 \put(14,3.2){\tiny $\beta'$}

\put(15,1){\line(0,1){2}}
\put(13,1){\line(1,0){2}}
\put(13,3){\line(1,0){2}}
 \put(15,1){\line(-1,1){2}}

%%%%%%%%%%%%%%%%%%%%%%%%%%%%

\put(15.8,0.5){\tiny $\beta'$}
\put(16,3.2){\tiny $\beta'$}
\put(14.1,1.9){\tiny $\beta$}
\put(16.1,1.9){\tiny $\beta$}
\put(15.1,1.9){\tiny $\beta$}
\put(17.1,1.9){\tiny $\beta$}

\put(15,1){\line(1,0){2}}
\put(15,3){\line(1,0){2}}
 \put(17,1){\line(-1,1){2}}
 \put(14.85,0.8){$\bullet$}
\put(14.85,2.8){$\bullet$}

%%%%%%%%%%%%%%%%%%%%%%%%%%%%%%%%%
 \put(17,1){\line(0,1){2}}
 \put(3,1){\line(0,1){2}}
 \put(16.85,0.8){$\bullet$}
\put(2.85,2.8){$\bullet$}
 %%%%%%%%%%%%%%%%%%%%%%%%%%%

\put(18.5, 2){%\footnotesize
 %\put(2.5,-2){\footnotesize
$\;\; = \;\; \int d^D x_1 d^D y_2
\dfrac{\langle x_1,x_2 | (\hat{Q}_{12}^{(\beta)})^{2N}| y_1,y_2
\rangle} {(x_1-x_2)^{2\beta}}\;$ ,}

\end{picture}

\vspace{-1cm}

\be
\lb{zgzg3}
{}
\ee

\vspace{0.5cm}

\noindent
for odd loops

\unitlength=4mm
\begin{picture}(25,4)(2,1)

\put(2.2,3.2){\footnotesize $x_1$}
\put(2.1,0.6){\footnotesize $x_2$}

\put(5,1){\line(0,1){2}}
\put(3,3){\line(1,0){2}}
 \put(5,1){\line(-1,1){2}}
 \put(3,1){\line(1,0){2}}

\put(3.8,0.5){\tiny $\beta'$}
 \put(4,3.2){\tiny $\beta'$}
 \put(5.8,0.5){\tiny $\beta'$}
 \put(6,3.2){\tiny $\beta'$}
 \put(7.8,0.5){\tiny $\beta'$}
 \put(8,3.2){\tiny $\beta'$}

 \put(4.6,1.9){\tiny $\beta$}
\put(3.5,1.8){\tiny $\beta$}
 \put(6.6,1.9){\tiny $\beta$}
\put(5.5,1.8){\tiny $\beta$}
 \put(8.6,1.9){\tiny $\beta$}
\put(7.5,1.8){\tiny $\beta$}
%%%%%%%%%%%%%%%%%%%%%%%%%%%%

\put(7,1){\line(0,1){2}}
\put(5,1){\line(1,0){2}}
\put(5,3){\line(1,0){2}}
 \put(7,1){\line(-1,1){2}}
 \put(4.85,0.8){$\bullet$}
\put(4.85,2.8){$\bullet$}

 %%%%%%%%%%%%%%%%%%%%%%%%%%%%%%

 \put(6.85,0.8){$\bullet$}
\put(6.85,2.8){$\bullet$}
 \put(9,1){\line(0,1){2}}
\put(7,1){\line(1,0){2}}
\put(7,3){\line(1,0){2}}
 \put(9,1){\line(-1,1){2}}

\put(8.85,0.8){$\bullet$}
\put(8.85,2.8){$\bullet$}
%%%%%%%%%%%%%%%%%%%%%%%%%%%%%%
 \put(10,2){$. \; .\; . \; . $}
%%%%%%%%%%%%%%%%%%%%%%%%%%%%%%

\put(14,1){\line(0,1){2}}
\put(12,1){\line(1,0){2}}
\put(12,3){\line(1,0){2}}
 \put(14,1){\line(-1,1){2}}

%%%%%%%%%%%%%%%%%%%%%%%%%%%%

\put(14.8,0.5){\tiny $\beta'$}
\put(15,3.2){\tiny $\beta'$}
\put(12.9,3.2){\tiny $\beta'$}
\put(12.7,0.5){\tiny $\beta'$}
\put(16.9,0.5){\tiny $\beta'$}
\put(16.1,1.9){\tiny $\beta$}
\put(17.2,2){\tiny $\beta$}
\put(14.6,1.7){\tiny $\beta$}
\put(13.6,1.9){\tiny $\beta$}
\put(12.6,1.7){\tiny $\beta$}

\put(16,1){\line(0,1){2}}
\put(14,1){\line(1,0){2}}
\put(14,3){\line(1,0){2}}
 \put(16,1){\line(-1,1){2}}
  %\put(11.85,0.8){$\bullet$}
  %\put(11.85,2.8){$\bullet$}
 \put(13.85,0.8){$\bullet$}
\put(13.85,2.8){$\bullet$}

\put(16,1){\line(1,0){2}}
 \put(15.85,0.8){$\bullet$}
\put(18.2,0.8){\footnotesize $y_1$}
\put(16.4,3){\footnotesize $y_2$}

%%%%%%%%%%%%%%%%%%%%%%%%%%%%%%%%%
 \put(3,1){\line(0,1){2}}
  \put(18,1){\line(-1,1){2}}
 \put(15.85,2.8){$\bullet$}
\put(2.85,2.8){$\bullet$}
 %%%%%%%%%%%%%%%%%%%%%%%%%%%

\put(19,2){
$\; = \;\; \int d^D x_1 d^D y_2
\dfrac{\langle x_1,x_2 | (\hat{Q}_{12}^{(\beta)})^{2N+1}| y_1,y_2
\rangle} {(x_1-x_2)^{2\beta}}\,$ ,}

\end{picture}

\vspace{-1cm}

\be
\lb{zgzg4}
{}
\ee

\vspace{0.6cm}
\noindent
and the corresponding two point correlation
function in $M$-loop order is given by
\begin{equation}\label{g-2}
    G^{(M)}_2(x_2,y_1) = \int d^D x_1 d^D y_2
    \dfrac{\langle x_1,x_2 | (\hat{Q}_{12}^{(\beta)})^{M}| y_1,y_2
    \rangle} {(x_1-x_2)^{2\beta}}\,.
\end{equation}

As we will see
 in the following sections,  the
 problem of calculating the
 diagrams of the zig-zag series
  is reduced to the
  construction of a complete
  set of eigenvectors (and the corresponding
  eigenvalues) for the graph building
 operator $Q_{12}^{(\beta)}$. In other
 words, we need to investigate the
 spectral problem
 for the operator $Q_{12}^{(\beta)}$.
 So our method for evaluating zig-zag
 diagrams will almost repeat
 the method for evaluating
 ladder diagrams applied in Sect. {\bf \ref{2ladder}}.

%%%%%%%%%%%%%%%%%%%%%%%%%%%%%%%%%%%%%
%%%%%%%%%%%%%%%%%%%%%%%%%%%%%%%%%%%%%

\section{Eigenfunctions for the
graph building operator $\hat{Q}_{12}$.
Scalar product and completeness\label{eigpro}}
\setcounter{equation}0

 \subsection{Conformal triangles\label{tricon}}

To find a complete set of eigenvectors for the
graph building operator (\ref{q-d}) we
firstly consider the 3-point correlation function
of three conformal
fields ${\cal O}_{\Delta_1}$, ${\cal O}_{\Delta_2}$
and ${\cal O}_{\Delta}^{\mu_1...\mu_n}$
 in the $D$-dimensional conformal field theory $(D>2)$.
 Here  ${\cal O}_{\Delta_1}$ and ${\cal O}_{\Delta_2}$
 are two scalar fields with conformal dimensions
 $\Delta_1$ and $\Delta_2$, while
 ${\cal O}_{\Delta}^{\mu_1...\mu_n}$
 is the rank-$n$ tensor field with conformal dimension
 $\Delta$. Recall that the rank-$n$ tensor
 ${\cal O}_{\Delta}^{\mu_1...\mu_n}$
 is symmetric and traceless. The single conformally-invariant 3-point
 correlation function is completely fixed
(up to a normalization constant
 factor) and is represented as
 \cite{ToMiPe}, \cite{DPPT},
 \cite{GKK},\cite{Polyakov1},\cite{Polyakov2}
\begin{multline}
\lb{cor01}
\langle  {\cal O}_{\Delta_1}(x_1) \;
{\cal O}_{\Delta_2}(x_2) \; {\cal O}_{\Delta}^{\mu_1...\mu_n}(x)
\rangle = \\
\frac{\left(\frac{x-x_1}{(x-x_1)^2}-\frac{x-x_2}{
(x-x_2)^2}\right)^{\mu_1\cdots\mu_n}}
{(x_1-x_2)^{\Delta_1+\Delta_2-\Delta+n}
(x-x_1)^{\Delta_1-\Delta_2+\Delta-n}
(x-x_2)^{\Delta_2-\Delta_1+\Delta-n}}\; ,
\end{multline}
where $x_1, x_2, x \in \mathbb{R}^D$ and
 %by power we mean
 $x^{\lambda} := (x^2)^{\frac{\lambda}{2}}$.
The index structure is the same as in \eqref{eigfun},
so $x^{\mu_1\ldots\mu_n}$ denotes the symmetric and traceless part of
$x^{\mu_1}\ldots x^{\mu_n}$.

As we will see in Section {\bf \ref{orth}} for generic
$\Delta_1, \Delta_2 \in \mathbb{R}$ and a special choice
of the parameter
\be
\lb{specc}
 \Delta = D/2 + 2i\nu \,, \;\;\;\; \nu \in \mathbb{R} \; ,
 \ee
 the
system of functions
\begin{align}
\lb{conf-tr}
\Psi^{\mu_1\cdots\mu_n}_{\nu,x}(x_1\,,x_2) =
\frac{\left(\frac{x-x_1}{(x-x_1)^2}-\frac{x-x_2}{
(x-x_2)^2}\right)^{\mu_1\cdots\mu_n}}
{(x_1-x_2)^{\Delta_1+\Delta_2-\Delta+n}
(x-x_1)^{\Delta_1-\Delta_2+\Delta-n}
(x-x_2)^{\Delta_2-\Delta_1+\Delta-n}},
\end{align}
is orthogonal and complete with respect to some
natural scalar product on the space of functions
of two variables $x_1,x_2$.
For the sake of simplicity, we
will sometimes not indicate the
dependence on the parameters $\Delta_1$ and
$\Delta_2$ in the notation
$\Psi^{\mu_1\cdots\mu_n}_{\nu,x}(x_1\,,x_2)$.

In Sections {\bf \ref{diaS}},
{\bf \ref{scprpsi}}, we show that the eigenvectors of
the operator $\hat{Q}^{(\beta)}_{12}$
are given by a special form of the 3-point function (\ref{cor01})
when three parameters $\Delta_1,\Delta_2,\Delta$ are expressed
in terms of dimension $D$
and two arbitrary numbers
$\alpha \in \mathbb{C}\,,\beta \in \mathbb{R}$
 \be
 \lb{cor00}
 \Delta_1 = \frac{D}{2} \; , \;\;\; \Delta_2 = \frac{D}{2}-\beta
 \; , \;\;\; \Delta = D-2\alpha-\beta+n \; ,
 \ee
 so that we define
 \be
 \lb{cor02}
 \Psi^{\mu_1...\mu_n}_{\alpha,\beta,x}(x_1\,,x_2) \;  : = \;
 \frac{\bigl( \frac{x-x_1}{(x-x_1)^2} -
\frac{x-x_2}{(x-x_2)^2} \bigr)^{\mu_1...\mu_n}}{(x_1-x_2)^{2\alpha}
(x-x_1)^{2\alpha'}(x-x_2)^{2(\alpha+\beta)'}}
 \; .
 \ee
where $\alpha' := \frac{D}{2} - \alpha$ and
$(\alpha+ \beta)' = \frac{D}{2} - (\alpha + \beta)$.

Note that (\ref{cor02}) is a formal eigenfunction for an arbitrary
complex parameter $\alpha$, but there exists some special choice
of the parameter $\alpha$ when the system of eigenfunctions is
orthogonal and complete. This choice was indicated
 in \eqref{specc} and for this choice we deduce
the relations
 \be
 \lb{condes}
 %\Delta_1, \Delta_2 \in \mathbb{R} \; , \;\;\;
 \Delta = D - 2\alpha - \beta +n = D/2 + 2i\nu \;\;\;\; \Rightarrow
 \;\;\;\;
 \alpha = \frac{1}{2}\left(n + D/2 - \beta\right) - i \nu \,.
 \ee
Finally, the orthogonal and complete system of eigenfunctions
of the operator $\hat{Q}^{(\beta)}_{12}$ is given by
\begin{equation}\label{d-eigenf}
 %\label{d-eigenfunc}
  \Psi^{\mu_1...\mu_n}_{\nu,\beta,x}(x_1\,,x_2)
  =\frac{\left(\frac{x-x_1}{
    (x-x_1)^2}-\frac{x-x_2}{
    (x-x_2)^2}\right)^{\mu_1...\mu_n}}
{(x_1-x_2)^{D/2 - \beta + n - 2i\nu}
(x-x_1)^{D/2 + \beta - n + 2i\nu}
(x-x_2)^{D/2 - \beta - n + 2i\nu}}\,.
\end{equation}

In \eqref{conf-tr} we consider
 $x \in \mathbb{R}^D$ as a parameter, while $x_1$ and $x_2$ are related
 to the Heisenberg algebras
 ${\cal H}_1$ and ${\cal H}_2$.
 In other words, we interpret
 the wave function (\ref{conf-tr}) as
 follows:
\begin{equation}
\label{wavef}
 \Psi^{\mu_1...\mu_n}_{\nu,x}(x_1\,,x_2)
  = \langle x_1, x_2 |
 \Psi^{\mu_1...\mu_n}_{\nu,x}\rangle \;
\end{equation}
and similarly
\begin{equation}
\label{wavefab}
 \Psi^{\mu_1...\mu_n}_{\alpha,\beta,x}(x_1\,,x_2)
  = \langle x_1, x_2 |
 \Psi^{\mu_1...\mu_n}_{\alpha,\beta,x}\rangle \ \ \ ;\ \ \
 \Psi^{\mu_1...\mu_n}_{\nu,\beta,x}(x_1\,,x_2)
  = \langle x_1, x_2 |
 \Psi^{\mu_1...\mu_n}_{\nu,\beta,x}\rangle \; .
\end{equation}

Following the paper \cite{VPH}, we call
the functions (\ref{conf-tr}),\eqref{cor02} and
\eqref{d-eigenf},
which are a special form of the three-point correlators (\ref{cor01}),
 {\it conformal triangles}.
In the next section we show that the conformal triangles (\ref{cor02})
 (3-point correlation functions (\ref{cor01})
 with a special choice of the conformal
  dimensions) are the eigenfunctions of the operator
 $\hat{Q}^{(\beta)}_{12}$ that
  was defined in (\ref{q-d}). So it is possible to write a spectral
  decomposition of the operator $\hat{Q}^{(\beta)}_{12}$ and
 evaluate\footnote{When we say analytically
 ''evaluate/calculate''
 multiple integrals of the
 type (\ref{zgzg44}) -- (\ref{zgzg22}),
 we mean reducing these multiple
 integrals to a
 single integral and/or a single infinite sum
 and then expressing the answer in terms of known special functions.}
  integrals \eqref{zgzg3} and \eqref{zgzg4}.

Before performing such calculations, we
first need to explore the properties of functions \eqref{cor02},
\eqref{conf-tr}.
The set of functions \eqref{conf-tr} forms a
complete orthogonal system (see \cite{DobMac},\cite{ToMiPe}
 and references therein).
In Section {\bf \ref{orth}}, we present our direct explicit proof of
orthogonality of these functions with a special measure and derive
from them a possible form of the completeness identity.

\subsection{Diagonalization of the operator
$\hat{Q}^{(\beta)}_{12}$\label{diaS}}

First we
prove that the functions \eqref{cor02}
are eigenfunctions for the operator $\hat{Q}^{(\beta)}_{12}$.

%{\bf Proposition 4.}
\begin{proposition}\label{prop2}
The vectors
 $| \Psi_{\alpha,\beta,y}^{\mu_1...\mu_n} \rangle \in V_1 \otimes V_2$,
 defined
 in (\ref{cor02}) and \eqref{wavefab}:

\unitlength=6mm
\begin{picture}(25,4)

\put(0,2){\footnotesize $\langle y_1,y_2 |
\Psi_{\alpha, \beta , y}^{\mu_1...\mu_n} \rangle \;\;
= \;$}

\put(5.5,3.2){\footnotesize $y_1$}
\put(5.6,1.9){\tiny $\alpha$}
\put(6.8,1.1){\tiny $\alpha'-\beta$}
\put(7,2.8){\tiny $\alpha'$}
\put(5.5,0.8){\footnotesize $y_2$}
\put(8.3,1.9){\footnotesize $y$}
\put(6,1){\line(0,1){2}}

 \put(6,3){\line(2,-1){2}}
 \put(6,1){\line(2,1){2}}
 %\put(7.9,1.8){$\bullet$}

\put(9,2){\footnotesize $\Bigl( \frac{y-y_1}{(y-y_1)^2} -
\frac{y-y_2}{(y-y_2)^2} \Bigr)^{\mu_1...\mu_n} =
\frac{\bigl( \frac{y-y_1}{(y-y_1)^2} -
\frac{y-y_2}{(y-y_2)^2} \bigr)^{\mu_1...\mu_n}}{(y_1-y_2)^{2\alpha}
(y-y_1)^{2\alpha'}(y-y_2)^{2(\alpha+\beta)'}}\, ,$ }

\end{picture}

\vspace{-2cm}
\be
\lb{psif}{}
\ee

\vspace{0.5cm}

 \noindent
 for any parameters $\alpha$
 and $\beta$, are eigenvectors for the
 graph building operator (\ref{q-d})
 \be
 \lb{cor03}
 \hat{Q}^{(\beta)}_{12} \;
 | \Psi_{\alpha,\beta,y}^{\mu_1...\mu_n} \rangle =
 \tau(\alpha,\beta,n)
 \; | \Psi_{\alpha,\beta,y}^{\mu_1...\mu_n} \rangle \; ,
 \ee
 with the eigenvalue
\be
 \lb{cor04}
 \tau(\alpha,\beta,n) = (-1)^n \,
\frac{\pi^{D/2} \Gamma(\beta) \Gamma(\alpha)
\Gamma(\alpha'-\beta+n)}{ \Gamma(\beta') \Gamma(\alpha'+n)
\Gamma(\alpha+\beta)}  \; .
 \ee
 \end{proposition}
{\bf Proof.} The statement of this
 Proposition follows from the chain
of equalities:
\begin{multline*}
\int d^D y_1 d^D y_2 \langle x_1,x_2 |
\hat{Q}^{(\beta)}_{12}  | y_1,y_2 \rangle \langle y_1,y_2 |
\Psi_{\alpha, \beta, y}^{\mu_1...\mu_n} \rangle = \\
\int d^D y_1 \;
\frac{1}{(x_2-y_1)^{2\beta'} (y_1-x_1)^{2(\alpha+\beta)}}
\frac{\bigl( \frac{y-y_1}{(y-y_1)^2} -
\frac{y-x_1}{(y-x_1)^2} \bigr)^{\mu_1...\mu_n}}{
(y-y_1)^{2\alpha'}(y-x_1)^{2(\alpha+\beta)'}}
\stackrel{(\ref{int04mu})}{=}
\\
\frac{\pi^{D/2} \Gamma(\beta) \Gamma(\alpha) \Gamma(\alpha'-\beta+n)}{
\Gamma(\beta') \Gamma(\alpha'+n)
\Gamma(\alpha+\beta)}\frac{1}{(x_1-x_2)^{2\alpha}}
  \frac{\bigl( \frac{y-x_2}{(y-x_2)^2} -
\frac{y-x_1}{(y-x_1)^2}
\bigr)^{\mu_1...\mu_n}}{(y-x_2)^{2(\alpha'-\beta)}
(y-x_1)^{2\alpha'}} = \\
(-1)^n \,
\frac{\pi^{D/2} \Gamma(\beta) \Gamma(\alpha) \Gamma(\alpha'-\beta+n)}{
\Gamma(\beta') \Gamma(\alpha'+n) \Gamma(\alpha+\beta)}
\langle x_1,x_2 | \Psi_{\alpha, \beta, y}^{\mu_1...\mu_n} \rangle
\end{multline*}
\hfill \qed

The proof is based on the tensor generalization of the
star-triangle identity:
\be
\lb{int04mu}
\int \frac{d^D z  \;\;\; \bigl( \frac{y-z}{(y-z)^2} -
\frac{y-x_1}{(y-x_1)^2} \bigr)^{\mu_1...\mu_n}}{(z-x_2)^{2\beta'}
(z-x_1)^{2(\alpha+\beta)}(z-y)^{2\alpha'}} =
\tau(\alpha,\beta,n)
\frac{ \bigl(
 \frac{y-x_2}{(y-x_2)^2} -
\frac{y-x_1}{(y-x_1)^2}
\bigr)^{\mu_1...\mu_n}}{(y-x_2)^{2(\alpha'-\beta)}
(x_2-x_1)^{2\alpha}(x_1-y)^{2\beta}}\,
\ee
(the standard star-triangle identity (\ref{startr}) is obtained for
$n=0$).
The derivation of this identity is given
 in Appendix {\bf C}.
We note that
identity (\ref{int04mu}) was presented in
\cite{FradPal} (see there eq. (A3.2) in Appendix 3).

{\bf Remark 1.}  In the scalar version of
 the star-triangle relation (\ref{startr2}) we
have an equivalent but simpler
  counterpart, the chain relation (\ref{chain}).
Of course, there exists a similar relative of
relation (\ref{int04mu})
\be
\lb{int05mu}
\int \frac{d^D z  \;\;\; \bigl( \frac{y-z}{(y-z)^2} -
\frac{y-x}{(y-x)^2} \bigr)^{\mu_1...\mu_n}}
{(z-x)^{2(\alpha+\beta)}(z-y)^{2\alpha'}} =
\tau(\alpha,\beta,n)\;
\frac{(x-y)^{\mu_1...\mu_n}}
{(x-y)^{2(n+\beta)}}\,.
\ee
In full analogy with the scalar case this relation
is obtained from (\ref{int04mu}) by sending $x_2 \to \infty$ and
changing the notation $x_1\to x$.
Relation (\ref{int05mu}) can be rewritten in an equivalent compact form
as follows:
\be
\lb{int06mu}
 \int d^D z \,
\frac{\langle z,x |\Psi^{\mu_1\cdots\mu_n}_{\alpha,\beta,y}\rangle}{
(z-x)^{2\beta}} = \tau(\alpha,\beta,n) \;
\frac{(x-y)^{\mu_1...\mu_n}}{(x-y)^{2(n+\alpha')}} \,
\ee
and it will be used later. Note that relation (\ref{int05mu})
generalizes the identity (\ref{dchain}).

\noindent
{\bf Remark 2.}  The analog of
Proposition {\bf \ref{prop2}}
 for fixed $D=4$ and $\beta=1$ was proven in \cite{GKK}.

 \subsection{Properties of the graph building operator and
 modified scalar product in $V_1 \otimes V_2$
 \label{scprpsi}}

Here we list properties of the graph building operator
$\hat{Q}_{12}^{(\beta)}$.
  Note that with respect to the standard Hermitian
 scalar product in $V_1 \otimes V_2$
  \be
 \lb{scpro}
 \langle \Psi | \Phi \rangle =
 \int d^D x_1 \, d^D x_2 \,
 \langle \Psi | x_1,x_2 \rangle
 \langle  x_1,x_2| \Phi \rangle =
  \int d^D x_1 d^D x_2 \, \Psi^*(x_1,x_2) \,\Phi(x_1,x_2) \; ,
 \ee
 the operator (\ref{q-d})
 is Hermitian, for $\beta \in \mathbb{R}$,  up to the equivalence
 transformation:
 \be
 \lb{hermc}
 (\hat{Q}_{12}^{(\beta)})^\dagger =
 \frac{1}{a(\beta)} \; (\hat{q}_{12})^{-2\beta}  \; (\p_1)^{-2\beta}
 \; {\cal P}_{12} = U \, \hat{Q}_{12}^{(\beta)} \, U^{-1}\, ,
 \;\;\;  U:= {\cal P}_{12} \, (\hat{q}_{12})^{-2\beta}=
 (\hat{q}_{12})^{-2\beta} \, {\cal P}_{12} \, .
 \ee
It means that the operator $\hat{Q}^{(\beta)}_{12}$
 is Hermitian with respect to the modified
  scalar product \cite{DISh}
   \begin{align}
 \lb{scprod}
\langle \overline{\Psi}|\Phi\rangle :=
\langle \Psi| \, U \, |\Phi\rangle =
\int d^D x_1 d^D x_2\frac{\Psi^{*}(x_2\,,x_1) \,\Phi(x_1\,,x_2)}{
(x_1-x_2)^{2\beta}} \; ,
\end{align}
where $\langle \Psi|$ is the standard Hermitian conjugation of the
vector $|\Psi \rangle$, and we introduce new conjugation
\begin{equation}
\label{nconj}
    \langle \overline{\Psi} | :=  \langle \Psi|U = \langle \Psi
    |\mathcal{P}_{12}(\hat{q}_{12})^{-2\beta}
    \; .
\end{equation}
 The operator $U$ in (\ref{scprod}),
 (\ref{nconj}) plays the role of
 the metric in the space $V_1 \otimes V_2$.
 Since the operator $\hat{Q}^{(\beta)}_{12}$
 is Hermitian with respect to the
 scalar product (\ref{scprod}),
  the eigenfunctions of $\hat{Q}^{(\beta)}_{12}$ with
  different eigenvalues should be orthogonal (we use this below).

  It is evident that the graph building operator
  (\ref{q-d}) commutes with the dilatation operator:
 \be
 \lb{qpd}
 \hat{\sf D} = \frac{i}{2} \sum_{a=1}^2 (\q_a \p_a + \p_a \q_a)
 + \frac{1}{2}(y^\mu \, \partial_{y^\mu}
 + \partial_{y^\mu} \, y^\mu) - \beta ,
 \ee
 which acts on the eigenvector
 $| \Psi_{\alpha,\beta,y}^{\mu_1...\mu_n}\rangle$ as following
 \be
\lb{qpd2}
 \hat{\sf D} \;
 | \Psi_{\alpha,\beta,y}^{\mu_1...\mu_n}\rangle =
 \Bigl(2\alpha+\beta-\frac{1}{2} D -n \Bigr) \;
 | \Psi_{\alpha,\beta,y}^{\mu_1...\mu_n}\rangle \; ,
 \ee
 and, for $\beta \in \mathbb{R}$, satisfies
 $\hat{\sf D}^\dagger = - U \, \hat{\sf D}\, U^{-1}$
 (to prove formula (\ref{qpd2}), one needs to act
 on both sides of (\ref{qpd2}) by the vector $\langle x_1,x_2|$).
 Thus,
 the operator $\hat{\sf D}$ is anti-Hermitian
 with respect to the scalar product (\ref{scprod}), and
 the corresponding condition on its eigenvalue gives (cf.
 (\ref{condes}))
 \be
 \lb{cor05}
 2(\alpha^* + \alpha) = 2n +D-2\beta \;\;\;\; \Rightarrow \;\;\;\;
 \alpha = \frac{1}{2}\left(n + D/2 - \beta\right) - i \nu \; ,
 \;\;\;\;
 \nu \in \mathbb{R} \; .
 \ee
 It is a remarkable fact that under this condition,
  the eigenvalue (\ref{cor04}) is real
 \be
 \lb{cor11}
 \tau(\alpha,\beta,n) =  (-1)^n \,
\frac{\pi^{D/2} \Gamma(\beta) \; \Gamma(\frac{D}{4} + \frac{n}{2} -
\frac{\beta}{2} + i \nu) \, \Gamma(\frac{D}{4} + \frac{n}{2} -
\frac{\beta}{2} - i \nu)}{ \Gamma(\beta') \;
\Gamma(\frac{D}{4} + \frac{n}{2} +\frac{\beta}{2} + i \nu) \,
\Gamma(\frac{D}{4} + \frac{n}{2} +\frac{\beta}{2} - i \nu)} \; ,
 \ee
 and the parameter $\Delta$ in (\ref{cor00})
 acquires the form $\Delta = \frac{D}{2} + 2 i \nu$ (see
 (\ref{condes})).

 Note that Proposition
 {\bf \ref{prop2}} holds for an arbitrary parameter $\alpha$, but the
 set of functions \eqref{conf-tr} forms a complete orthogonal system
 only in case
 when $\Delta = D/2 + 2i\nu$, which imposes restrictions on the
 parameter $\alpha$.

\subsection{Orthogonality and completeness
for the eigenfunctions $| \Psi_{\nu,x}^{\mu_1...\mu_n}\rangle$}
\label{orth}

Since the eigenvalue (\ref{cor11}) is real (it is invariant under
the transformation $\nu \to - \nu$), two eigenvectors
$| \Psi_{\nu,\beta,x}^{\mu_1...\mu_n}\rangle$ and
$| \Psi_{\lambda,\beta,y}^{\nu_1...\nu_m}\rangle$,
having different eigenvalues (\ref{cor11})
(e.g. $n \neq m$ and $\lambda \neq \pm \nu$), should be
orthogonal to each other with respect to
the scalar product (\ref{scprod}).

It is indeed the fact and moreover,
we have the following more general statement --
the orthogonality condition
for two general conformal triangles (see, e.g.,
\cite{DobMac}, \cite{ToMiPe}, \cite{DPPT}, \cite{GKK}).
The scalar product in this more general situation is defined
in a similar to \eqref{scprod} way
\begin{equation}
\lb{Cscprod}
\langle \overline{\Psi}|\Phi\rangle :=
\langle \Psi| \, U \, |\Phi\rangle =
\int d^D x_1 d^D x_2\frac{\Psi^{*}(x_2\,,x_1) \,\Phi(x_1\,,x_2)}{
(x_1-x_2)^{2(D-\Delta_1-\Delta_2)}} \; ,
 %\;\;\;\;\;
 %(\beta \equiv D-\Delta_1 - \Delta_2) \; ,
\end{equation}
where $\langle \Psi|$ is the standard Hermitian conjugation of the
vector $|\Psi \rangle$ and we introduce new conjugation
\begin{equation}
\label{Cnconj}
    \langle \overline{\Psi} | :=  \langle \Psi|U = \langle \Psi
    |\mathcal{P}_{12}\,\hat{q}_{12}^{-2(D-\Delta_1-\Delta_2)}
    \; .
\end{equation}

\begin{proposition}\label{prop3} Eigenfunctions \eqref{conf-tr} form an
orthogonal system of functions and the following orthogonality relation
holds
\begin{multline}
\label{trig-prod}
\langle \overline{\Psi_{\lambda,y}^{\nu_1...\nu_m}}|
\Psi_{\nu,x}^{\mu_1...\mu_n}\rangle =
    \int \! d^D x_1 \, d^D x_2
\frac{\Bigl(\Psi^{\nu_1...\nu_m}_{\lambda,y}(x_2\,,x_1)
\Bigr)^{*} \Psi^{\mu_1...\mu_n}_{\nu,x}(x_1\,,x_2)}
{(x_1-x_2)^{2(D-\Delta_1-\Delta_2)}} = \\ C_1(n\,,\nu)\,\delta_{n
m}\,\delta(\nu -\lambda)\,\delta^D(x-y)\,P^{\mu_1 ... \mu_n}_{\nu_1 ...
\nu_n} +
C_2(n\,,\nu)\,\delta_{n m}\,\delta(\nu +\lambda)\,
\frac{S^{\mu_1 ... \mu_n}_{\nu_1 ... \nu_n}(x,y)}{(x-y)^{D+4i\nu}},
\end{multline}
where $P^{\mu_1 ... \mu_n}_{\nu_1 ... \nu_n}$ is the projector on
the symmetric traceless tensors
(see \eqref{orthog} Appendix {\bf A}) and
\begin{align*}
S^{\mu_1 ... \mu_n}_{\nu_1 ... \nu_n}(x,y) =
P^{\mu_1 ... \mu_n}_{\alpha_1 ... \alpha_n}
\left(\delta^{\alpha_1}_{\beta_1} -
2\frac{(x-y)^{\alpha_1}(x-y)_{\beta_1}}{(x-y)^2}\right)\cdots
\left(\delta^{\alpha_n}_{\beta_n} -
2\frac{(x-y)^{\alpha_n}(x-y)_{\beta_n}}{(x-y)^2}\right)
P^{\beta_1 ... \beta_n}_{\nu_1 ... \nu_n}.
\end{align*}
The explicit form of the coefficients $C_1$ and $C_2$ is
\begin{align}\label{C1}
C_1(n\,,\nu)  =
\frac{(-1)^n\,2^{1-n}\,\pi^{3D/2 + 1}\,n!\,
\Gamma\left(2i\nu\right)\Gamma\left(-2i\nu\right)}{
\Gamma\left(\frac{D}{2} + n\right)
\left(\left(\frac{D}{2}+n-1\right)^2 + 4\nu^2\right)\,
\Gamma\left(\frac{D}{2}+2i\nu-1\right)\,
\Gamma\left(\frac{D}{2}-2i\nu-1\right)}\\
\label{C2}
C_2(n\,,\nu) =
\pi^{D+1}\,\frac{n!}{2^{n-1}}\,C_{\Delta_1\Delta_2}(n\,,\nu)\,
\frac{\Gamma\left(2i\nu\right)\Gamma\left(\frac{D}{2}+2i\nu-1+n\right)}
{\Gamma\left(\frac{D}{2}+n-2i\nu\right)
\Gamma\left(\frac{D}{2}+2i\nu-1\right)
\Gamma\left(\frac{D}{2}+n\right)}\,,
\end{align}
where for simplicity we use special notation
for the following combination of $\Gamma$-functions
\begin{align}
\label{C}
C_{\Delta_1\Delta_2}(n\,,\nu) =
\frac{\Gamma\left(\frac{D}{4}-\frac{\Delta_1-\Delta_2}{2}+
\frac{n}{2}-i\nu\right)}
{\Gamma\left(\frac{D}{4}-\frac{\Delta_1-\Delta_2}{2}+
\frac{n}{2}+i\nu\right)}\,
\frac{\Gamma\left(\frac{D}{4}+\frac{\Delta_1-\Delta_2}{2}+
\frac{n}{2}-i\nu\right)}
{\Gamma\left(\frac{D}{4}+\frac{\Delta_1-\Delta_2}{2}+
\frac{n}{2}+i\nu\right)}\,
\end{align}

\end{proposition}

\textbf{Proof.} The proof of this statement is very technical and
cumbersome,
 and we move it to Appendix {\bf D}. \hfill \qed

%%%%%%%%%%%%%%%%%%%%%%%%%%%
\vspace{0.2cm}

\noindent
\textbf{Remark 1.} Formula (\ref{trig-prod}) for the scalar product
was obtained and discussed without its detailed proof in many papers
(see e.g.  \cite{Lipatov}, \cite{DobMac,ToMiPe,DPPT}, \cite{GKK}).
That is why we decided
to give here our direct
derivation of this formula
(see Appendix {\bf D}).
For example, the two-dimensional analog of (\ref{trig-prod}) is
given by L.Lipatov in \cite{Lipatov} and in a completely
different context of the representation theory
of $SL(2,\mathbb{C})$ \cite{Gelfand} by M.Naimark in \cite{Naimark}.
Our consideration in Appendix {\bf D} is very similar to
the two-dimensional derivation given in \cite{DerSpir}.\\

\noindent
\textbf{Remark 2.}
The general conformal triangle
$| \Psi_{\nu,x}^{\mu_1...\mu_n}\rangle$  \eqref{conf-tr}
depends on two arbitrary real parameters $\Delta_1$ and $\Delta_2$.
The eigenfunction $| \Psi_{\nu,\beta,x}^{\mu_1...\mu_n}\rangle$
is obtained from the general conformal triangle
$| \Psi_{\nu,x}^{\mu_1...\mu_n}\rangle$  by reduction
when the parameters
are fixed in a special way
\be
\Delta_1 = \frac{D}{2} \; , \;\;\; \Delta_2 = \frac{D}{2}-\beta \,.
\ee
The choice of a specific scalar product for the functions $|
\Psi_{\nu,\beta,x}^{\mu_1...\mu_n}\rangle$ was motivated in
Section {\bf \ref{scprpsi}} (see also
Section {\bf \ref{algf}}  below), and indeed due to the relation
$D - \Delta_1 - \Delta_2 = \beta$ it coincides with the
scalar product from Proposition {\bf \em \ref{prop3}}.
Thus, the scalar product of two eigenfunctions
$| \Psi_{\nu,\beta,x}^{\mu_1...\mu_n}\rangle$ is  taken
as in the left-hand side of (\ref{trig-prod}).
We also note that the coefficient $C_1$ does not depend
on $\Delta_1,\Delta_2$ and plays an important role
as the inverse of the Plancherel measure used
in the completeness condition (see below). In contrast to this,
the coefficient $C_2$ in (\ref{C2})
depends on $\Delta_1,\Delta_2$, but the explicit form for $C_2$
 will not be important for us.

\vspace{0.2cm}

Now we turn over to a consideration
 of the completeness relation. Since
 the orthogonality condition (\ref{trig-prod})
 is known, the natural conjecture
 for the completeness relation
 (resolution of unity) has the
 following form
\begin{equation}
 \label{complet2}
    \sum_{n = 0}^{\infty}\int\limits_{-\infty}^{+\infty} d\nu \mu(n,
    \nu) \int d^Dx |\Psi_{\nu,x}^{\mu_1\ldots\mu_n}\rangle\langle
    \overline{\Psi_{\nu,x}^{\mu_1\ldots\mu_n}}| = \operatorname{I},
\end{equation}
where $\operatorname{I}$ is the unity operator
 in $V_1 \otimes V_2$, and the conjugated vector
 $\langle\overline{\Psi}|$ is defined in (\ref{Cnconj}).

%\textbf{Proposition 3.}
\begin{proposition}\label{prop4}
The integration measure $\mu(n,\nu)$
in (\ref{complet2}) has the following form:
\begin{equation}
\label{measure}
    \mu(n\,,\nu)= \frac{1}{2}\frac{1}{C_1(n\,,\nu)},
\end{equation}
where $C_1$ is a constant \eqref{C1}.
\end{proposition}
\textbf{Proof.} Let us rewrite relation (\ref{complet2})
in terms of integral kernels applying the
vectors $\langle x_1, x_2|$ from the left
and $|x_3, x_4 \rangle$ from the right
\begin{multline*}
\delta^D(x_{13})\delta^D(x_{24}) = \sum_{n=0}^{\infty}
\int_{-\infty}^{\infty} d \nu \mu(n\,,\nu)
\frac{1}{x_{12}^{2\left(D-\Delta_1-\Delta_2\right)}}\,
\int d^D x\, \Psi^{\mu_1...\mu_n}_{n,-\nu,x}(x_1\,,x_2)\,
\Psi^{\mu_1...\mu_n}_{n, \nu,x}(x_4\,,x_3),
\end{multline*}
where we used $\langle \Psi^{\mu_1...\mu_n}_{\nu,x} |x_1,x_2\rangle=
(\Psi^{\mu_1...\mu_n}_{\nu, x}(x_1, x_2))^* =
\Psi^{\mu_1...\mu_n}_{-\nu, x}(x_1, x_2)$. For further convenience, we
change the
 variables and represent the previous relation
 in a slightly different form
\begin{multline*}
\delta^D(x_{13})\delta^D(x_{24}) = \sum_{m=0}^{\infty}
\int_{-\infty}^{\infty} d \lambda\, \mu(m\,,\lambda)
\frac{1}{x_{12}^{2\left(D-\Delta_1-\Delta_2\right)}}\,
\int d^D y \,\Psi^{\nu_1...\nu_m}_{m,-\lambda,y}(x_1\,,x_2)\,
\Psi^{\nu_1...\nu_m}_{m,\lambda,y}(x_4\,,x_3)
\end{multline*}
Then we multiply both sides of this relation by
$\Psi^{\mu_1...\mu_n}_{\nu,x}(x_3\,,x_4)$,
integrate over $x_3$ and $x_4$ and use formula
(\ref{trig-prod}) for the scalar product of two
conformal triangles written in the form
\begin{multline*}
\langle\overline{
\Psi^{\nu_1...\nu_m}_{-\lambda,y}}|
\Psi^{\mu_1...\mu_n}_{\nu,x}\rangle =
C_1(n\,,\nu)\,\delta_{n m}\,\delta(\nu +\lambda)\,\delta^D(x-y)\,
P^{\mu_1 ... \mu_n}_{\nu_1 ... \nu_n} + \\
C_2(n\,,\nu)\,\delta_{n m}\,\delta(\nu -\lambda)\,
\frac{S^{\mu_1 ... \mu_n}_{\nu_1 ... \nu_n}(x,y)}
{(x-y)^{2\left(\frac{D}{2}+2i\nu\right)}}
\end{multline*}
and obtain
\begin{multline*}
\Psi^{\mu_1...\mu_n}_{\nu,x}(x_1\,,x_2) =
\int_{-\infty}^{\infty} d \lambda\, \mu(n\,,\lambda) \int d^D y
\Psi^{\nu_1...\nu_n}_{-\lambda,y}(x_1\,,x_2)\,\\
\left(
C_1(n\,,\nu)\,\delta(\nu +\lambda)\,\delta^D(x-y)\,
P^{\mu_1 ... \mu_n}_{\nu_1 ... \nu_n} +
C_2(n\,,\nu)\,\delta(\nu -\lambda)\,
\frac{S^{\mu_1 ... \mu_n}_{\nu_1 ... \nu_n}(x,y)}
{(x-y)^{2\left(\frac{D}{2}+2i\nu\right)}}\right) = \\
\mu(n\,,-\nu)\,C_1(n\,,\nu)\,\Psi^{\mu_1...\mu_n}_{n,\nu,x}(x_1\,,x_2)
+\mu(n\,,\nu)\,C_2(n\,,\nu)\,
\int d^D y\, \Psi^{\nu_1...\nu_n}_{-\nu,y}(x_1\,,x_2)\,
\frac{S^{\mu_1 ... \mu_n}_{\nu_1 ... \nu_n}(x,y)}
{(x-y)^{2\left(\frac{D}{2}+2i\nu\right)}}
\end{multline*}
Now we need relation \eqref{amp1} with change $\nu \to -\nu$
\begin{multline*}
\int \mathrm{d}^D\, z\,
\frac{S^{\mu_1...\mu_n}_{\nu_1...\nu_n}(x,z)}
{(x-z)^{2\left(\frac{D}{2}+2i\nu\right)}}
\Psi^{\nu_1...\nu_n}_{-\nu,z}(x_1\,,x_2) = \\
\Psi^{\mu_1...\mu_n}_{\nu,x}(x_1\,,x_2)
(-1)^n\pi^{\frac{D}{2}}\,
C_{\Delta_1\Delta_2}(n\,,-\nu)\,
\frac{\Gamma(-2i\nu)\Gamma\left(\frac{D}{2}-2i\nu+n-1\right)}
{\Gamma\left(\frac{D}{2}-2i\nu-1\right)\Gamma\left(\frac{D}{2}+2i\nu+n\right)}
\end{multline*}
so that
\begin{multline*}
\Psi^{\mu_1...\mu_n}_{\nu,x}(x_1\,,x_2) =
\mu(n\,,-\nu)\,C_1(n\,,\nu)\,\Psi^{\mu_1...\mu_n}_{n,\nu,x}(x_1\,,x_2)
\, + \\
\mu(n\,,\nu)\,C_2(n\,,\nu)\,\Psi^{\mu_1...\mu_n}_{n,\nu,x}(x_1\,,x_2)
(-1)^n\pi^{\frac{D}{2}}\,
C_{\Delta_1\Delta_2}(n\,,-\nu)
\frac{\Gamma(-2i\nu)\Gamma\left(\frac{D}{2}-2i\nu+n-1\right)}
{\Gamma\left(\frac{D}{2}-2i\nu-1\right)\Gamma\left(\frac{D}{2}+2i\nu+n\right)}
= \\
= \mu(n\,,-\nu)\,C_1(n\,,\nu)\,\Psi^{\mu_1...\mu_n}_{\nu,x}(x_1\,,x_2)
+\mu(n\,,\nu)\,C_1(n\,,-\nu)\,\Psi^{\mu_1...\mu_n}_{\nu,x}(x_1\,,x_2)
\end{multline*}
where at the last step we used the consistency relation \eqref{2}.
First of all, note that $C_1(n\,,-\nu) = C_1(n\,,\nu)$.
Secondly, we should have
$\mu(n\,,-\nu) = \mu(n\,,\nu)$ due to the fact that the convolution of
conformal triangles inside the completeness relation
is invariant under the change $\nu \to -\nu$.
The last statement should be checked with the help of two relations
\eqref{convolution1} and \eqref{amp1}.
Finally, we obtain relation (\ref{measure})
and arrive at the completeness relation in the form (\ref{complet2}).
\hfill \qed

\vspace{0.2cm}

\textbf{Remark 1.} Using the
 above mentioned statement that the convolution of conformal triangles
inside the completeness relation is invariant
under the change $\nu \to -\nu$, it is possible to rewrite
the completeness relation in an equivalent form
 (see e.g. \cite{DobMac}, \cite{ToMiPe}, \cite{DPPT}, \cite{GKK})
\begin{multline}\label{complet1}
\delta^D(x_{13})\delta^D(x_{24}) = \sum_{n=0}^{\infty}
\int_{0}^{\infty} \frac{d \nu}{C_1(n\,,\nu)}
\frac{1}{x_{12}^{2\left(D-\Delta_1-\Delta_2\right)}}\,
\int d^D x\, \Psi^{\mu_1...\mu_n}_{-\nu,x}(x_1\,,x_2)\,
\Psi^{\mu_1...\mu_n}_{\nu,x}(x_4\,,x_3)
\end{multline}

 \vspace{0.2cm}

\textbf{Remark 2.}
 The concise version of
 the completeness condition (\ref{complet1})
 (or resolution of unity $I$)
 for the basis eigenfunctions (\ref{wavef}) is written as (cf.
 (\ref{complet2}))
 \be
 \lb{cor08}
 \begin{array}{c}
 \displaystyle
 I = \sum_{n=0}^{\infty}
\int_{0}^{\infty} \frac{d \nu}{C_1(n\,,\nu)} \int d^D x\,
| \Psi^{\mu_1...\mu_n}_{\nu,x}\rangle
\langle \overline{\Psi^{\mu_1...\mu_n}_{\nu,x} } |
 = \\ [0.3cm]
\displaystyle
= \sum_{n=0}^{\infty}
\int_{0}^{\infty} \frac{d \nu}{C_1(n\,,\nu)} \int d^D x\,
| \Psi^{\mu_1...\mu_n}_{\nu,x}\rangle
\langle \Psi^{\mu_1...\mu_n}_{\nu,x}
|\, U \; .
 \end{array}
 \ee

\vspace{0.2cm}

\textbf{Remark 3.} Due to the degeneracy $\nu \to -\nu$,
 it is possible to restrict everything to the case
 of positive $\nu$ from the very beginning.
In this case, the term with $C_2(n\,,\nu)$ disappears from the
orthogonality relation \eqref{trig-prod} and the form of the measure in
the
completeness relation \eqref{complet1} can be fixed using the
orthogonality relation for the positive $\nu$. We used the
 orthogonality relation  for
arbitrary $\nu$ for the sake of completeness
of the presentation and for additional
nontrivial crosschecks of the derived constants
$C_1(n,\nu)$ and $C_2(n,\nu)$ and the consistency relation \eqref{2}.

\vspace{0.2cm}

\textbf{Remark 4.} Note that we have fixed the measure
in the completeness relation but the whole completeness
relation needs independent proof which is given in
\cite{DobMac,ToMiPe,DPPT}.
In the two-dimensional case, the completeness is proved in the context
of representation theory of $SL(2,\mathbb{C})$ in
\cite{Gelfand, Naimark}.
The simple and direct proof of the completeness in the two-dimensional
case is presented in \cite{BelDer}.

\vspace{0.2cm}

\begin{figure}[h]
\center{\includegraphics[width=0.9\linewidth]{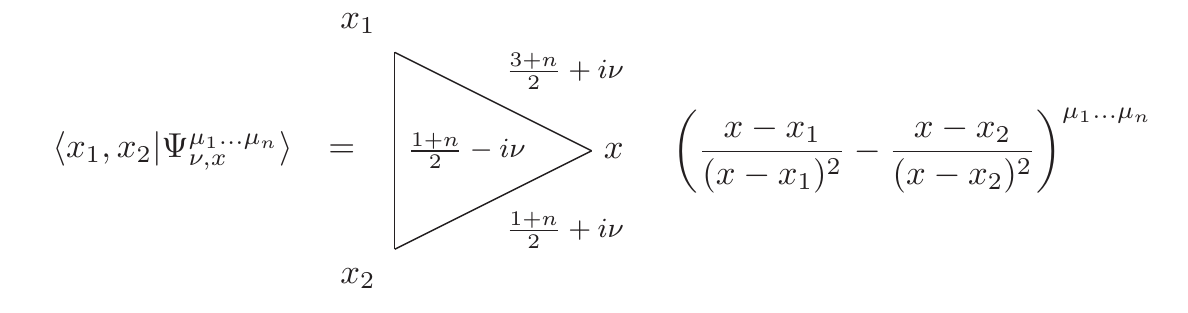}}
\caption{Set of eigenfunctions in case $D = 4$ and $\beta = 1$}
\label{eigen}
\end{figure}

Let us point out the four-dimensional case,
i.e. $D = 4, \beta = 1$. In this case,
 a complete orthogonal set of functions has parameters $\alpha =
 \frac{n + 1}{2} - i\nu$ and the following function is shown in
 Fig.\ref{eigen}.

\noindent
For these functions we have
\begin{equation}
\hat{Q}_{12}|\Psi_{\nu,x}^{\mu_1...\mu_n}\rangle =
\tau(\nu, n)|\Psi_{\nu,x}^{\mu_1...\mu_n}\rangle,
\end{equation}
where
\begin{equation}\label{4-egvalue}
\tau(\nu, n) = (-1)^n\dfrac{4\pi^2}{(1+n)^2 + 4\nu^2}.
\end{equation}
Integration measure \eqref{measure} in the completeness relation is
\begin{equation}\label{4-measure}
    \dfrac{1}{\mu(n, \nu)} = \dfrac{\pi^5}{2^{n + 2}(n +
    1)\nu^2}\,\tau(\nu, n)
\end{equation}
and we can write spectral decomposition for the operator
$\hat{Q}_{12}$
\begin{equation}
    \hat{Q}_{12} = \sum_{n =
    0}^{\infty}\int\limits_{-\infty}^{+\infty}
    d\nu\,\mu(n, \nu)\,\tau(\nu, n)\,\int d^4x\,
    |\Psi^{\mu_1...\mu_n}_{\nu,x}\rangle\langle
    \overline{\Psi^{\mu_1...\mu_n}_{\nu,x}}|
\end{equation}

 \section{Four-point and two-point correlation
 functions for zig-zag diagrams\label{BK24}}
\setcounter{equation}0

Substitution of the resolution of unity (\ref{cor08}) into
 expressions (\ref{zgzg1}), (\ref{zgzg2}) for zig-zag 4-point
 Feynman graphs gives
\begin{multline}\label{cor09}
 G^{(M)}_4(x_1,x_2;y_1,y_2) =
 \langle x_1,x_2 | \bigl(\hat{Q}_{12}^{(\beta)}\bigr)^{M}| y_1,y_2
 \rangle
(y_1-y_2)^{2\beta} = \\
= \sum\limits_{n=0}^{\infty}
\int\limits_{0}^{\infty} \frac{d \nu}{C_1(n\,,\nu)} \int d^D x\,
\langle x_1,x_2 | \bigl(\hat{Q}_{12}^{(\beta)}\bigr)^{M}|
\Psi^{\mu_1\cdots\mu_n}_{\nu,\beta,x}\rangle
\langle \Psi^{\mu_1\cdots\mu_n}_{\nu,\beta,x} |\, U  | y_1,y_2 \rangle
(y_1-y_2)^{2\beta}  = \\
= \sum\limits_{n=0}^{\infty}
\int\limits_{0}^{\infty} d \nu \,
\frac{\bigl(\tau(\alpha,\beta,n)\bigr)^M}{
C_1(n\,,\nu)}  \int d^D x\,
\langle x_1,x_2 |\Psi^{\mu_1\cdots\mu_n}_{\nu,\beta,x}\rangle
\langle \Psi^{\mu_1\cdots\mu_n}_{\nu,\beta,x} |y_2,y_1 \rangle \, ,
\end{multline}
where the integral over $x$ in the right-hand side
of (\ref{cor09}) is evaluated
in terms of conformal blocks \cite{DolOsb1}, \cite{DolOsb2},
\cite{DPPT}
(in the four-dimensional case, this integral was considered in detail
in \cite{GKK}).

Making use of the standard relations between the
4-point zig-zag functions $G^{(M)}_4(x_1,x_2;y_1,y_2)$ constructed
 in (\ref{cor09})
and 2-point zig-zag functions
$G^{(M)}_2(x_2,y_1)$ (the graphs for these functions are presented
in (\ref{zgzg1}) -- (\ref{zgzg4})),
 we write explicit expressions
for the 2-point $M$-loop zig-zag diagrams as follows:
\begin{multline}\label{cor12}
G^{(M)}_2(x_2,y_1)
= \sum\limits_{n=0}^{\infty}
\int\limits_{0}^{\infty} d \nu
\frac{(\tau(\alpha,\beta,n))^M}{C_1(n\,,\nu)}
 \, \int d^D x_1 d^D y_2 \, d^D x\,
\frac{\langle x_1,x_2 |\Psi^{\mu_1\cdots\mu_n}_{\nu,\beta,x}\rangle
\langle \Psi^{\mu_1\cdots\mu_n}_{\nu,\beta,x} |y_2,y_1 \rangle}{
(x_1-x_2)^{2\beta}(y_1-y_2)^{2\beta}} = \\
= \frac{1}{(x_2-y_1)^{2\beta}} \frac{\Gamma(D/2-1)}{\Gamma(D-2)}
\sum\limits_{n=0}^{\infty}
\frac{(-1)^n \Gamma(n+D-2)}{2^{n} \Gamma(n+D/2-1)}
\int\limits_{0}^{\infty} d \nu \;
\frac{\tau^{M+3}(\alpha,\beta,n)}{C_1(n\,,\nu)} \, ,
\end{multline}
  where we apply the two-point master integral
\begin{multline}\label{cor13}
 \int d^D x_1 d^D y_2 \, d^D x\,
\frac{\langle x_1,x_2 |\Psi^{\mu_1\cdots\mu_n}_{\nu,\beta,x}\rangle
\langle \Psi^{\mu_1\cdots\mu_n}_{\nu,\beta,x} |y_2,y_1 \rangle}{
(x_1-x_2)^{2\beta}(y_1-y_2)^{2\beta}} = \\
 = \frac{(-1)^n \Gamma(n+D-2)\Gamma(D/2-1)}{2^{n}
 \Gamma(n+D/2-1)\Gamma(D-2)}
\, \frac{\tau^3(\alpha,\beta,n)}{(x_2-y_1)^{2\beta}} \; .
\end{multline}
To calculate this integral, we apply
 identity \eqref{int06mu} in the form
\be
\lb{mast3}
 \int d^D x_1 \,
\frac{\langle x_1,x_2 |\Psi^{\mu_1\cdots\mu_n}_{\nu,\beta,x}\rangle}{
(x_1-x_2)^{2\beta}} = \tau(\alpha,\beta,n) \;
\frac{(x_2-x)^{\mu_1...\mu_n}}{(x_2-x)^{2(n+\alpha')}} \,
\ee
and the corresponding complex conjugate relation
\be
\lb{mast4}
 \int d^D y_2 \,
\frac{\langle \Psi^{\mu_1\cdots\mu_n}_{\nu,\beta,x} |y_2,y_1 \rangle}{
(y_1-y_2)^{2\beta}} = \tau(\alpha,\beta,n) \;
\frac{(y_1-x)^{\mu_1...\mu_n}}{(y_1-x)^{2(n+\bar{\alpha}')}} \, .
\ee
where we take into account $\langle x_1,x_2
|\Psi^{\mu_1\cdots\mu_n}_{\nu,\beta,x}\rangle^* =
\langle \Psi^{\mu_1\cdots\mu_n}_{\nu,\beta,x} |x_1,x_2\rangle$
and $\tau(\bar{\alpha},\beta,n) = \tau(\alpha,\beta,n)$.
Finally, for the left-hand side of \eqref{cor13} we obtain
$$
\begin{array}{c}
 \int d^D x_1 d^D y_2 \, d^D x\,
\frac{\langle x_1,x_2 |\Psi^{\mu_1\cdots\mu_n}_{\nu,\beta,x}\rangle
\langle \Psi^{\mu_1\cdots\mu_n}_{\nu,\beta,x} |y_2,y_1 \rangle}{
(x_1-x_2)^{2\beta}(y_1-y_2)^{2\beta}} =
\tau^2(\alpha,\beta,n) \; \int  d^D x\,
\frac{(x-x_2)^{\mu_1...\mu_n}}{(x-x_2)^{2(n+\alpha')}}
\frac{(x-y_1)^{\mu_1...\mu_n}}{(x-y_1)^{2(n+\bar{\alpha}')}}
\; \stackrel{(\ref{zig04})}{=} \\ [0.3cm]
= \tau^2(\alpha,\beta,n)
\frac{\pi^{D/2} \Gamma(\alpha)\Gamma(\bar{\alpha})\Gamma(\beta)}{
 \Gamma(\alpha'+n)\Gamma(\bar{\alpha}'+n)\Gamma(\beta')} \cdot
 \frac{\Gamma(n+D-2)\Gamma(D/2-1)}{2^{n} \Gamma(n+D/2-1)\Gamma(D-2)}
\cdot \frac{1}{(x_2-y_1)^{2\beta}} \stackrel{(\ref{cor04})}{=}
\\ [0.3cm]
= (-1)^n \tau^3(\alpha,\beta,n) \cdot
 \frac{\Gamma(n+D-2)\Gamma(D/2-1)}{2^{n} \Gamma(n+D/2-1)\Gamma(D-2)}
\cdot \frac{1}{(x_2-y_1)^{2\beta}}
\end{array}
$$
where $\alpha = \frac{1}{2}(n+\frac{D}{2}-\beta)+i\nu$ and
we have used the identity (\ref{zig04})
 derived in Appendix {\bf C}.

The integral over $\nu$ in the right-hand side of (\ref{cor12})
for $\beta=1$ and even $D >2$ can be evaluated explicitly
and gives a linear combination of $\zeta$-values with
rational coefficients.
We will publish the explicit formula for (\ref{cor12}) elsewhere.
Here we consider only one special case
$\beta=1$ and $D=4$, which is needed to prove the
Broadhurst and Kreimer conjecture \cite{BrKr}
for zig-zag diagrams (we use the
approach of the paper \cite{DISh}).
In this case we have
 $\alpha=\frac{n+1}{2}-i\nu$
 and the master integral (\ref{cor13})
 is simplified
  \be
 \lb{cor14}
 %\begin{array}{c}
 \int d^4 x_1 d^4 y_2 \, d^4 x\,
\frac{\langle x_1,x_2 |\Psi^{\mu_1\cdots\mu_n}_{\nu,x}\rangle
\langle \Psi^{\mu_1\cdots\mu_n}_{\nu,x} |y_2,y_1 \rangle}{
(x_1-x_2)^{2}(y_1-y_2)^{2}}
 =  (-1)^n \frac{(n+1)}{2^n} \, \tau^3(\nu,n)
\frac{1}{(x_2-y_1)^2} \; ,
 %\end{array}
 \ee
 where
 $\langle x_1,x_2 |\Psi^{\mu_1\cdots\mu_n}_{\nu,x}\rangle :=
 \Psi^{\mu_1\cdots\mu_n}_{\nu,\beta,x}(x_1,x_2)
 |_{D=4,\beta=1}$ and (see (\ref{cor11}))
 \be
 \lb{cor15}
 \tau(\nu,n) := \left. \tau(\alpha,\beta,n)\right|_{D=4,\beta=1} =
 \frac{(-1)^n (2\pi)^2}{(1+n)^2+4\nu^2} \; .
 \ee
 The coefficient $C_1$ in (\ref{C1}) for $D=4$ and $\beta=1$
 is reduced to
 \be
 \lb{cor16}
 C_1(n,\nu) =
 \frac{\pi^5}{2^{n+3}(1+n)\, \nu^2}\, \tau(\nu,n)  \; .
 \ee
 Finally we substitute (\ref{cor14}) -- (\ref{cor16})
 into (\ref{cor12}) for $D=4$ and obtain
 %\marginpar{\bf Is.19.10 \tiny Развернуть
 %и сделать подробнее выкладки (\ref{cor17}) }
\begin{multline}\label{cor17}
G_2(x_2,y_1)|_{_{D=4,\beta=1}}
 %=\frac{1}{(x_2-y_1)^2} \frac{2^3}{\pi^5}
 %\sum\limits_{n=0}^{\infty}(-1)^n\, (n+1)^2\,
 %\int\limits_{0}^{\infty} d \nu \,\nu^2\, (\tau(\nu,n))^{M+2}
  = \\
= \frac{(2\pi)^{2(M+2)}}{(x_2-y_1)^2} \frac{2^3}{\pi^5}
  \sum\limits_{n=0}^{\infty}(-1)^{n(M+1)} (n+1)^2
\int\limits_{0}^{\infty} d \nu \,
\frac{\nu^2}{\left((1+n)^2+4\nu^2\right)^{M+2}}  = \\
= \frac{4 \pi^{2M}}{(x_2-y_1)^2} \, {\sf C}_M  \,
%\frac{4 \pi^{2M}}{(M+1)}\binom{2M}{M}
\sum\limits_{n=0}^{\infty}(-1)^{n(M+1)}
\frac{1}{(n+1)^{2M-1}} \; ,
 %= \\= \frac{4 \pi^{2M}}{(x_2-y_1)^2} C_M
 %\sum\limits_{p=1}^{\infty}(-1)^{(p-1)(M+1)}
 %\frac{1}{p^{2(M+1)-3}} \; ,
\end{multline}
where ${\sf C}_M = \frac{1}{(M+1)} \binom{2M}{M}$ is the Catalan
number.
In the last equality in (\ref{cor17}) we used the integral
\begin{align}
\int_{0}^{+\infty} d\nu\,
\frac{\nu^2}{\left(4\nu^2+\left(1+n\right)^2\right)^{M+2}} =
\frac{1}{2^5(n+1)^{2M+1}}\,
\frac{\Gamma\left(\frac{1}{2}\right)\Gamma\left(M+\frac{1}{2}\right)}
{\Gamma\left(M+2\right)}
\end{align}
 and applied the identity
$\frac{\Gamma(M+\frac{1}{2})\Gamma(\frac{1}{2})}{\Gamma(M+2)}
= \frac{(2M)!\; \pi}{2^{2M} M!(M+1)!}$.

Thus, we get the result (\ref{cor17}) for the two point correlation
function related to the zig-zag series,
which can be written in the form
 \be
 \lb{BKr}
 G^{(M + 1)}_2(x,y) = \frac{\pi^{2M}}{(x-y)^2} \; Z(M+1) \; ,
 \ee
where
\begin{align}\label{BK}
Z(M+1) &= 4 {\sf C}_M
\sum\limits_{p=1}^{\infty}
\frac{(-1)^{(p-1)(M+1)}}{p^{2(M+1)-3}} =\footnotesize
\left\{
\begin{array}{l}
4 \, C_M \, \zeta_{2M-1} \;\; {\rm for} \;\; M =2N+1 \, ,\\ [0.2cm]
4\, C_M \, (1 - 2^{2(1-M)}) \, \zeta_{2M-1} \;\;
{\rm for} \;\; M=2N \, .
\end{array}
\right.
\end{align}

Finally we note
that D.Broadhurst and D.Kreimer fixed
in their paper \cite{BrKr}
the loop measure for each integration over
loop momenta $k$ as $\frac{d^4 k}{\pi^2}$.
Expression (\ref{cor17}) is related
to the $M$ loop zig-zag diagram
(it corresponds to the $n=(M+1)$ loop contribution to
the $\beta$-function of $\phi^4_{D=4}$ theory). Therefore,
we have to divide our answer in (\ref{cor17})
by $(\pi^{2})^M$.
In this case our result (\ref{cor17}) justifies the
normalization factor $(\pi^{2})^M$ in relation (\ref{BKr})
which together with (\ref{BK})
states the Broadhurst and Kreimer conjecture \cite{BrKr}.

\section{General graph building operator and Yang-Baxter
equation\label{YB}}
\setcounter{equation}0

\subsection{Conformal algebra and scalar conformal
fields\label{algf}}

In this Subsection we prove that the modified
scalar product (\ref{scprod}) is needed
also for the hermitian property
of the Casimir operator (see the definition below) for
the conformal Lie algebra $\textsf{conf}(\mathbb{R}^{D})$
in the representation which acts in the space $V_1 \otimes V_2$.
For this purpose, we
summarize some known facts about the
$D$-dimensional conformal Lie algebra
and its field representations \cite{MackSalam}
(see also
\cite{DobMac},\cite{ToMiPe},\cite{FradPal},\cite{Osb2},\cite{FMS},
 \cite{IsRub1} and references therein).
Here we denote by $\textsf{conf}(\mathbb{R}^{D}) = so(1,D+1)$
the Lie algebra
of the conformal group in $\mathbb{R}^{D}$ with the basis elements
$T_{AB} = - T_{BA}$ $(A,B = 0,1,\dots,D,D+1)$,
 and use the notation:
$$
T_{\mu \nu} = L_{\mu \nu} \; , \;\;\;
T_{0 \mu} = \frac{1}{2}(P_{\mu}+K_{\mu})\; , \;\;\;
T_{D+1,\mu} = \frac{1}{2}(P_{\mu}-K_{\mu}) \; , \;\;\;
{\cal D} = T_{D+1,0} \; ,
$$
where $\mu,\nu = 1,\dots,D$ and $\mathbb{R}^{1,D+1}$-metric is
$\eta_{00}=\eta_{11} =...=\eta_{_{DD}}=-\eta_{_{D+1,D+1}}=1$.
The defining relations
for $\textsf{conf}(\mathbb{R}^{D})$ are:
$$
[ {\cal D}\, ,\, P_{\mu} ] = i\, P_{ \mu }\ \,, \ \ \  [ {\cal D}\, ,\,
K_{\mu} ]  = - i\, K_{\mu}\ \,,
\ \ \  [ L_{\mu \nu}\, ,\, L_{\rho \sigma} ]  = i\,( \delta_{\nu
\rho}\, L_{\mu
\sigma} + \delta_{\mu \sigma}\, L_{\nu \rho} - \delta_{\mu \rho}\,
L_{\nu
\sigma} - \delta_{\nu \sigma}\, L_{\mu \rho} )
$$
 \begin{equation}
 \label{cnfA}
[ K_{\rho}\, ,\, L_{\mu \nu} ] = i\,( \delta_{\rho \mu}\, K_{\nu} -
\delta_{\rho
\nu}\, K_{\mu} ) \; , \;\;\;  [ P_{\rho}\, ,\, L_{\mu \nu} ]  = i\, (
\delta_{\rho
\mu}\, P_{\nu} - \delta_{\rho \nu}\, P_{\mu} ) \; ,
 \end{equation}
$$
[ K_{\mu}\, ,\, P_{\nu} ] = 2 i\, ( \delta_{\mu \nu}\, {\cal D} -
L_{\mu \nu} ) \; , \;\;\;
[ P_{\mu}\, ,\, P_{\nu} ] = 0 \; , \;\;\; [ K_{\mu}\, ,\, K_{\nu} ] = 0
\; , \;\;\;
[ L_{\mu \nu}\, ,\, {\cal D} ] = 0 \; .
$$
Note that the elements $L_{\mu \nu}$ generate the Lie
subalgebra $so(D)$ in the conformal algebra
 $\textsf{conf}(\mathbb{R}^{D}) = so(1,D+1)$.
The quadratic Casimir operator for
$\textsf{conf}(\mathbb{R}^{D}) = so(1,D+1)$ is
 \begin{equation}
\label{cnfA2b}
\widehat{C}_2  = \frac{1}{2} T_{AB} \, T^{AB} =
\frac{1}{2}\left( L_{\mu \nu} L^{\mu \nu} +  P_{\mu}
K^{\mu} + K_{\mu} P^{\mu} \right) - {\cal D}^2  \; .
\end{equation}

The standard
 realization $\rho$ of the elements
$T_{AB}=\{L_{\mu \nu} , P_{\mu},K_{\mu},{\cal D} \}$ of the algebra
(\ref{cnfA}) is \cite{MackSalam}:
 \begin{equation}
\label{cnfA1}
\begin{array}{c}
 \rho(P_{\mu}) =
  %- i \partial_{x_\mu}  \equiv
  \hat{p}_{\mu} \; , \;\;\;\;
 \rho({\cal D}) = \q^{\mu} \, \hat{p}_{\mu} - i \Delta \; ,
 \;\;\;
 \rho(L_{\mu \nu}) = \hat{\ell}_{\mu \nu} + S_{\mu \nu}   \; ,
 \\ [0.2cm]
  \rho(K_{\mu}) = 2 \, \q^{\nu} \, (\hat{\ell}_{\nu \mu}
 + S_{\nu \mu})
+ \q^2 \,  \hat{p}_{\mu}
- 2 i \Delta \, \q_{\mu} \; = \;
2 \, \q^{\nu} \, S_{\nu \mu} + 2 \q_\mu (\q^\nu \p_\nu)
- \q^2 \, \p_{\mu} - 2i \Delta \, \q_\mu \; ,
\end{array}
\end{equation}
 $$
\hat{\ell}_{\mu \nu} := (\q_{\nu} \hat{p}_{\mu} -  \q_{\mu}
\hat{p}_{\nu}) \; ,
$$
where
$\q_\mu$, $\p_\mu$ are the generators of the
Heisenberg algebra (\ref{gr001}),
$\Delta \in \mathbb{R}$ is the conformal parameter,
$S_{\mu \nu}=-S_{\nu \mu}$ are the spin generators
of $so(D)$ with the same
commutation relations as for the generators
 $L_{\mu \nu}$ (see (\ref{cnfA})):
 \begin{equation}
\lb{spin}
[ S_{\mu \nu} , S_{\rho \sigma} ] =
 i( \delta_{\nu \rho} S_{\mu \sigma}
+ \delta_{\mu \sigma} S_{\nu \rho} - \delta_{\mu \rho} S_{\nu \sigma}
-
\delta_{\nu \sigma} S_{\mu \rho} )  \; ,
 \end{equation}
and $[ S_{\mu \nu} , \q_\rho] = 0 = [ S_{\mu \nu} , \hat{p}_\rho]$.
Note that the quadratic
 Casimir operator (\ref{cnfA2b})
 in the representation (\ref{cnfA1})  acquires the form:
\begin{equation}
\label{cnfA2z}
\rho(\widehat{C}_2) =
\frac{1}{2} \left( S_{\mu \nu} \, S^{\mu \nu}
- \hat{\ell}_{\mu \nu} \, \hat{\ell}^{\mu \nu} \right) +
C_{(2)} + \frac{1}{4}(4-D)D+\Delta(\Delta - D) =
\frac{1}{2}  S_{\mu \nu}S^{\mu \nu} +\Delta(\Delta - D) \; ,
\end{equation}
 where we use
 $$
 \frac{1}{2} \hat{\ell}_{\mu \nu} \, \hat{\ell}^{\mu \nu} =
 \left(\q^2 \, \p^2 + i \,  (D-2) (\q^\nu \, \p_\nu) -
 (\q^\nu \, \p_\nu)^2 \right) =  C_{(2)}
 + \frac{1}{4}(4-D)D \; ,
 $$
 and the operator $C_{(2)}$ was introduced
  in (\ref{cas2}).
We will only use the
scalar representations $S_{\mu \nu} = 0$ of
$\textsf{conf}(\mathbb{R}^{D}) = so(1,D+1)$.
In this case, we have
\be
\lb{hcaz2}
\rho(\widehat{C}_2) = \Delta(\Delta - D) \, I_\rho \; ,
\ee
and $I_\rho$ is the unit operator in the space of
the representation $\rho|_{S_{\mu\nu}=0}$
(we will omit such unit operators below).
Since we deal with a scalar case of the
conformal algebra
representation $\rho$ parameterized by only one parameter,
conformal dimension $\Delta$, we denote the representation by
$\rho^{\Delta}:=  \rho|_{S_{\mu\nu}=0}$.

We recall that
the scalar representation $\rho^{\Delta}$ of the conformal
group acts in the space of the scalar
conformal fields $\Phi_\Delta(x) =
\langle x | \Phi_\Delta \rangle$. An infinitesimal form
of this action is $\delta \Phi_\Delta(x) =
\omega^{AB} \langle x | \rho^{\Delta}(T_{AB})
\Phi_\Delta \rangle$, where $\omega^{AB}$ are the parameters and
$$
\begin{array}{c}
\langle x | \rho^{\Delta}(P_{\mu})
\Phi_\Delta \rangle = - i \partial_{x_\mu} \Phi_\Delta(x)
\; , \;\;\;\;
\langle x | \rho^{\Delta}({\cal D})
\Phi_\Delta \rangle
  = - i (x^{\mu} \, \partial_{\mu} +\Delta)
  \Phi_\Delta(x) \; , \\ [0.2cm]
  \langle x | \rho^{\Delta}(L_{\mu \nu})
\Phi_\Delta \rangle = i (x_\mu \partial_{\nu}-
x_\nu \partial_{\mu})\Phi_\Delta(x) \; ,
 \\ [0.2cm]
 \langle x | \rho^{\Delta}(K_{\mu})
\Phi_\Delta \rangle  \; = \;
- 2 i \bigl(x_\mu (x^\nu \partial_{\nu})
- \frac{1}{2} x^2 \, \partial_{\mu} + \Delta \, x_\mu\bigr)
\Phi_\Delta(x) \; ,
\end{array}
$$
where $\partial_{\mu} = \frac{\partial}{\partial x_\mu}$.

Let us introduce the operator
\be
\lb{splcaz2}
\widehat{C}_{12} :=
\Bigl( {\bf \Delta}(\widehat{C}_2) -
\widehat{C}_2 \otimes 1 -1 \otimes  \widehat{C}_2 \Bigr)
= T_{AB} \otimes T^{AB} \; ,
\ee
where
${\bf \Delta}$ is the comultiplication
in the universal enveloping algebra of
$\textsf{conf}(\mathbb{R}^{D}) = so(1,D+1)$, i.e.
${\bf \Delta}(T_{AB}) = 1 \otimes T_{AB} +
T_{AB} \otimes 1$.
Operator (\ref{splcaz2}) is called split (or polarized) Casimir
operator
and plays an important role in the representation theory
of Lie algebras and Lie groups (see e.g. \cite{IsRub1}).
It is instructive to present the explicit
form of (\ref{cnfA2b}) and (\ref{splcaz2}) in the
representation $\rho^{\Delta_1} \otimes \rho^{\Delta_2}$.
We have\footnote{In view of (\ref{hcaz2}),  in the
representation $\rho^{\Delta_1} \otimes \rho^{\Delta_2}$,
 the operator
(\ref{cnfA2b}) is equal to (\ref{splcaz2})
up to an additional constant.}
 \be
 \lb{splcaz}
\begin{array}{c}
(\rho^{\Delta_1} \otimes \rho^{\Delta_2}) \, \widehat{C}_{12} =
(\rho^{\Delta_1} \otimes \rho^{\Delta_2})
\Bigl(\hat{\ell}_{\mu \nu} \otimes \hat{\ell}^{\mu \nu} +
P_\mu \otimes K^\mu + K_\mu \otimes P^\mu
- 2 {\cal D} \otimes {\cal D}  \Bigr) = \\ [0.3cm]
 = -\left(\q_{12}\right)^2 (\p_1\p_2) +
2 \q_{12}^{\mu} \, \q_{12}^{\nu}\, \p_{1\mu}\, \p_{2\nu}
+2i\Delta_2\, (\q_{12} \, \p_{1}) -
2i\Delta_1\, (\q_{12} \, \p_{2}) + 2 \Delta_1 \Delta_2 \; ,
\end{array}
 \ee
where we use a more compact
notation $\q^{\mu}_{12} := \left(\q_1-\q_2\right)^{\mu}$,
$\;\left(\q_{12}\right)^{2\alpha} :=
(\q_{12}^\mu\, \q_{12\, \mu})^\alpha$,
$\;(\q_{12} \, \p_{1}):= \q_{12}^\mu\, \p_{1\, \mu}$, etc,
and subscripts $1$ and $2$ indicate the
generators of the first and second Heisenberg algebras in
the product ${\cal H} \otimes {\cal H}$.
 Now we introduce the notation
\begin{align}
\lb{hcaz5}
\widehat{C}_{\Delta_1\Delta_2} :=
(\rho^{\Delta_1} \otimes \rho^{\Delta_2})\,
 {\bf \Delta}(\widehat{C}_2) \equiv
 (\rho^{\Delta_1} \otimes \rho^{\Delta_2})\,\widehat{C}_{12}
 + \Delta_1\,(\Delta_1-D) + \Delta_2\,(\Delta_2-D) \; ,
\end{align}
and, using (\ref{splcaz}),
we write the operator $\widehat{C}_{\Delta_1\Delta_2}$
 in the equivalent form
\be
 \lb{splcaz3}
\begin{array}{r}
\widehat{C}_{\Delta_1\Delta_2}=
(\q_{12})^{-\Delta_1-\Delta_2}\left[ -\left(\q_{12}\right)^2 (\p_1\p_2)
+
2 \q_{12}^{\mu}\, \q_{12}^{\nu} \, \p_{1\,\mu} \, \p_{2\, \nu}
\right] (\q_{12})^{\Delta_1+\Delta_2} \; - \\ [0.2cm]
- \; i(\Delta_1-\Delta_2)\q_{12}^{\mu}
\left(\p_{1\, \mu}+\p_{2\,\mu}\right)
 \; .
\end{array}
\ee
The equivalence can be checked by direct calculations
with the help of the main formula
\be
\lb{mainf}
\begin{array}{c}
(\q_{12})^{-2\alpha} \, \widehat{\sf D}_{12} \,
 (\q_{12})^{2\alpha} = \widehat{\sf D}_{12} +
2i\,\alpha\,\q_{12}^{\mu}
\left(\p_{1\, \mu}-\p_{2\,\mu}\right)+
2\alpha\left(2\alpha-D\right) \; , \\ [0.2cm]
\widehat{\sf D}_{12} := -(\q_{12})^2\, (\p_1 \p_2) +
2 \q_{12}^{\mu}\,\q_{12}^{\nu}\, \p_{1\,\mu}\,\p_{2\,\nu}
\equiv \left( -(\q_{12})^2\, \delta^{\mu\nu} +
2 \q_{12}^{\mu}\,\q_{12}^{\nu}\right)
 \p_{1\,\mu}\,\p_{2\,\nu} \; .
\end{array}
\ee
Using identity (\ref{mainf}) for $\alpha = D$,
we obtain
\be
\lb{mainf2}
\left[\p_{1\,\mu}\,\p_{2\,\nu} , \;
\bigl(-\delta^{\mu\nu} (\q_{12})^2 +
2 \q_{12}^\mu\, \q_{12}^\nu \bigr)(\q_{12})^{-2D}\right] = 0
 \;\;\;\;\; \Rightarrow  \;\;\;\;\;
 \widehat{\sf D}_{12}^\dagger
 = (\q_{12})^{-2D} \, \widehat{\sf D}_{12}\, (\q_{12})^{2D} \; ,
\ee
where $\dagger$ is the standard Hermitian conjugation with
respect  to the scalar product (\ref{scpro}).
Then, we calculate with the help of (\ref{mainf2})
 and obvious relation
$[\left(\p_1+\p_2\right)^{\nu}, \, f(\q_{12})] = 0$
the conjugation of the operator
(\ref{splcaz3})
\be
\lb{mainf4}
\begin{array}{c}
\widehat{C}_{\Delta_1\Delta_2}^{\; \dagger} =
\q_{12}^{\Delta_1+\Delta_2} \;
\widehat{\sf D}_{12}^\dagger \; \q_{12}^{-\Delta_1-\Delta_2} \; +\;
i(\Delta_1-\Delta_2)\left(\p_1+\p_2\right)_{\mu}\,\q_{12}^{\mu}
= \\ [0.2cm]
 = \q_{12}^{\Delta_1+\Delta_2-2D} \; \widehat{\sf D}_{12}
 \; \q_{12}^{2 D -\Delta_1-\Delta_2}
+i(\Delta_1-\Delta_2)\,\q_{12}^{\mu}\,\left(\p_1+\p_2\right)^{\mu} = \\
[0.2cm]
= \q_{12}^{-2(D-\Delta_1-\Delta_2)} \left( {\cal P}_{12}
\, \widehat{C}_{\Delta_1\Delta_2} \, {\cal P}_{12}\right)
\q_{12}^{2(D-\Delta_1-\Delta_2)} \; .
\end{array}
\ee
Thus, the operator $\widehat{C}_{\Delta_1\Delta_2}$
 is Hermitian for $\Delta_1\,,\Delta_2 \in \mathbb{R}$
 up to the equivalence transformation (cf. (\ref{hermc})):
\begin{align}
\lb{Chermc}
\widehat{C}_{\Delta_1\Delta_2}^\dagger =
U \, \widehat{C}_{\Delta_1\Delta_2} \, U^{-1}\, ,
 \;\;\;\;\;  U:= {\cal P}_{12} \,
 (\hat{q}_{12})^{-2(D-\Delta_1-\Delta_2)}=
 (\hat{q}_{12})^{-2(D-\Delta_1-\Delta_2)} \, {\cal P}_{12} \, .
\end{align}
It means that the operator $\widehat{C}_{\Delta_1\Delta_2}=
(\rho^{\Delta_1} \otimes \rho^{\Delta_2})\,
 {\bf \Delta}(\widehat{C}_2)$
is Hermitian with respect
to the modified scalar product (\ref{Cscprod}).

\subsection{General graph building operator}

 The generalization of the graph building operator
  (\ref{q-d}) is:
 \be
 \lb{Qgen}
Q_{12}^{(\zeta,\kappa,\gamma)}  : =
\frac{1}{a(\kappa)a(\gamma)}\, {\cal P}_{12} \,
 \hat{q}_{12}^{-2\zeta} \; \p_1^{-2\kappa} \; \p_2^{-2\gamma} \;
 \hat{q}_{12}^{-2\beta}  \; ,
 \ee
 where the parameters are restricted by only one
 condition\footnote{This condition guarantees
 the scale invariance of (\ref{Qgen}) under transformations
 $\q_i \to \lambda \, \q_i$, $\p_i \to \lambda^{-1} \, \p_i$.}
  $\zeta + \beta = \kappa + \gamma$ and
 we have $\hat{Q}_{12}^{(\beta)} =
 a(\gamma)Q_{12}^{(\zeta,\kappa,\gamma)}
 \bigr|_{\zeta,\gamma=0}$ .
 We depict the integral kernel of the
 $D$-dimensional operator
 $Q_{12}^{(\zeta,\kappa,\gamma)} $
 as follows ($\kappa':=D/2-\kappa$,
 $\gamma':=D/2-\gamma$):

 \unitlength=5.5mm
\begin{picture}(25,4)(-1,0)

\put(0.2,3){\footnotesize $x_1$}
\put(3.1,1.9){\scriptsize $\beta$}
\put(0.6,1.9){\scriptsize $\zeta$}
\put(1.5,1.2){\scriptsize $\kappa'$}
\put(1.5,2.6){\scriptsize $\gamma'$}
\put(0.2,0.8){\footnotesize $x_2$}
\put(3.2,0.8){\footnotesize $y_2$}
\put(3.2,3){\footnotesize $y_1$}

\put(1,1){\line(0,1){2}}
\put(3,1){\line(0,1){2}}
 \put(3,3){\line(-1,-1){2}}
 %\multiput(3,1)(-0.18,0.18){12}{\circle*{0.1}}
  \put(1,3){\line(1,-1){0.9}}
  \put(2.1,1.9){\line(1,-1){0.9}}

 \put(4,1.8){$=$}
 %%%%%%%%%%%%%%%%%%%%%%%%%

 \put(5.2,3){\footnotesize $x_2$}
\put(8.1,1.9){\scriptsize $\beta$}
\put(5.6,1.9){\scriptsize $\zeta$}
\put(6.7,2.5){\scriptsize $\kappa\,'$}
\put(6.7,0.5){\scriptsize $\gamma\,'$}
\put(5.2,0.8){\footnotesize $x_1$}
\put(8.2,0.8){\footnotesize $y_2$}
\put(8.2,3){\footnotesize $y_1$}

\put(6,1){\line(0,1){2}}
\put(8,1){\line(0,1){2}}
 \put(8,3){\line(-1,0){2}}
 \put(6,1){\line(1,0){2}}
 %\multiput(8,1)(-0.25,0){9}{\circle*{0.1}}
%%%%%%%%%%%%%%%%%%%%

 \put(10,2){%\footnotesize
 $= \; \langle x_1,x_2 |
Q_{12}^{(\zeta,\kappa,\gamma)}  | y_1,y_2 \rangle \, =$}

\put(11,0.3){%\footnotesize
$= \, \frac{1}{a(\kappa)a(\gamma)}
  \cdot  \langle x_1,x_2 |\;
{\cal P}_{12} \,\hat{q}_{12}^{-2\zeta} \; \p_1^{-2\kappa}
\,\p_2^{-2\gamma} \,
\hat{q}_{12}^{-2\beta} \, | y_1,y_2 \rangle \, = $}

 %\put(15,1){$= \;
 %\frac{a(\gamma)}{(x_2-y_1)^{2\kappa'} \,
 %(x_1-y_2)^{2\gamma'} \,(y_1-y_2)^{2(\kappa+\gamma)}} $ .}

\end{picture}

 {\footnotesize $$
= \;
\frac{1}{(x_1-x_2)^{2\zeta}
(x_2-y_1)^{2\kappa\,'} \, (x_1-y_2)^{2\gamma\,'} \,
(y_1-y_2)^{2\beta}} \; .
$$}
Thus, the operator $Q_{12}^{(\zeta,\kappa,\gamma)}$
is the graph building operator for the ladder diagrams
(these diagrams are Fourier dual to the diagrams in Fig.\ref{fig7})

\unitlength=4.2mm
\begin{picture}(25,5)(3,-0.5)

 %\put(3.5,1.8){\tiny $\beta$}
 \put(3.7,1.9){\tiny $\beta\!+\!\zeta$}
 %\put(7.5,1.8){\tiny $\beta$}
\put(7.7,1.9){\tiny $\beta\!+\!\zeta$}
 %\put(5.5,1.8){\tiny $\beta$}
 \put(5.7,1.9){\tiny $\beta\!+\!\zeta$}

 \put(3.8,3.2){\tiny $\kappa'$}
 \put(3.8,0.5){\tiny $\gamma'$}
  \put(5.8,3.2){\tiny $\gamma'$}
 \put(5.8,0.5){\tiny $\kappa'$}
 \put(7.8,0.5){\tiny $\gamma'$}
 \put(8,3.2){\tiny $\kappa'$}

\put(2.1,3){\footnotesize $x_2$}
\put(2.1,1){\footnotesize $x_1$}
\put(5,1){\line(0,1){2}}
\put(3,1){\line(1,0){2}}
\put(3,3){\line(1,0){2}}
 %\put(5,1){\line(-1,1){2}}

%%%%%%%%%%%%%%%%%%%%%%%%%%

\put(17.2,0.8){\footnotesize $y_2$}
\put(17.2,3){\footnotesize $y_1$}
\put(7,1){\line(0,1){2}}
\put(5,1){\line(1,0){2}}
\put(5,3){\line(1,0){2}}
 %\put(7,1){\line(-1,1){2}}
 \put(4.85,0.8){$\bullet$}
\put(4.85,2.8){$\bullet$}

 %%%%%%%%%%%%%%%%%%%%%%%%%%%%%%

 \put(6.85,0.8){$\bullet$}
\put(6.85,2.8){$\bullet$}
 \put(9,1){\line(0,1){2}}
\put(7,1){\line(1,0){2}}
\put(7,3){\line(1,0){2}}
 %\put(9,1){\line(-1,1){2}}

\put(8.85,0.8){$\bullet$}
\put(8.85,2.8){$\bullet$}
%%%%%%%%%%%%%%%%%%%%%%%%%%%%%%
 \put(9.8,2){$. \; .\; . \; . \; .\; . $}
%%%%%%%%%%%%%%%%%%%%%%%%%%%%%%

 \put(13.8,0.5){\tiny $\gamma'$}
 \put(14,3.2){\tiny $\kappa'$}

\put(15,1){\line(0,1){2}}
\put(13,1){\line(1,0){2}}
\put(13,3){\line(1,0){2}}
 %\put(15,1){\line(-1,1){2}}

%%%%%%%%%%%%%%%%%%%%%%%%%%%%

\put(15.8,0.5){\tiny $\kappa'$}
\put(13.7,1.9){\tiny $\beta\!+\!\zeta$}
\put(16,3.2){\tiny $\gamma'$}
 %\put(16.1,1.9){\tiny $\beta$}

 %\put(17,1){\line(0,1){2}}
\put(15,1){\line(1,0){2}}
\put(15,3){\line(1,0){2}}
 %\put(17,1){\line(-1,1){2}}
 \put(14.85,0.8){$\bullet$}
\put(14.85,2.8){$\bullet$}

\put(19,2){
$= \;\; (x_1-x_2)^{2\zeta} \,
\langle x_1,x_2 | (\hat{Q}_{12}^{(\zeta,\kappa,\gamma)})^{2N}
| y_1,y_2 \rangle (y_1-y_2)^{2\beta}$,}

\end{picture}

\noindent
where four parameters on the lines are restricted
by the condition $\kappa'+\gamma'+\beta+\zeta=D$.

\begin{proposition}\label{prop5}
The eigenfunction for the operator
$Q_{12}^{(\zeta,\kappa,\gamma)}$ is
given by the conformal 3-point correlation
function (\ref{cor01})

\unitlength=5mm
\begin{picture}(25,4)

\put(0,2){\footnotesize
$\langle y_1,y_2 | \Psi_{\delta,\rho}^{\mu_1\cdots\mu_n}(y)
\rangle = $}

 %\put(3,1){\line(1,1){2}}
 %\put(3,2){\line(2,-1){2}}
\put(6.5,3.2){\footnotesize $y_1$}
\put(6.6,1.9){\scriptsize $\alpha$}
\put(8,1){\scriptsize $\rho$}
\put(8,2.8){\scriptsize $\delta$}
\put(6.5,0.7){\footnotesize $y_2$}
\put(9.2,1.9){\footnotesize $y$}
\put(7,1){\line(0,1){2}}

 \put(7,3){\line(2,-1){2}}
 \put(7,1){\line(2,1){2}}

\put(10,2){$\Bigl( \frac{y-y_1}{(y-y_1)^2} -
\frac{y-y_2}{(y-y_2)^2} \Bigr)^{\mu_1\cdots\mu_n}\equiv$ }

\put(19.5,3.2){\footnotesize $y_1$}
\put(19.6,1.9){\scriptsize $\alpha$}
\put(21,1){\scriptsize $\rho,n$}
\put(21,2.8){\scriptsize $\delta,n$}
\put(19.5,0.7){\footnotesize $y_2$}
\put(22.2,1.9){\footnotesize $y$}
\put(20,1){\line(0,1){2}}

 \put(20,3){\line(2,-1){2}}
 \put(20,1){\line(2,1){2}}

 \thicklines
\put(21.2,2.4){\vector(-2,1){0.4}}
\put(20.8,1.4){\vector(2,1){0.4}}

\end{picture}

\vspace{-2cm}

 \be
 \lb{gwavef}
 {}\ee

 \vspace{0.5cm}

 \noindent
 where we depict the nontrivial rank-$n$ tensor numerator
 as arrows on the lines (the rank is fixed by the modification
 of indices on the lines: $\rho \to (\rho,n)$, etc) and denote
 \be
 \lb{conds}
 2\alpha = \Delta_1 + \Delta_2 -(\Delta -n) \; , \;\;\;
 2\delta = \Delta_1 - \Delta_2 +(\Delta -n) \; , \;\;\;
 2\rho = \Delta_2 - \Delta_1 +(\Delta -n) \; ,
 \ee
i.e., conformal dimensions $\Delta,\Delta_1,\Delta_2$ are
arbitrary parameters in this case. Thus, we have
 \be
 \lb{cor33}
Q_{12}^{(\zeta,\kappa,\gamma)}  \;
| \Psi_{\delta,\rho}^{\mu_1\cdots\mu_n}(y)\rangle =
\bar{\tau}(\kappa,\gamma;\delta,\alpha;n) \;
| \Psi_{\delta,\rho}^{\mu_1\cdots\mu_n}(y) \rangle \; ,
 \ee
where the parameters of (\ref{gwavef}) are
connected to the parameters of $Q_{12}^{(\zeta,\kappa,\gamma)}$
as follows:
\be
\lb{conds3}
\alpha +\rho = \kappa' + \zeta  \; , \;\;\;\;\;
\alpha +\delta =\gamma' + \zeta \; ,
\ee
and $\bar{\tau}(\kappa,\gamma;\delta,\alpha;n)$
is the eigenvalue
 \be
 \lb{factor}
\begin{array}{c}
\bar{\tau}(\kappa,\gamma;\delta,\alpha;n) = (-1)^n \cdot
 \tau(\delta',\kappa,n)\cdot
\tau\bigl((\rho+\kappa)',\gamma,n\bigr)  \, , %\\ [0.3cm]
\end{array}
\ee
\be
 \lb{factor1}
\tau(\alpha,\beta,n) = (-1)^n \,
\frac{\pi^{D/2} \Gamma(\beta) \Gamma(\alpha) \Gamma(\alpha'-\beta+n)}{
\Gamma(\beta') \Gamma(\alpha'+n) \Gamma(\alpha+\beta)}  \; .
\ee
\end{proposition}
{\bf Proof.} The proof is based on the tensor
 generalization of the star-triangle identity
  (\ref{int04mu}) that is depicted as

  \unitlength=4.8mm
\begin{picture}(20,5.3)(-6,-0.5)

\put(2,1){\line(2,1){2}}
\put(4,2){\line(2,-1){2}}
\put(4,1.9){\line(0,1){2}}
\put(4,4){\line(-2,-3){2}}
\put(3.8,1.8){$\bullet$}

{ \thicklines
\put(4,3){\vector(0,-1){0.4}}
\put(2.8,2.2){\vector(2,3){0.4}}
}

\put(5.1,1.6){\tiny $\beta'$}
\put(2.8,1.1){\tiny $\alpha+\beta$}
\put(4.3,3){\tiny $\alpha',n$}
\put(1.9,2.4){\tiny $0,n$}

\put(4.2,2.3){\scriptsize $z$}
\put(3.7,4.3){\scriptsize $y$}
\put(1.8,0.5){\scriptsize $x_1$}
\put(6,0.5){\scriptsize $x_2$}

\put(7.5,2.3){$= \; \tau(\alpha,\beta,n)$}

%%%%%%%%%%%%%%%%%%%%%%%%%%%%%%%%%%%%

\put(12.95,0.9){\line(1,0){4.1}}
\put(15,4){\line(-2,-3){2.05}}
\put(15,4){\line(2,-3){2.05}}

{ \thicklines
\put(15.8,2.8){\vector(2,-3){0.4}}
\put(13.8,2.2){\vector(2,3){0.4}}
}

\put(15,1.1){\tiny $\alpha$}
\put(16,2.7){\tiny $(\alpha+\beta)',n$}
\put(12.9,2.4){\tiny $\beta,n$}

\put(14.7,4.3){\scriptsize $y$}
\put(12.8,0.5){\scriptsize $x_1$}
\put(17,0.5){\scriptsize $x_2$}

\end{picture}

\vspace{-2cm}
\be
\lb{sttr5}
{}
\ee
\vspace{0.4cm}

\noindent
where the function $\tau(\alpha,\beta,n)$ is defined
in (\ref{cor04}) and (\ref{factor1}).
We act by the operator (\ref{Qgen}) on the
 wave function (\ref{gwavef}) and deduce

 \unitlength=6mm
\begin{picture}(25,4)(1.5,0)

 \put(2,1){\line(1,0){2}}
 \put(2,3){\line(1,0){2}}

\put(1.7,2){\tiny $\zeta$}
\put(2.8,1.9){\tiny $\alpha+\beta$}
\put(5,1){\scriptsize $\rho,n$}
\put(5,2.8){\scriptsize $\delta,n$}

\put(2.5,0.5){\scriptsize $\gamma'$}
\put(2.5,2.6){\scriptsize $\kappa'$}
\put(1.5,3.3){\footnotesize $x_2$}
\put(1.5,0.6){\footnotesize $x_1$}

\put(3.5,3.3){\footnotesize $y_1$}
\put(3.5,0.6){\footnotesize $y_2$}
\put(6.2,1.9){\footnotesize $y$}
\put(2,1){\line(0,1){2}}
\put(4,1){\line(0,1){2}}
 \put(4,3){\line(2,-1){2}}
 \put(4,1){\line(2,1){2}}

 \put(3.8,2.8){$\bullet$}
 \put(3.8,0.9){$\bullet$}

 {\thicklines
\put(5.2,2.4){\vector(-2,1){0.4}}
\put(4.8,1.4){\vector(2,1){0.4}}
 }

\put(6.9,2){\footnotesize $= \tau(\delta',\kappa,n)\cdot$ }

%%%%%%%%%%%%%%%%%%%%%%%%%%%%%%%%%%%%

\put(10.9,1){\line(0,1){2.2}}
 \put(10.9,1){\line(1,0){2.1}}
 \put(10.9,3.2){\line(3,-1){3.9}}
 \put(10.9,3.2){\line(1,-1){2}}
 \put(13,1){\line(2,1){1.8}}

  {\thicklines
\put(13,2.5){\vector(-3,1){0.4}}
\put(13.8,1.4){\vector(2,1){0.4}}
 }

\put(10.5,1.9){\scriptsize $\zeta$}
\put(11.5,1.8){\scriptsize $\delta'$}
\put(14,1){\scriptsize $\rho+\kappa,n$}
\put(12.7,2.8){\scriptsize $(\alpha+\beta)',n$}

\put(11.5,0.5){\scriptsize $\gamma'$}
\put(10.5,3.4){\footnotesize $x_2$}
\put(10.5,0.6){\footnotesize $x_1$}

 %\put(12.5,3.3){\footnotesize $y_1$}
\put(12.5,0.6){\footnotesize $y_2$}
\put(15,1.9){\footnotesize $y$}

 \put(12.8,0.9){$\bullet$}

%%%%%%%%%%%%%%%%%%%%%%%%%%%%%%%%%%%

\put(15.7,1.9){\footnotesize $\; = \;
\tau(\delta',\kappa,n)\cdot
\tau\bigl((\rho+\kappa)',\gamma,n\bigr)\cdot$ }

\put(23.5,3.2){\footnotesize $x_2$}
\put(24.1,1.9){\scriptsize $\alpha$}
\put(25,1){\scriptsize $\delta,n$}
\put(25,2.8){\scriptsize $\rho,n$}
\put(23.5,0.8){\footnotesize $x_1$}
\put(26.2,1.9){\footnotesize $y$}
\put(24,1){\line(0,1){2}}

 \put(24,3){\line(2,-1){2}}
 \put(24,1){\line(2,1){2}}

 \thicklines
\put(25.2,2.4){\vector(-2,1){0.4}}
\put(24.8,1.4){\vector(2,1){0.4}}

\end{picture}

\vspace{-2cm}
\be
\lb{qqkg04}
{}
\ee
\vspace{0.4cm}

\noindent
where we apply the star-triangle identity (\ref{sttr5}) twice
(to the vertices $y_1$ and $y_2$)
and fix the index $\alpha$ on the line $(x_1,x_2)$
in the right-hand side of (\ref{qqkg04}) to obtain
the eigenfunction (\ref{gwavef}), which
give the conditions
\be
\lb{conds4}
\begin{array}{c}
\alpha+\beta+\delta-\kappa=D/2 \; , \;\;\;
\rho+\kappa=\delta+\gamma \; , \;\;\;
\alpha+\rho+\kappa-\zeta=D/2 \;\;\;\;\;\; \Longrightarrow
\end{array}
\ee
\be
\lb{conds2}
\beta+\zeta=\kappa+\gamma \; , \;\;\;
  \gamma - \zeta= D/2 - \alpha- \delta  \; , \;\;\;
 \kappa - \zeta = D/2 - \alpha- \rho  \; .
\ee
The first two relations in (\ref{conds4})
are the conditions for applying (\ref{sttr5}), and
the third relation in (\ref{conds4}) is the result
of fixing $\alpha$ in the right-hand side of (\ref{qqkg04}).
We can see that (\ref{gwavef}) is the eigenfunction of (\ref{Qgen})
only if $\beta+\zeta=\kappa+\gamma$, which was
chosen at the beginning,
and the two last relations in (\ref{conds2}) are equivalent to
(\ref{conds3}). An additional
factor $(-1)^n$ appears in (\ref{factor})
since we have to change the arrows to the opposite
in the right-hand side of  (\ref{qqkg04}).
Thus, we have proved (\ref{cor33})
and (\ref{factor}). \hfill \qed

\vspace{0.3cm}

\noindent
{\bf Remark 1.} We multiply both sides of
(\ref{cor33}) by $a(\gamma)$ and take the limit
$\zeta,\gamma \to 0$; as a result,
we reproduce (\ref{cor03}).

\vspace{0.3cm}

\noindent
{\bf Remark 2.} Substitution of (\ref{conds}) in
 (\ref{conds2}) gives
 $$
 \beta - \zeta = D -\Delta_1 - \Delta_2 \; , \;\;\;
 \gamma - \zeta = D/2 -\Delta_1 \; , \;\;\;
  \kappa - \zeta = D/2 -\Delta_2 \; ,
 $$
 and after introducing a new notation
 $\beta + \zeta = - 2 u$, we deduce
 \be
 \lb{rybe4}
 \beta = \frac{1}{2} (D - \Delta_1 -\Delta_2) - u \; , \;\;\;
  \zeta  = - \frac{1}{2} (D - \Delta_1 -\Delta_2) - u \; , \;\;\;
 \kappa = \frac{1}{2}(\Delta_1 -\Delta_2) - u \; , \;\;\;
 \gamma = \frac{1}{2}(\Delta_2 -\Delta_1) - u  \; .
 \ee
 In this case, we obtain the expression for (\ref{Qgen})
  \be
 \lb{rybe2}
 \begin{array}{c}
   Q_{12}^{(\zeta,\kappa,\gamma)}  =
  \frac{1}{a(\kappa)a(\gamma)} \; {\cal P}_{12} \;
 \hat{q}_{12}^{2(u+ \frac{D - \Delta_1 -\Delta_2}{2})}
 \p_1^{2(u+\frac{\Delta_2 -\Delta_1}{2})} \;
 \p_2^{2(u+\frac{\Delta_1 -\Delta_2}{2})} \;
 \hat{q}_{12}^{2(u+\frac{\Delta_1 +\Delta_2-D}{2})} \; .
 \end{array}
 \ee

\subsection{Conformal R-operator\label{roper}}

We multiply
 the graph building operator (\ref{rybe2}) by
 the factor $a(\kappa)a(\gamma)$ and introduce a set of
 the $so(D)$-invariant operator functions
\be
\lb{Rdelta}
R_{\Delta_i\Delta_k}(u) = {\cal P}_{ik}\;
\hat{q}_{ik}^{2(u+\frac{D-\Delta_i-\Delta_k}{2})}\,
\p_k^{2(u+\frac{\Delta_i-\Delta_k}{2})}\,
\p_i^{2(u+\frac{\Delta_k-\Delta_i}{2})} \;
 \hat{q}_{ik}^{2(u+\frac{\Delta_i+\Delta_k-D}{2})}
\ee
that are expressed in terms of the generators of two Heisenberg
algebras ${\cal H}_i$ and ${\cal H}_k$ (see (\ref{deff})).
A remarkable fact is that the operators (\ref{Rdelta})
satisfy the Yang-Baxter equation
\begin{align}
\lb{RRR}
R_{\Delta_1\Delta_2}(u-v)\,
R_{\Delta_1\Delta_3}(u) R_{\Delta_2\Delta_3}(v) =
R_{\Delta_2\Delta_3}(v)\,
R_{\Delta_1\Delta_3}(u) R_{\Delta_1\Delta_2}(u-v)
\end{align}
Indeed, in \cite{ChDeIs} we constructed
the $so(D)$-invariant operator
 \be
 \lb{rybe3}
 \begin{array}{c}
 \check{R}_{12}^{(\Delta_1\Delta_2)}(u)  =
 \hat{q}_{12}^{2(u+ \frac{D - \Delta_1 -\Delta_2}{2})}
 \p_1^{2(u+\frac{\Delta_2 -\Delta_1}{2})} \;
 \p_2^{2(u+\frac{\Delta_1 -\Delta_2}{2})} \;
 \hat{q}_{12}^{2(u+\frac{\Delta_1 +\Delta_2-D}{2})} =
 {\cal P}_{12} \, R_{\Delta_1\Delta_2}(u)  \; ,
 \end{array}
 \ee
 as
 a solution of the Yang-Baxter equation
\be
\lb{RRRhat}
\check{R}^{(\Delta_1\Delta_2)}_{23}(u-v)\,
\check{R}^{(\Delta_1\Delta_3)}_{12}(u)
\check{R}^{(\Delta_2\Delta_3)}_{23}(v) =
\check{R}^{(\Delta_2\Delta_3)}_{12}(v)\,
\check{R}^{(\Delta_1\Delta_3)}_{23}(u)
\check{R}^{(\Delta_1\Delta_2)}_{12}(u-v) \; .
\ee
Substitution of the right-hand side of (\ref{rybe3})
into (\ref{RRRhat}) gives (\ref{RRR}). There are other
 equivalent
forms of the $R$-operator (\ref{rybe3}), which are useful  in many
applications
 \be
 \lb{rybe0}
 \begin{array}{rl}
 \check{R}_{12}^{(\Delta_1\Delta_2)}(u)  & =
 \p_1^{-2\Delta_2'}\,
 \hat{q}_{12}^{2(u - \Delta_{-})}\,
 \p_1^{2(u+ \Delta_{+}')} \,
 \p_2^{2(u-\Delta_{+}')} \,
 \hat{q}_{12}^{2(u+ \Delta_{-})}  \, \p_2^{2\Delta_2'} = \\ [0.2cm]
 & = \p_2^{-2\Delta_1'}\,
 \hat{q}_{12}^{2(u + \Delta_{-})}\,
 \p_2^{2(u+ \Delta_{+}')} \,
 \p_1^{2(u-\Delta_{+}')} \,
 \hat{q}_{12}^{2(u- \Delta_{-})}  \, \p_1^{2\Delta_1'} \; ,
 \end{array}
 \ee
 where $\Delta_{\pm} = \frac{1}{2}(\Delta_1 \pm \Delta_2)$
 and $\Delta_i'=D/2 - \Delta_i$.
 These forms demonstrate the
 symmetry $\check{R}_{12}^{(\Delta_1\Delta_2)}(u)  =
 \check{R}_{21}^{(\Delta_2\Delta_1)}(u)$ and can be
  deduced from (\ref{rybe3}) by means
 of the generalization
  of the star-triangle identity (\ref{startr}):
 $$
 \p_i^{2\alpha} \, \q_{ij}^{2(\alpha + \beta)}\, \p_i^{2\beta} =
 \q_{ij}^{2\beta}\, \p_i^{2(\alpha + \beta)}\, \q_{ij}^{2\alpha} \; .
 $$

\begin{proposition}\label{prop7}
The operators (\ref{Rdelta}) are conformal invariant,
i.e. they are invariant under the adjoint action of the
 group $\textsf{Conf}(\mathbb{R}^{D}) = SO(1,D+1)$
 in the representation $\rho^{\Delta_i} \otimes \rho^{\Delta_k}$.
\end{proposition}
\noindent
{\bf Proof.} The
operator (\ref{Rdelta}) is evidently invariant under translations
$\hat{q}_j^\mu \to \hat{q}_j^\mu + a^\mu$,
$\hat{p}_j^\mu \to \hat{p}_j^\mu$,
  $SO(D)$-rotations and dilatations
 $\hat{q}_j^\mu \to \lambda \hat{q}_j^\mu$,
 $\hat{p}_j^\mu \to \frac{1}{\lambda} \hat{p}_j^\mu$,
 which are generated by the adjoint action of translation, rotation
 and dilatation elements of
 $\textsf{Conf}(\mathbb{R}^{D}) = SO(1,D+1)$.
 Thus, it remains to prove the invariance of (\ref{Rdelta})
  under inversions
\begin{align}
\lb{RIIR}
R_{\Delta_1\Delta_2}(u)\,
{\cal I}^{(1)}_{\Delta_1}\,
{\cal I}^{(2)}_{\Delta_2}\, =
{\cal I}^{(1)}_{\Delta_1}\,
{\cal I}^{(2)}_{\Delta_2}\,
R_{\Delta_1\Delta_2}(u) \; ,
\end{align}
where ${\cal I}^{(k)}_{\Delta_i}
:= {\cal I}_k\,\q_k^{2\Delta_i}$ are the shifted
inversion operators (\ref{shin})
 for the Heisenberg algebras ${\cal H}_k$.
The shifted inversion
operator ${\cal I}_\Delta$ is interpreted
as the $\rho^\Delta$-representation of the inversion
element in the conformal group. In particular,
the
special conformal generators $K_\mu^{(\Delta)}$
 in this representation
are given in (\ref{shK}) and they are the same as
$- \rho(K_{\mu})$ in (\ref{cnfA1})
for $S_{\mu\nu}=0$.

To prove (\ref{RIIR}), we note that
the operator $R_{\Delta_1\Delta_2}(u)$ admits a
factorization in the product of simpler operators \cite{Der,ChDeIs}
\begin{align}
\lb{QQ5}
R_{\Delta_1\Delta_2}(u) = Q_{_{\Delta_1\Delta_2}}^{-1}(-u)\;
{\cal P}_{12}\; Q_{_{\Delta_1\Delta_2}}(u)
\end{align}
where
\be
\lb{QQ}
Q_{_{\Delta_1\Delta_2}}(u) =
\hat{q}_{12}^{2\left(\frac{D}{2}-\Delta_1\right)}
\p_1^{2\left(u+\frac{\Delta_2-\Delta_1}{2}\right)} \;
 \hat{q}_{12}^{2\left(u+\frac{\Delta_1+\Delta_2-D}{2}\right)}
 \; .
\ee
Operator $Q_{_{\Delta_1\Delta_2}}(u)$ has the following transformation
properties
\begin{align}
\label{confR}
Q_{_{\Delta_1\Delta_2}}(u)\,
{\cal I}^{(1)}_{\Delta_1}\,
{\cal I}^{(2)}_{\Delta_2}\, =
{\cal I}^{(1)}_{\frac{\Delta_1+\Delta_2}{2}+u}\,
{\cal I}^{(2)}_{\frac{\Delta_1+\Delta_2}{2} -u}
\,Q_{_{\Delta_1\Delta_2}}(u) \; ,
\end{align}
or equivalently
\begin{align}
\lb{QIIQ}
{\cal I}^{(1)}_{\Delta_1}\,
{\cal I}^{(2)}_{\Delta_2}\; Q_{_{\Delta_1\Delta_2}}^{-1}(u) =
Q_{_{\Delta_1\Delta_2}}^{-1}(u)\;
{\cal I}^{(1)}_{\frac{\Delta_1+\Delta_2}{2}+u}\,
{\cal I}^{(2)}_{\frac{\Delta_1+\Delta_2}{2} -u} \; .
\end{align}
The proof of (\ref{confR})
 is based on the following identities
\be
\lb{III}
{\cal I}^{(1)}_{\alpha}\,
{\cal I}^{(2)}_{\beta}\; \hat{q}_{12}^{2\lambda} =
\hat{q}_{12}^{2\lambda} \;
{\cal I}^{(1)}_{\alpha+\lambda}\,
{\cal I}^{(2)}_{\beta+\lambda}\, , \;\;\;\;\;\;
 %{\cal I}^{(k)}_{\alpha}\,{\cal I}^{(k)}_{\beta} = \,
 %\hat{q}_{k}^{2(\beta-\alpha)}\, , \;\;\;\;
{\cal I}^{(k)}_{\frac{D}{2}+\alpha}\,
\hat{p}_{k}^{2\alpha} = \hat{p}_{k}^{2\alpha}\,
{\cal I}^{(k)}_{\frac{D}{2}-\alpha}\, ,
\ee
which follow from (\ref{inver2}).
Then we have
{\footnotesize
\begin{align*}
{\cal I}^{(1)}_{\frac{\Delta_1+\Delta_2}{2}+u}\,
{\cal I}^{(2)}_{\frac{\Delta_1+\Delta_2}{2} -u}\,
Q_{_{\Delta_1\Delta_2}}(u) & =
{\cal I}^{(1)}_{\frac{\Delta_1+\Delta_2}{2}+u}\,
{\cal I}^{(2)}_{\frac{\Delta_1+\Delta_2}{2}
-u}\,\hat{q}_{12}^{2\left(\frac{D}{2}-\Delta_1\right)}
\p_1^{2\left(u+\frac{\Delta_2-\Delta_1}{2}\right)} \;
 \hat{q}_{12}^{2\left(u+\frac{\Delta_1+\Delta_2-D}{2}\right)}
 \\[0.2cm]
 %& = \hat{q}_{12}^{2\left(\frac{D}{2}-\Delta_1\right)}
 %{\cal I}^{(1)}_{\frac{D}{2}+\frac{\Delta_2-\Delta_1}{2}+u}\,
 %{\cal I}^{(2)}_{\frac{D}{2}+\frac{\Delta_2-\Delta_1}{2}-u}\,
 %\p_1^{2\left(u+\frac{\Delta_2-\Delta_1}{2}\right)} \;
 %\hat{q}_{12}^{2\left(u+\frac{\Delta_1+\Delta_2-D}{2}\right)}\\
 %& = \hat{q}_{12}^{2\left(\frac{D}{2}-\Delta_1\right)}\,
 %\p_1^{2\left(u+\frac{\Delta_2-\Delta_1}{2}\right)}\,
 %{\cal I}^{(1)}_{\frac{D}{2}-\frac{\Delta_2-\Delta_1}{2}-u}\,
 %{\cal I}^{(2)}_{\frac{D}{2}+\frac{\Delta_2-\Delta_1}{2}-u}\,
 %\hat{q}_{12}^{2\left(u+\frac{\Delta_1+\Delta_2-D}{2}\right)}
 %\\[0.2cm]
& = \hat{q}_{12}^{2\left(\frac{D}{2}-\Delta_1\right)}\,
\p_1^{2\left(u+\frac{\Delta_2-\Delta_1}{2}\right)}\,
\hat{q}_{12}^{2\left(u+\frac{\Delta_1+\Delta_2-D}{2}\right)}\,
{\cal I}^{(1)}_{\Delta_1}\,
{\cal I}^{(2)}_{\Delta_2} = Q_{_{\Delta_1\Delta_2}}(u)\,
{\cal I}^{(1)}_{\Delta_1}\,
{\cal I}^{(2)}_{\Delta_2} \; .
\end{align*}
}
Now it is not difficult to check conformal invariance
of the whole operator $R_{\Delta_1\Delta_2}(u)$:
\begin{align*}
Q_{_{\Delta_1\Delta_2}}^{-1}(-u)\,{\cal P}_{12}\,
Q_{_{\Delta_1\Delta_2}}(u)\,
{\cal I}^{(1)}_{\Delta_1}\,
{\cal I}^{(2)}_{\Delta_2}\, =
Q_{_{\Delta_1\Delta_2}}^{-1}(-u)\,{\cal P}_{12}\, {\cal
I}^{(1)}_{\frac{\Delta_1+\Delta_2}{2}+u}\,
{\cal I}^{(2)}_{\frac{\Delta_1+\Delta_2}{2} -u}\,
Q_{_{\Delta_1\Delta_2}}(u) = \\
Q_{_{\Delta_1\Delta_2}}^{-1}(-u)\,
{\cal I}^{(1)}_{\frac{\Delta_1+\Delta_2}{2}-u}\,
{\cal I}^{(2)}_{\frac{\Delta_1+\Delta_2}{2} + u}\,
{\cal P}_{12}\,Q_{_{\Delta_1\Delta_2}}(u) =
{\cal I}^{(1)}_{\Delta_1}\,
{\cal I}^{(2)}_{\Delta_2}\,
Q_{_{\Delta_1\Delta_2}}^{-1}(-u)\,{\cal P}_{12}
\,Q_{_{\Delta_1\Delta_2}}(u)\,  ,
\end{align*}
which is equivalent to (\ref{RIIR}).
\hfill \qed

\vspace{0.2cm}

\noindent
{\bf Remark 1.} It is instructive to prove (\ref{RIIR}) by using
the representations of the $R$-operators
(\ref{Rdelta}) in terms of the integral kernels.
The proof of the main identity (\ref{confR})
in this representation is given in
Appendix {\bf E}.

\vspace{0.2cm}

\noindent
{\bf Remark 2.} Since the
operator $R_{\Delta_1\Delta_2}(u)$ is conformal
invariant, i.e. commutes with all elements  of the conformal group
in the representation $\rho^{\Delta_1}\otimes\rho^{\Delta_2}$,
it is natural to consider $R_{\Delta_1\Delta_2}(u)$
as an operator acting in the tensor product of two spaces
$V_{\Delta_1} \otimes V_{\Delta_2}$
of scalar conformal fields with conformal dimensions
$\Delta_1$ and $\Delta_2$.
Moreover, the conformal
invariance of the operator $R_{\Delta_1\Delta_2}(u)$
guarantees its commutativity
with the quadratic Casimir operator
$(\rho^{\Delta_1}\otimes\rho^{\Delta_2})\,
{\bf \Delta} (\widehat{C}_2)$
 acting in $V_{\Delta_1} \otimes V_{\Delta_2}$.
This means that the operator $R_{\Delta_1\Delta_2}(u)$
(given in (\ref{Rdelta})) and
quadratic Casimir operator (\ref{hcaz5}) for the
conformal algebra ${\sf conf}(\mathbb{R}^D)$
in the representation $\rho^{\Delta_1}\otimes\rho^{\Delta_2}$
(see (\ref{splcaz3}))
have a common set of eigenvectors.

\vspace{0.2cm}

In particular, this Remark {\bf 2}
 explains  why conformal 3-point correlators (conformal
triangles)
 should be eigenfunctions of the general
 graph building operator (\ref{rybe2}).
Indeed, the
eigenvectors of the Casimir operator $\widehat{C}_2$ acting in
$\rho^{\Delta_1}\otimes\rho^{\Delta_2}$
are conformal 3-point
 correlation functions \cite{DobMac},\cite{Polyakov1},\cite{Polyakov2}
 which we considered in Subsection {\bf \ref{tricon}}.
 This system of functions (\ref{conf-tr}) is defined by
$\Delta_1, \Delta_2 \in \mathbb{R}$ and  $\Delta$
\begin{equation}
\lb{cnftri}
\langle x_1\,,x_2 |
\Psi^{\mu_1\cdots\mu_n}_{\Delta_1,\Delta_2,\Delta,x}
\rangle =
\frac{\left(\frac{x-x_1}{(x-x_1)^2}-\frac{x-x_2}{
(x-x_2)^2}\right)^{\mu_1\cdots\mu_n}}
{(x_1-x_2)^{\Delta_1+\Delta_2-\Delta+n}
(x-x_1)^{\Delta_1-\Delta_2+\Delta-n}
(x-x_2)^{\Delta_2-\Delta_1+\Delta-n}},
\end{equation}
(here we restore the dependence of $\Psi$
in $\Delta_1, \Delta_2$)
and for the special choice (\ref{specc})
of the parameter $\Delta$ this system
is orthogonal \eqref{trig-prod}
and complete \eqref{complet2} with respect to the
modified scalar product \eqref{Cscprod} which we defined
 in the space of functions of two variables $x_1,x_2$.
The form of the scalar product \eqref{Cscprod}
is dictated  by the
requirement that the graph building operator
 $\hat{Q}_{12}^{(\beta)}|_{\beta=D-\Delta_1-\Delta_2}$
(see Subsection {\bf \ref{scprpsi}}) and the
Casimir operator $\widehat{C}_2$ (defined in (\ref{cnfA2b}))
in the representation $(\rho^{\Delta_1}\otimes\rho^{\Delta_2})$
(see (\ref{splcaz}), (\ref{hcaz5}),
Subsection {\bf \ref{algf}}) are Hermitian with
respect to this scalar product.

\vspace{0.5cm}

At the end of this subsection we show that the factorization
of the eigenvalue (\ref{factor}) of the
general graph building operator (or $R$-matrix)
(\ref{rybe2}), (\ref{Rdelta})
follows from the factorization (\ref{QQ5}) of the $R$-matrix.

\begin{proposition}\label{prop8}
Due to the special conformal
transformation properties, the operator
$Q_{_{\Delta_1\Delta_2}}(u)$
converts the conformal triangle (\ref{cnftri})
 to another conformal triangle
\begin{align}
\label{qtpsi}
Q_{_{\Delta_1\Delta_2}}(u)\,
|\Psi^{\mu_1\cdots\mu_n}_{\Delta_1\,,\Delta_2,\,\Delta}\rangle =
a(\kappa)\,\tau\left(\delta',\kappa,n\right)\,
\Big|\Psi^{\mu_1\cdots\mu_n}_{\frac{\Delta_1+\Delta_2}{2} +
u\,,\frac{\Delta_1+\Delta_2}{2} - u,\,\Delta} \Big\rangle
\end{align}
where we omit $x$ in the notation
of $|\Psi\rangle$ for simplicity and define
\begin{align}
\delta = \frac{1}{2}(\Delta_1-\Delta_2+\Delta-n) \
\ \ \ , \ \ \
\kappa := \frac{1}{2}(\Delta_1 -\Delta_2) - u \; , \;\;\;\;
\delta' = D/2 - \delta \; .
\end{align}
\end{proposition}
\noindent
{\bf Proof.} The integral kernel for the
 building operator (\ref{QQ}) is
 \be
 \lb{intqq}
 \begin{array}{c}
 \langle x_1,x_2| Q_{_{\Delta_1\Delta_2}} | y_1,y_2 \rangle =
 \langle x_1,x_2| (\q_{12})^{2(\gamma-\zeta)}
 (\p_1)^{-2\kappa} (\q_{12})^{-2\beta} | y_1,y_2 \rangle =
 \\[0.2cm]  \displaystyle
 = a(\kappa) \frac{\delta^D(x_2-y_2)}{
 (x_1-x_2)^{2(\zeta-\gamma)} (x_1-y_1)^{2\kappa'}
 (y_1-y_2)^{2\beta}}
 \end{array}
 \ee
 where we use the parameters (\ref{rybe4})
 to make the notation concise. Here we use the diagram techniques
 developed in the proof of Proposition {\bf \em \ref{prop5}}
 to deduce  (\ref{qtpsi}). The action of the
 building operator (\ref{QQ}) to the function (\ref{cnftri})
 gives (cf. (\ref{qqkg04}))

  \unitlength=6mm
\begin{picture}(25,4)

 \put(2,3){\line(1,0){2}}
 %\put(2,1){\line(1,0){2}}

 \multiput(2,1)(0.21,0){9}{\circle*{0.08}}

\put(0.8,2){\tiny $\zeta-\gamma$}
\put(2.8,1.9){\tiny $\alpha+\beta$}
\put(5,1){\scriptsize $\rho,n$}
\put(5,2.8){\scriptsize $\delta,n$}

\put(2.6,2.55){\scriptsize $\kappa'$}
\put(1.5,3.3){\footnotesize $x_1$}
\put(1.5,0.6){\footnotesize $x_2$}

\put(3.5,3.3){\footnotesize $y_1$}
\put(3.5,0.6){\footnotesize $y_2$}
\put(6.2,1.9){\footnotesize $x$}
\put(2,1){\line(0,1){2}}
\put(4,1){\line(0,1){2}}
 \put(4,3){\line(2,-1){2}}
 \put(4,1){\line(2,1){2}}

 \put(3.8,2.8){$\bullet$}
 \put(3.8,0.9){$\bullet$}

 {\thicklines
\put(5.2,2.4){\vector(-2,1){0.4}}
\put(4.8,1.4){\vector(2,1){0.4}}
 }

\put(7.3,2){\footnotesize $= \; \tau(\delta',\kappa,n)\; \cdot$ }

%%%%%%%%%%%%%%%%%%%%%%%%%%%%%%%%%%%%

 \put(11.9,3.2){\line(3,-1){3.9}}
 \put(11.9,3.2){\line(1,-1){2.18}}
 \put(14.06,1.03){\line(2,1){1.75}}

  {\thicklines
\put(14,2.5){\vector(-3,1){0.4}}
\put(14.8,1.4){\vector(2,1){0.4}}
 }

\put(11.5,1.7){\scriptsize $\zeta$--$\gamma$}
\put(12.3,1.7){\scriptsize $+\delta'$}
\put(15,1){\scriptsize $\rho+\kappa,n$}
\put(13.7,2.8){\scriptsize $(\alpha+\beta)',n$}

\put(11.5,3.4){\footnotesize $x_1$}

\put(13.5,0.6){\footnotesize $x_2$}
\put(16,1.9){\footnotesize $x$}

%%%%%%%%%%%%%%%%%%%%%%%%%%%%%%%%%%%

\put(17.3,2){\footnotesize $= \; \tau(\delta',\kappa,n)\; \cdot \;
|\Psi^{\mu_1\cdots\mu_n}_{\widetilde{\Delta}_1\,,
\widetilde{\Delta}_2,\,\widetilde{\Delta}}\rangle$ \; ,}

\end{picture}

\vspace{-2cm}
\be
\lb{intqq2}
{}
\ee
\vspace{0.4cm}

\noindent
where, according to the definition (\ref{cnftri}), we fix
the parameters
 \be
 \lb{rybe5}
 \delta = \frac{1}{2}(\Delta_1-\Delta_2+\Delta-n) \, , \;\;\;
  \rho = \frac{1}{2}(\Delta_2-\Delta_1+\Delta-n) \, , \;\;\;
  \alpha = \frac{1}{2}(\Delta_1+\Delta_2-\Delta+n) \, ,
 \ee
 and
 \be
 \lb{rybe6}
 (\alpha+\beta)' =
 \frac{1}{2}(\widetilde{\Delta}_1-\widetilde{\Delta}_2
 +\widetilde{\Delta}-n) \, , \;\;\;
  \rho + \kappa =
  \frac{1}{2}(\widetilde{\Delta}_2-\widetilde{\Delta}_1
  +\widetilde{\Delta}-n) \, , \;\;\;
  \zeta-\gamma +\delta' =
  \frac{1}{2}(\widetilde{\Delta}_1+\widetilde{\Delta}_2
  -\widetilde{\Delta}+n) \, .
 \ee
 Substitution of (\ref{rybe4}) and (\ref{rybe5})
 in equations (\ref{rybe6}) leads to relations
 $$
 \widetilde{\Delta}_1 = \frac{\Delta_1 + \Delta_2}{2} +u \, , \;\;\;
  \widetilde{\Delta}_2 = \frac{\Delta_1 + \Delta_2}{2} -u \, , \;\;\;
  \widetilde{\Delta} = \Delta \; ,
 $$
 that finish the proof of (\ref{qtpsi}). \hfill \qed

\vspace{0.5cm}

{\bf Remark 3.} Another derivation of
(\ref{qtpsi}) is based on conformal transformations.
We are going to demonstrate that relation
(\ref{qtpsi})
is dictated by the commutation relations of the operator
$Q_{_{\Delta_1\Delta_2}}(u)$
with transformations from the conformal group and it remains
to prove that the coefficient $\tau(\delta',\kappa,n)$
is given by formula (\ref{factor1}).

\vspace{0.2cm}

The system of functions (\ref{conf-tr}),
(\ref{cnftri}) is defined by
$\Delta_1, \Delta_2 \in \mathbb{R}$ and
by the parameter $\Delta$ and can
be represented in the following form:
\begin{align}
\Psi^{\mu_1\cdots\mu_n}_{\Delta_1,\Delta_2,\Delta,x}
(x_1\,,x_2) =
e^{-x(\partial_1+\partial_2)}\,
\frac{\left(\frac{x_1}{x_1^2}-
\frac{x_2}{x_2^2}\right)^{\mu_1\cdots\mu_n}}
{(x_1-x_2)^{\Delta_1+\Delta_2-\Delta+n}
x_1^{\Delta_1-\Delta_2+\Delta-n}
x_2^{\Delta_2-\Delta_1+\Delta-n}} = \\
e^{-i x(\p_1+\p_2)}\,
{\cal I}^{(1)}_{\Delta_1}\,
{\cal I}^{(2)}_{\Delta_2}\circ
\frac{\left(x_1-x_2\right)^{\mu_1\cdots\mu_n}}
{(x_1-x_2)^{\Delta_1+\Delta_2-\Delta+n}}\, ,
\end{align}
where the circle $\circ$ denotes the action of
operators on the wave function and we apply formulas (\ref{inver2}).
Now we use the commutation relations (\ref{confR}) and
 $[Q_{_{\Delta_1\Delta_2}}(u), \,\left(\p_1+\p_2\right)] =0$
to obtain
\be
\lb{qqpsi}
\begin{array}{c}
Q_{_{\Delta_1\Delta_2}}(u)\circ
\Psi^{\mu_1\cdots\mu_n}_{\Delta_1,\Delta_2,\Delta,x}
(x_1\,,x_2)  = Q_{_{\Delta_1\Delta_2}}(u)\,
e^{-ix(\p_1+\p_2)}\,
{\cal I}^{(1)}_{\Delta_1}\,
{\cal I}^{(2)}_{\Delta_2}\circ
\frac{\left(x_{12}\right)^{\mu_1\cdots\mu_n}}
{(x_{12})^{\Delta_1+\Delta_2-\Delta+n}} = \\[0.3cm]
= e^{-ix(\p_1+\p_2)}\,
{\cal I}^{(1)}_{\frac{\Delta_1+\Delta_2}{2}+u}\,
{\cal I}^{(2)}_{\frac{\Delta_1+\Delta_2}{2} -u}
\,Q_{_{\Delta_1\Delta_2}}(u)\circ
\frac{\left(x_{12}\right)^{\mu_1\cdots\mu_n}}
{(x_{12})^{\Delta_1+\Delta_2-\Delta+n}}\, ,
\end{array}
\ee
where
 $x_{12}^\mu :=x_{1}^\mu - x_{2}^\mu$.
Thus, the problem is reduced to the action of the operator
$Q_{_{\Delta_1\Delta_2}}(u)$ on the function
$\frac{x_{12}^{\mu_1\cdots\mu_n}}
{x_{12}^{\Delta_1+\Delta_2-\Delta+n}}$.
In view of the spectral relation (\ref{spect}),
any function of the form $\frac{x_{12}^{\mu_1\cdots\mu_n}}
{x_{12}^{2\sigma}}$ must be an eigenfunction
of the operator $Q_{_{\Delta_1\Delta_2}}(u)$.
Indeed, we obtain
\be
\lb{qqxx}
\begin{array}{c}
Q_{_{\Delta_1\Delta_2}}(u)\circ
\frac{x_{12}^{\mu_1\cdots\mu_n}}
{x_{12}^{2\sigma}} =
{\q}_{12}^{2\left(\frac{D}{2}-\Delta_1\right)}\,
\p_1^{2\left(u+\frac{\Delta_2-\Delta_1}{2}\right)}\,
\q_{12}^{2\left(u+\frac{\Delta_1+\Delta_2-D}{2}\right)}\circ
\frac{x_{12}^{\mu_1\cdots\mu_n}}
{x_{12}^{2\sigma}} = \\ [0.3cm]
= e^{-x_2\partial_1}{x}_{1}^{2\left(\frac{D}{2}-\Delta_1\right)}\,
\Bigl( \p_1^{2\left(u+\frac{\Delta_2-\Delta_1}{2}\right)}\,
\q_1^{2\left(u+\frac{\Delta_2-\Delta_1}{2}\right)}\Bigr)
\circ \frac{x_{1}^{\mu_1\cdots\mu_n}}{
x_1^{2(\sigma+\frac{D}{2}-\Delta_1)}} =
 q_{\Delta_1\Delta_2}(u,\,\sigma)\,
 \frac{x_{12}^{\mu_1\cdots\mu_n}} {x_{12}^{2\sigma}} \, ,
\end{array}
\ee
where, according to (\ref{spect}), the eigenvalue is
 \be
 \lb{qqdd}
 q_{\Delta_1\Delta_2}(u,\,\sigma) =
 4^{-\kappa} \,
 \frac{\Gamma(D/2-\chi-\kappa+n)\, \Gamma(\chi)}{
 \Gamma(\chi+\kappa)\,  \Gamma(D/2-\chi+n)} \; , \;\;\;\;\;
 \kappa := \frac{\Delta_1-\Delta_2}{2} - u \; , \;\;\;
 \chi := \sigma+\frac{D}{2}-\Delta_1 \; .
 \ee
 Substitution of (\ref{qqxx}) into (\ref{qqpsi}) gives
 \be
 \lb{qqpsi2}
\begin{array}{c}
Q_{_{\Delta_1\Delta_2}}(u)\circ
\Psi^{\mu_1\cdots\mu_n}_{\Delta_1,\Delta_2,\Delta,x}
(x_1\,,x_2)  =  \\[0.3cm]
= q_{\Delta_1\Delta_2}(u,\sigma)\, e^{-x(\partial_1+\partial_2)}\,
{\cal I}^{(1)}_{\frac{\Delta_1+\Delta_2}{2}+u}\,
{\cal I}^{(2)}_{\frac{\Delta_1+\Delta_2}{2} -u}\circ
\frac{\left(x_1-x_2\right)^{\mu_1\cdots\mu_n}}
{(x_1-x_2)^{\Delta_1+\Delta_2-\Delta+n}} = \\[0.3cm]
= q_{\Delta_1\Delta_2}\,(u,\sigma )\,
\Psi^{\mu_1\cdots\mu_n}_{\frac{\Delta_1+\Delta_2}{2}+u,
\frac{\Delta_1+\Delta_2}{2}-u,\Delta,x}
(x_1\,,x_2)
\end{array}
\ee
where $\sigma = \frac{1}{2}(\Delta_1+\Delta_2-\Delta+n)$,
and we have $\chi :=
\frac{1}{2}(\Delta_2-\Delta_1-\Delta+n)+\frac{D}{2}$.
Thus, the explicit form of the eigenvalue
(\ref{qqdd}) is
 $$
\begin{array}{c}
\displaystyle
q_{\Delta_1\Delta_2}\,(u) = q_{\Delta_1\Delta_2}\,(u,\sigma ) =
 4^{-\kappa} \,
 \frac{\Gamma(D/2-\chi-\kappa+n)\,  \Gamma\left(\chi\right)}{
 \Gamma(\chi+\kappa)\,  \Gamma(D/2-\chi+n)} = \\ [0.4cm]
 \displaystyle
 = 4^{\frac{\Delta_2-\Delta_1}{2} + u} \,
 \frac{\Gamma\left(\frac{1}{2}(\Delta + n) +u\right)\,
 \Gamma\left(\frac{1}{2}(D+n-\Delta+\Delta_2-\Delta_1)\right)}{
 \Gamma\left(\frac{1}{2}(D + n-\Delta)-u \right)\,
 \Gamma\left(\frac{1}{2}(\Delta + n+\Delta_1-\Delta_2)\right)}
 \; .
\end{array}
$$
where we omit for simplicity $\sigma$ in the notation
for the eigenvalue $q_{\Delta_1\Delta_2}\,(u) =
q_{\Delta_1\Delta_2}\,(u,\sigma )$.
It is easy to check that
$q_{\Delta_1\Delta_2}\,(u) =
a(\kappa)\,\tau\left(\delta',\kappa,n\right)$
so that this expression is compatible to (\ref{qtpsi}).

Finally we
note that relation (\ref{qqpsi2}) is equivalently written as
\begin{align}
\nonumber
q^{-1}_{\Delta_1\Delta_2}\,(u)\,
|\Psi^{\mu_1\cdots\mu_n}_{\Delta_1\,,\Delta_2}\rangle =
 %\tau\left(\alpha,\beta-u,n\right)\,
 Q_{_{\Delta_1\Delta_2}}^{-1}(u)\,
|\Psi^{\mu_1\cdots\mu_n}_{\frac{\Delta_1+\Delta_2}{2} +
u\,,\frac{\Delta_1+\Delta_2}{2} - u}\rangle \; ,
\end{align}
Now for the whole R-operator one obtains
\begin{multline*}
R_{\Delta_1\Delta_2}(u)\,|\Psi^{\mu_1\cdots\mu_n}_{\Delta_1\,,\Delta_2}\rangle
= Q_{_{\Delta_1\Delta_2}}^{-1}(-u)\,{\cal P}_{12}\,
Q_{_{\Delta_1\Delta_2}}(u)\,
|\Psi^{\mu_1\cdots\mu_n}_{\Delta_1\,,\Delta_2}\rangle = \\
q_{\Delta_1\Delta_2}\,(u)\,
Q_{_{\Delta_1\Delta_2}}^{-1}(-u)
\,{\cal P}_{12}\,|
\Psi^{\mu_1\cdots\mu_n}_{\frac{\Delta_1+\Delta_2}{2} +
u\,,\frac{\Delta_1+\Delta_2}{2} - u}\rangle = \\
q_{\Delta_1\Delta_2}\,(u)\,(-1)^n\,
Q_{_{\Delta_1\Delta_2}}^{-1}(-u)\,|
\Psi^{\mu_1\cdots\mu_n}_{\frac{\Delta_1+\Delta_2}{2} -
u\,,\frac{\Delta_1+\Delta_2}{2} + u}\rangle =
(-1)^n\, \frac{q_{\Delta_1\Delta_2}\,(u)}
{q_{\Delta_1\Delta_2}\,(-u)}\,
|\Psi^{\mu_1\cdots\mu_n}_{\Delta_1\,,\Delta_2}\rangle
\end{multline*}
so that the eigenvalue $r_{\Delta,n}(u)$ of the operator
$R_{\Delta_1\Delta_2}(u)$ can be represented in a
compact form
\begin{align}
\label{eigR}
r_{\Delta,n}(u) = (-1)^n\, \frac{q_{\Delta_1\Delta_2}\,(u)}
{q_{\Delta_1\Delta_2}\,(-u)} = (-1)^n\,4^{2u}\,
\frac{\Gamma\left(\frac{\Delta+n}{2}+u\right)}{\Gamma\left(\frac{\Delta+n}{2}-u\right)}
\frac{\Gamma\left(\frac{D-\Delta+n}{2}+u\right)}{\Gamma\left(\frac{D-\Delta+n}{2}-u\right)}
\end{align}

One can check directly that this eigenvalue is equal to
the eigenvalue (\ref{factor}), (\ref{factor1})
up to the factor $a(\kappa)a(\gamma)$ which appears
in the relation of the graph building operator
$Q_{12}^{(\zeta,\kappa,\gamma)}$, given in (\ref{rybe2}),
and the $R$-operator (\ref{Rdelta})
$$
Q_{12}^{(\zeta,\kappa,\gamma)} =
\frac{1}{a(\kappa)a(\gamma)} R_{\Delta_1,\Delta_2}(u) \; ,
$$
where the parameters $\zeta$, $\kappa$, $\gamma$
are expressed via the parameters $\Delta_1$, $\Delta_2$, $u$
by means of equations (\ref{rybe4}).

\vspace{0.5cm}

\section{Conclusion}

 In this paper, we have proposed an operator approach
 to evaluating
 multiple integrals for special classes of
 the multiloop Feynman massless diagrams.

 We have considered classes of
 diagrams of iterative type containing
 some repeating elementary building blocks. These blocks are
 represented
 as special operators (graph-building operators)
 being elements of the
 direct product of several Heisenberg algebras.
 For the  ladder diagrams, we have identified
 (see also \cite{Isa}) a set of
 commutative graph building operators $H_\alpha$. Here
 the parameters $\alpha$ are related to the indices of the
 massless propagators. Then the
 multiple integrals for $L$-loop ladder diagrams are given
 by the product of $L$ commuting operators $H_{\alpha_i}$
 $(i=1,...,L)$
 and therefore expressed as a product of $L$ eigenvalues
 of $H_{\alpha_i}$. A single complete set of eigenfunctions
 and corresponding eigenvalues for all commuting
 operators $H_\alpha$ $(\forall \alpha)$ were found.
 This enables us to explicitly express a wide
 class of $2$- and $4$-point ladder diagrams  as
 Mellin-Barnes type integrals. Special cases of
 these integrals are explicitly evaluated, that demonstrated
 the advantages of the method.

 As the second class of repeating type diagrams, we considered
 the zig-zag 2-point and
4-point planar Feynman graphs relevant to
 the bi-scalar $D$-dimensional ''fishnet'' field theory
 \cite{KG}, \cite{KO}.
The graph building operators for zig-zag diagrams were identified
 and convenient operator representations
for these Feynman diagrams were obtained. An amusing fact was that
 the complete set of eigenfunctions for
 zig-zag graph building operators was given
 by special 3-point correlation functions
 in $D$-dimensional conformal field theories. By making
 use of all these facts, we were enable to exactly evaluate
 the multiple integrals for the special zig-zag diagrams.
In particular, we found a fairly simple derivation of values for
zig-zag $L$-loop two-point diagrams for $D=4$, that was the
rationale of the Broadhurst-Kreimer conjecture \cite{BrKr}.

The role of conformal symmetry in our approach, and
especially the
surprising appearance of 3-point conformal correlators as
eigenfunctions for the graph building operators were
 explained by the connection of the graph
 building operators with the conformal
invariant solution $R_{\Delta_1,\Delta_2}(u)$
of the Yang-Baxter equation.
 The operator $R_{\Delta_1,\Delta_2}(u)$ is interpreted as an
 intertwining operator
 of two spaces $V_{\Delta_1}$ and
 $V_{\Delta_2}$ of scalar conformal fields with
 conformal dimensions $\Delta_1$ and $\Delta_2$.

Finally, we have to mention the relation of the
graph building operator (\ref{Qgen})
(and $R$-matrix (\ref{rybe3}), (\ref{rybe0}))
to the BFKL equation which describes compound states of two
reggeized gluons \cite{Lipatov},\cite{Lip1},
\cite{Lip2},\cite{Lip3},\cite{FK}
(for recent review about BFKL equation see \cite{Kirschner}).
Indeed, the special case
{{$\Delta_1=\Delta_2 \equiv \Delta_+$}}
of the $R$-operator (\ref{rybe3}), (\ref{rybe0})
underlies the Lipatov integrable
model of the high-energy asymptotics of
multicolor QCD
$$
\begin{array}{c}
\check{R}_{12}^{(\Delta_1,\Delta_2)}(u)
 \;\; \stackrel{u \to 0}{\rightarrow} \;\;
 1 + u \, H^{(\Delta_+)}_{12} + u^2 \dots \, , \;\;\;\;\;\;
 H^{(\Delta_+)}_{12} = 2 \ln q_{12}^2 +
 \hat{q}_{12}^{2\Delta_+'} \ln(  \p_1^{2} \;
 \p_2^{2}) \; \hat{q}_{12}^{-2\Delta_+'} \, \equiv  \\ [0.3cm]
 \equiv \ln \p_1^{2} +
 \ln \p_2^{2} + \p_{1}^{-2\Delta_+'} \ln  (\hat{q}_{12}^{2})
 \p_{1}^{2\Delta_+'} + \p_{2}^{-2\Delta_+'} \ln  (\hat{q}_{12}^{2})
 \p_{2}^{2\Delta_+'} \; ,
 \end{array}
 $$
where as usual $\Delta_+' = D/2 -\Delta_+$
 and $H^{(\Delta_+)}_{12}$ gives the Hamiltonian of the
 model.
In two dimensions $D=2$ and in the special case
$\Delta_1=\Delta_2=\Delta_+= 0$
we obtain
\begin{align*}
H^{(\Delta_+)}_{12}|_{\Delta_+=0,D=2} = 2 \ln q_{12}^2 +
 \hat{q}_{12}^{2} \ln(  \p_1^{2} \;
 \p_2^{2}) \; \hat{q}_{12}^{-2} = \ln \p_1^{2} +
 \ln \p_2^{2} + \p_{1}^{-2} \ln  (\hat{q}_{12}^{2})
 \p_{1}^{2} + \p_{2}^{-2} \ln  (\hat{q}_{12}^{2})
 \p_{2}^{2}\,.
\end{align*}
There exists (in complex coordinates $\hat{q}_{12}^{2} \to
z_{12} \, \bar{z}_{12}$)
a holomorphic factorization when the initial
operator can be represented as the sum of operators acting on
holomorphic
and antiholomorphic coordinates separately
\begin{align*}
H^{(\Delta_+)}_{12}|_{\Delta_+=0,D=2} \to H_{12} + \bar{H}_{12}\,,\\
H_{12} = \ln {\sf p}_{1} + \ln {\sf p}_{2} +
{\sf p}^{-1}_{1}\,\ln(z_{12})\, {\sf p}_{1} +
{\sf p}_{2}^{-1}\,\ln(z_{12})\, {\sf p}_{2} \; ,
\end{align*}
where ${\sf p}_k = -i\frac{\partial}{\partial z_k}$,
 and the same story
occurs with the
Casimir operator (\ref{splcaz}) in two dimensions
\begin{align*}
\widehat{C}_{\Delta_1\Delta_2}|_{\Delta_k=0,D=2}=
-\left(\q_{12}\right)^2 (\p_1^\mu \, \p_{2\, \mu}) +
2 \q_{12}^{\mu}\, \q_{12}^{\nu} \, \p_{1\,\mu} \, \p_{2\, \nu} =
- 2 \bigl( z_{12}^2\,\partial_{z_1}\partial_{z_2}+
\bar{z}_{12}^2\,\partial_{\bar{z}_1}\partial_{\bar{z}_2}\bigr) \; .
\end{align*}

The operator $\mathcal{H}_{12} \equiv
H^{(\Delta_+)}_{12}|_{\Delta_+=0,D=2}$
coincides (up to an additional constant)
with the local (pair) Hamiltonian
of the Lipatov model \cite{Lipatov}.
For the Lipatov model, the BFKL equation takes the form of the
Schr\"{o}dinger equation
$\mathcal{H}_{12} \, \psi(x_1,x_2) = E \, \psi(x_1,x_2)$,
where two-dimensional vectors $x_1,x_2$ are the
eigenvalues of the position operators $\q_1,\q_2$.
The ground state eigenvalue $E$ is related
to the pomeron intercept and
eigenvalues $E$ can be extracted from
(\ref{eigR}) in the limit $u \to 0$.
The complete and orthogonal set of eigenfunctions
\cite{Lipatov} is given by \eqref{cnftri}.
In two dimensions the
symmetric traceless tensor $x^{\mu_1\cdots\mu_n}$
has two independent components which,
in complex coordinates, are: $z^n$ and $\bar{z}^n$.
The $z^n$-component of the function \eqref{cnftri} written in
complex coordinates (and under the conditions
$\Delta_1 =\Delta_2 \to 0$ and
$\Delta = \frac{D}{2}+2 i\nu \to 1+2 i\nu$)
is reduced to the form \cite{Lipatov}
\begin{multline*}
\langle x_1\,,x_2 |
\Psi^{\mu_1\cdots\mu_n}_{\Delta_1,\Delta_2,\Delta,x}
\rangle \to \frac{\left(\frac{z-z_1}{|z-z_1|^2}-\frac{z-z_2}{
|z-z_2|^2}\right)^{n}}
{|x_1-x_2|^{-(1-n+2i\nu)}
|x-x_1|^{1-n+2 i\nu}
|x-x_2|^{1-n+2 i\nu}} = \\
\left(\frac{\bar{z}_1-\bar{z}_2}{(\bar{z}-\bar{z}_1)(\bar{z}-\bar{z}_2)}\right)^n
\left(\frac{|x_1-x_2|}{|x-x_1||x-x_2|}\right)^{1-n+2i\nu}
\end{multline*}
The compound states of $N$
reggeized gluons are described by the Lipatov model with the
Hamiltonian ${\cal H} = \sum_{i=1}^N H_{i,i+1}$ with the periodic
boundary condition $x_{i+N}=x_i$. The Hamiltonian ${\cal H}$
is deduced by the standard method \cite{Lip2},\cite{FK} from the
transfer-matrix which is constructed
as the product of $N$ copies of
$R$-operators (\ref{rybe3}), (\ref{rybe0}).

%%%%%%%%%%%%%%%%%%%%%%%%%%%%%%%%%%%%%%%%%%%%%%%%%%%%%%%%%%%%%%%%%%%%%%%%%%%%%%%%5
\section*{Acknowledgment}

We thank G. Arutyunov, K. Chetyrkin, M. Kompaniets, A. Kotikov,
A. Kataev and V. Kazakov for valuable discussions.
 API is grateful to the L. Euler
International Mathematical Institute
in Saint Petersburg for kind hospitality
during his visit at the end of 2021
when this work was started.
The work of SED and LAS was supported by the Theoretical Physics and
Mathematics
Advancement Foundation «BASIS».

\vspace{1cm}

%%%%%%%%%%%%%%%%%%%%%%%%%%%%%%%%%%%%%%%%%%%%%%%%%%%%%%%%%%%%%%%%%%%%%%%%%%%%%%%%%%%%%%%%%%%%

\section{Appendix A\label{project}}
\setcounter{equation}0

The projector $P^{\mu_1 ... \mu_n}_{\nu_1 ... \nu_n}
\equiv (\Pi_{1\to n})^{\mu_1 ... \mu_n}_{\nu_1 ... \nu_n}$ in
$(\mathbb{R}^D)^{\otimes n}$,
which was introduced in (\ref{orthog}), satisfies
 $$
 \begin{array}{c}
 \Pi_{1 \to n} P_{k,r} = \Pi_{1 \to n} =  P_{k,r} \Pi_{1 \to n} \;,
 \;\;\;
 \Pi_{1 \to n} K_{k,r} = 0 = K_{k,r} \, \Pi_{1 \to n} \;, \;\;\;\;\;
 \forall k < r \leq n \; ,
 \end{array}
 $$
 where
 $$
 (P_{kr})^{\mu_1 ... \mu_n}_{\nu_1 ... \nu_n} =\delta^{\mu_1}_{\nu_1}
 \cdots  \delta^{\mu_k}_{\nu_k} \!\!\!\!\!\!\!\! \diagup \;\;
  \cdots \delta^{\mu_r}_{\nu_r} \!\!\!\!\!\!\!\! \diagup \;\;
   \cdots \delta^{\mu_n}_{\nu_n}\, P^{\mu_k\mu_r}_{\nu_k\nu_r} \, ,
 \;\;\; (K_{kr})^{\mu_1 ... \mu_n}_{\nu_1 ... \nu_n}
  =\delta^{\mu_1}_{\nu_1}
 \cdots  \delta^{\mu_k}_{\nu_k} \!\!\!\!\!\!\!\! \diagup \;\,
  \cdots \delta^{\mu_r}_{\nu_r} \!\!\!\!\!\!\!\! \diagup \;\,
   \cdots \delta^{\mu_n}_{\nu_n}\, K^{\mu_k\mu_r}_{\nu_k\nu_r} \, ,
  $$
and $P^{\mu_1\mu_2}_{\nu_1\nu_2} =
\delta^{\mu_1}_{\nu_2}  \delta^{\mu_2}_{\nu_1}$ is the permutation
matrix
while $K^{\mu_1\mu_2}_{\nu_1\nu_2} =
\delta^{\mu_1\mu_2}  \delta_{\nu_1\nu_2}$ is the rank-1 matrix;
$P$ and $K$ are the operators in $(\mathbb{R}^{D})^{\otimes 2}$.
The projector $\Pi_{1 \to n}$ can be defined
recurrently \cite{IsRub2} and we write this definition in the form
\be
  \lb{bra01}
 \Pi_{1\to n} = \Pi_{1\to n-1} \cdot
 \frac{(y_n +1)\cdot (y_n +\omega + n-3)}{n \cdot
 (2n -4+\omega)} \cdot \Pi_{1\to n-1} \; ,
 \ee
 where $\omega$ is a parameter (in our case $\omega=D$) and
 $y_n = \sum\limits_{k=1}^{n-1} (P_{k,n} - K_{k,n})$ is the
 commuting set of Jucyc-Murphy elements.  Relation
 (\ref{bra01}) is also written as
(cf. Remark 3.8 in \cite{IsMol}; \cite{IsPod})
\be
 \lb{bra04}
  \begin{array}{c}
 \displaystyle \Pi_{1\to n} = \Pi_{1\to n-1} \cdot
  R_{n-1}(n-1) \cdot \Pi_{1\to n-1}
  = \Pi_{1 \to n-1} \cdot R_{n-1}(n-1) \cdots  R_{2}(2) \cdot R_{1}(1)
  =
  \\ [0.2cm]
 = R_{1}(1) \cdot R_{2}(2) \cdots R_{n-1}(n-1) \cdot \Pi_{1\to n-1}  \;
 .
 \end{array}
 \ee
 Here
 $R_{k}(u) = I^{\otimes k-1} \otimes R(u) \otimes I^{\otimes n-k-1}$
 and
 $$
 R(u) = \frac{1}{1+u}(I + u P + \frac{u}{1-D/2-u}K) \;
 \;\;\;\;\; \Bigl(R(u) \cdot R(-u) =  I \Bigr)\; ,
 $$
 is the $so(D)$-invariant
  Zamolodchikov solution of the Yang-Baxter equation.
 The last expression in (\ref{bra04}) is simplified as
 follows:
  \be
\lb{Pin01}
\begin{array}{c}
\Pi_{1 \to n} = S_{1 \to n} \Bigl(1 - \frac{n-1}{D  + 2(n-2)}
 \, K_{n-1,n} \Bigr) \Pi_{1 \to n-1} =
 S_{1 \to n} \Bigl(1 - \frac{1}{D  + 2(n-2)}
 (\sum\limits_{i=1}^{n-1} K_{in})\Bigr)
\Pi_{1\to n-1} = \\ [0.2cm]
 = S_{1 \to n} \Bigl(1 - \frac{1}{D  + 2(n-2)}
 \sum\limits_{i=1}^{n-1} K_{in}\Bigr) \cdots
\Bigl(1 - \frac{1}{D  + 2}
 \sum\limits_{i=1}^{2} K_{i3}\Bigr)
 \Bigl(1 - \frac{1}{D } K_{12})\Bigr)
 \; \in \; {\rm End}( (\mathbb{R}^D)^{\otimes n})  \; ,
 \end{array}
 \ee
 where $S_{1\to n}$ is the standard symmetrizer
  (see e.g. \cite{IsRub2}) in
 the group algebra $\mathbb{C}[S_n]$ of the symmetric group $S_n$.
 In fact, the projector (\ref{bra04}), (\ref{Pin01})
 is the image of the complete symmetrizer in the Brauer algebra
 (see details in \cite{IsMol}, \cite{IsRub2}).
 From the first equality in
 (\ref{bra04}) we deduce
 \be
 \lb{Pin02}
 \begin{array}{c}
{\rm Tr}_n(\Pi_{1 \to n}) = \Pi_{1\to n-1} \cdot
  {\rm Tr}_n\bigl( R_{n-1}(n-1)\bigr)  \cdot \Pi_{1\to n-1} =
\frac{(D+2n-2)(D+n-3)}{n(D+2n-4)} \Pi_{1 \to n-1} \; \Rightarrow
\; \\ [0.3cm]
{\rm Tr}_{1...n}(\Pi_{1 \to n}) =
 \frac{(D+2n-2)\, \Gamma(D+n-2)}{n!\; \Gamma(D-1)} .
 \end{array}
 \ee
 It is clear that ${\rm Tr}_{1...n}(\Pi_{1 \to n})$ is equal to
 the dimension of the totally symmetric irreducible representation
 $[n]={\scriptsize  \begin{array}{|c|c|c|c|}
\hline
   \,   &    \,  & $\dots$ &  \,   \\ [0.1cm]
\hline
\end{array} }$ of $so(D)$.
 Now we define the
 generating function  of the projector (\ref{bra04}), (\ref{Pin01}):
 \be
 \lb{genF}
 \Pi_{n}(x;u):=
 x_{\mu_1} \cdots  x_{\mu_n}
  (\Pi_{n})^{\mu_1 ... \mu_n}_{\;\; \nu_1 ... \nu_n}
  u^{\nu_1} \cdots u^{\nu_n} \equiv
   x^{\mu_1 ... \mu_n} u^{\mu_1 ... \mu_n} \; ,
 \ee
  where we use the convenient notation
  $x^{\mu_1 ... \mu_n} := x^{\nu_1} \cdots  x^{\nu_n}
  \; (\Pi_{n})^{\mu_1 ... \mu_n}_{\;\; \nu_1 ... \nu_n}$.
  After substitution of (\ref{Pin01}) in (\ref{genF}),
  this generating function is written
  as\footnote{For details see, e.g., an
 analogous statement for the Behrends-Fronsdal
 projector in \cite{IsPod1}, \cite{IsPod}, where it is necessary to
 make
 a shift
 $D \to D+1$ to obtain (\ref{Pin02b});
 see also \cite{DerShum} and references therein.}
 \be
 \lb{Pin02b}
 \Pi_{n}(x;u) =  \sum_{A=0}^{[n/2]} a^{(n)}_A \;
  u^{2A} \, x^{2A} \, (ux)^{n-2A} \; , \;\;\;\;\;\;
 a^{(n)}_A  := \frac{(-1)^A \; n!}{(n-2A)!\; A!} \;
 \frac{\Gamma(n+D/2-1-A)}{\Gamma(n+D/2-1)} \; .
 \ee
 We take $u=x$ and use the first equality in (\ref{Pin01}),
 which gives an
 equation for the generating function $\Pi_{n}(x;x) = F(n) \, x^{2n}$:
 $$
 \Pi_n(x;x)  = \frac{(D  + n-3)}{D  + 2(n-2)}\;
 x^2 \,  \Pi_{n-1}(x;x) \;\;\;\; \Rightarrow \;\;\;\;
 F(n) = \frac{(D  + n-3)}{D  + 2(n-2)} F(n-1) \; /
 $$
 This equation is immediately solved as
$$
 x^{\mu_1 ... \mu_n} x^{\mu_1 ... \mu_n} \equiv \Pi_n(x;x)  =
 \frac{\Gamma(D +n-2)\Gamma(D /2-1)}{
 2^{n}\Gamma(D -2)\Gamma(D /2+n-1)} \; x^{2n} \; .
 $$

  Finally, we prove the orthogonality
  relation (\ref{orthog}). For the left-hand side of (\ref{orthog})
 we have
 \be
 \lb{Pin03}
 \int d^D x \frac{x^{\mu_1 ... \mu_n}}{x^{2(D/4+n/2-i v)}}
 \frac{x_{\nu_1 ... \nu_m}}{x^{2(D/4+m/2+i v')}} =
 \delta_{n,m} C_n \Pi^{\mu_1 ... \mu_n}_{\nu_1 ... \nu_n}
  \int d^D x \frac{1}{x^{2(D/2+i (v'-v))}} =
  \ee
  \be
 \lb{Pin04}
  = \pi \, \Omega_D\,  C_n \, \Pi^{\mu_1 ... \mu_n}_{\nu_1 ... \nu_n}
  \, \delta(v'-v) \delta_{n,m}  \; ,
 \ee
 where in the  last equality we use (\ref{gori}).
 The constant $C_n$ in (\ref{Pin03}) is found if we put $m=n$,
 contract indices $\mu_1 ... \mu_n$ with $\nu_1 ... \nu_n$
 and use relations (\ref{conv}) and (\ref{Pin02}).
 As a result we deduce $C_n = \frac{n! \Gamma(D/2)}{2^n
 \Gamma(D/2+n)}$.
 Substitution of this constant into (\ref{Pin04}) gives
 (\ref{orthog}).

 \section{Appendix B\label{TkaChe}}
\setcounter{equation}0

For the coefficient function $C(\alpha_1,\alpha_2,\alpha_6;
 \alpha_3,\alpha_4,\alpha_5)$ of the
two-loop master diagram in Fig.\ref{fig3}
 we deduced formula (\ref{2loop2}).  We write the  integrand in
 (\ref{2loop2}) in terms of the $\Gamma$-functions
 \be
 \lb{shum01}
 \begin{array}{c}
 \frac{\Gamma(\alpha_{1...5}-5D/4 + n/2-i\nu)
 \Gamma(\alpha_{125}-3D/4 + n/2-i\nu)
 \Gamma(\alpha_{1}-D/4 + n/2-i\nu)}{
 \Gamma(\alpha_{1235}-3D/4 + n/2-i\nu)
 \Gamma(\alpha_{12}-D/4 + n/2-i\nu)
 \Gamma(D/4 + n/2-i\nu)} \; \times \\ [0.3cm]
 \times \;
 \frac{\Gamma(5D/4 - \alpha_{1...5} + n/2+i\nu)
 \Gamma(3D/4-\alpha_{12} + n/2+i\nu)
 \Gamma(D/4 + n/2+i\nu)}{
 \Gamma(7D/4 - \alpha_{1...5} + n/2 + i\nu)
 \Gamma(5D/4 - \alpha_{125} + n/2+i\nu)
 \Gamma(3D/4 - \alpha_{1} + n/2+i\nu)}.
 \end{array}
 \ee
 Zeros and poles of expression (\ref{shum01})
 in the upper half-plane
 are defined respectively by the $\Gamma$-functions
 in the denominator and numerator in (\ref{shum01}).
 Thus, in the upper half-plane we have 3 sets of poles
 \be
 \lb{shum02}
 \left\{
  \begin{array}{l}
 \nu = i (D/4 +n/2+k) \; ; \\
 \nu = i (3D/4 -\alpha_{12} +n/2+k) \; ; \\
 \nu = i (5D/4 -\alpha_{1235} +n/2+k) \;;
 \end{array}
 \right.
 \ee
 and 3 sets of zeros
 \be
 \lb{shum03}
 \left\{
  \begin{array}{l}
 \nu = i (3D/4 -\alpha_{1} +n/2+k) \; ; \\
 \nu = i (5D/4 -\alpha_{125} +n/2+k) \; ; \\
 \nu = i (7D/4 -\alpha_{1...5} +n/2+k) \; ;
 \end{array}
 \right.
 \ee
  where $k \in \mathbb{Z}_{>0}$.

 For the Chetyrkin-Tkachov diagram we have
 $$
 \alpha_1 = \alpha_4 = \alpha_5 = D/2 - 1 \; , \;\;\;\;
  \alpha_3 = \alpha\; , \;\;\;\; \alpha_2 = \beta \; ,
 $$
  and the sets of poles (\ref{shum02})
  and zeros (\ref{shum03}) are written respectively as
   \be
 \lb{shum04}
 \left\{
  \begin{array}{l}
 \nu = i (D/4 +n/2+k) \; ; \\
 \nu = i (D/4 +1 -\beta +n/2+k) \; ; \\
 \nu = i (D/4 +2 -\alpha -\beta +n/2+k) \; ;
 \end{array}
 \right.
 \ee
 and
 \be
 \lb{shum05}
 \left\{
  \begin{array}{l}
 \nu = i (D/4 +1 +n/2+k) \; ; \\
 \nu = i (D/4 +2 -\beta +n/2+k) \; ; \\
 \nu = i (D/4 + 3 -\alpha-\beta +n/2+k) \; ,
 \end{array}
 \right.
 \ee
where $k \in \mathbb{Z}_{>0}$. We see that poles and zeros cancel each
other out, except the three poles for every $n \in \mathbb{Z}_{>0}$.
Thus, we can easily evaluate the integral~\eqref{2loop2} for
the Chetyrkin-Tkachov diagram by closing the contour in the upper
half-plane. Summing the residues, we obtain
\begin{align}\label{ChT-sum}
    ChT(\alpha, \beta) &= \dfrac{\pi^{D/2}}{(\Gamma(D/2 - 1))^2\Gamma(D
    - 2)}\dfrac{1}{(\alpha - 1)(\beta - 1)(\alpha + \beta - 2)}\sum_{n
    = 0}^{\infty}\dfrac{\Gamma(n + D - 2)}{n!}\times\nonumber \\
    \times&\left(\dfrac{\beta - 1}{(\alpha - D/2 - n)(\alpha + \beta -
    1 - D/2 - n)} + \dfrac{\alpha - 1}{(D/2 - 2 + \beta + n)(D/2 - 3 +
    \alpha + \beta + n)} + \nonumber\right.\\ + &\left.\dfrac{\alpha +
    \beta - 2}{(\beta - D/2 - n)(D/2 - 2 + \alpha + n)}\right),
\end{align}
where we use the notation $ChT(\alpha, \beta) = C(D/2 - 1, \beta, 3 -
\alpha - \beta; \alpha, D/2 - 1, D/2 - 1)$.

The sum over $n$ can be calculated analytically. Let us give an
example
 of how one can calculate the third term in~\eqref{ChT-sum}
\begin{equation}
    \sum_{n = 0}^{\infty}\dfrac{\Gamma(n + D - 2)}{n!}\dfrac{1}{(D/2 -
    \beta + n)(D/2 - 2 + \alpha + n)}.
\end{equation}
First of all, we write the summand in the form
\begin{align}\label{how-to-sum-3}
    \dfrac{1}{(D/2 - \beta +   n)(D/2 - 2 + \alpha + n)}& =
    \dfrac{1}{\alpha + \beta - 2}\times\nonumber \\
    &\times\left(\dfrac{1}{D/2 - \beta + n} - \dfrac{1}{\alpha + D/2 -2
    + n}\right).
\end{align}
Each term represents the definition of the
hypergeometric function, so we have
\begin{align}\label{how-to-sum-2}
   \sum_{n = 0}^{\infty}\dfrac{\Gamma(n + D - 2)}{n!}\dfrac{\Gamma(D/2
   - \beta + n)}{\Gamma(D/2 - \beta + n + 1)}& =  \dfrac{\Gamma(D -
   2)\Gamma(D/2 - \beta)}{\Gamma(D/2 - \beta + 1)}\times\nonumber \\
   &\times {}_2F_1(D - 2, D/2 - \beta;D/2 - \beta + 1; 1).
\end{align}
For a special value of the
hypergeometric function at the point $1$, we have the relation
\begin{equation}
    {}_2F_1(D - 2, D/2 - \beta;D/2 - \beta + 1; 1) = \dfrac{\Gamma(D -
    2)\Gamma(D/2 - \beta)\Gamma(3 - D)}{\Gamma(3 - D/2 -
    \beta)\Gamma(1)}.
\end{equation}
All other terms can be calculated in the same way, which lead us to the
well-known answer
\begin{align}\label{cht-final}
    ChT(\alpha,\beta) = \dfrac{\pi^{D/2}a_0(D - 2)}{\Gamma(D/2 -
    1)}&\left(\dfrac{a_0(\alpha)a_0(2 - \alpha)}{(1 - \beta)(\alpha +
    \beta - 2)} + \dfrac{a_0(\beta)a_0(2 - \beta)}{(1 - \alpha)(\alpha
    + \beta - 2)} +\right.\nonumber \\ &\left. + \dfrac{a_0(\alpha +
    \beta - 1)a_0(3 - \alpha - \beta)}{(\alpha - 1)(\beta -
    1)}\right).
\end{align}

 %\newpage

 \section{Appendix C}
 \label{B}
 \setcounter{equation}0

In this Appendix we collect some formulae
which we have used in the main text.
These formulae are useful in calculations so that for completeness
we have inserted the needed proofs and derivations.

\begin{itemize}
\item
\end{itemize}

We start with the identity
 \be
 \lb{magic}
\left.\partial_t^n \frac{1}{(1-tx)^A (1-ty)^{1-A-n}} \right|_{t=0} =
\frac{\Gamma(A+n)}{\Gamma(A)} (x-y)^n \; ,
 \ee
 (here $x,y,t,A$ are the parameters and $n \in \mathbb{Z}_{\geq 0}$)
 which can be easily proved by induction.
Let $\vec{u} \in \mathbb{C}^D$ be a complex vector and
$\vec{x},\vec{y} \in \mathbb{R}^D$ such that $\vec{u}\cdot\vec{u}=0$.
The application of the identity (\ref{magic}) is
 \be
 \lb{magic02}
 \begin{array}{c}
 \displaystyle
\left. \partial_t^n \frac{1}{(\vec{x} -t \vec{u})^{2A}
(\vec{y} -t \vec{u})^{2(1-A-n)}} \right|_{t=0} =
\left. \partial_t^n \frac{1}{(\vec{x}^2 - 2 t (\vec{u},\vec{x}) )^{A}
\;
(\vec{y}^2 -2 t (\vec{u},\vec{y}))^{(1-A-n)}} \right|_{t=0} = \\
[0.5cm]
 \displaystyle
=  2^n \; \frac{\Gamma(A+n)}{\Gamma(A)} \;
\frac{1}{x^{2A} y^{2(1-A-n)}} \;
\Bigl(\frac{(ux)}{x^2} -
\frac{(uy)}{y^2}\Bigr)^n \; ,
\end{array}
 \ee
where in the last expression (and below) we write $x,y,u$
instead of $\vec{x},\vec{y},\vec{u}$ and
$$
\left(\frac{(ux)}{x^2} -\frac{(uy)}{y^2}\right)^n = u^{\mu_1} \cdots
u^{\mu_n}
\left(\frac{x^{\mu_1}}{x^2} -\frac{y^{\mu_1}}{y^2}\right) \cdots
\left(\frac{x^{\mu_n}}{x^2} -\frac{y^{\mu_n}}{y^2}\right) =
u^{\mu_1...\mu_n} \; \left(\frac{x}{x^2}
-\frac{y}{y^2}\right)^{\mu_1...\mu_n} \; ,
$$
and as usual $z^{\mu_1...\mu_n}$ means
a traceless symmetric tensor. The structure
$(\frac{x}{x^2} -\frac{y}{y^2})^{\mu_1...\mu_n}$ appears in
the conformal triangles

\unitlength=6mm
\begin{picture}(25,4)(-5,0)

\put(3,2){\line(2,1){2}}
\put(3,2){\line(2,-1){2}}
\put(3.8,2.8){\footnotesize $\alpha$}
\put(3.8,1){\footnotesize $\gamma$}
\put(4.6,1.8){\footnotesize $\beta$}
\put(2.5,2){\footnotesize $z$}
\put(5.2,0.8){\footnotesize $y$}
\put(5.2,3){\footnotesize $x$}
\put(5,1){\line(0,1){2}}

 %\put(2.9,0.9){$\bullet$}
 %\put(2.9,2.9){$\bullet$}
 %\put(3,3){\line(2,-1){2}}
 %\put(4,2.8){$\alpha_5$}
 %\put(3,1){\line(2,1){2}}
 %\put(4,1){$\alpha_1$}

\put(6,2){$\displaystyle
\cdot \;\; \left(\frac{x}{x^2} -\frac{y}{y^2}\right)^{\mu_1...\mu_n}$}

 %\put(8,0){\bf Fig.2}

\end{picture}
\\and identity \eqref{magic02} allows one
to mimic this tensor structure by applying an
usual derivative with respect to an auxiliary parameter.

\begin{itemize}
\item
\end{itemize}

There exists an additional very useful identity which allows
one to
mimic the same tensor structure by different derivatives
$$
\left. \partial_t^n \frac{1}{(x-z)^{2B} (x-z -t u )^{2A}
(y-z -t u )^{2(1-A-n)}} \right|_{t=0} =
$$
$$
=  \frac{2^n\Gamma(A+n)}{\Gamma(A)} \; \frac{1}{(x-z)^{2(A+B)}
 (y-z)^{2(1-A-n)}} \,u^{\mu_1} \cdots u^{\mu_n}\, \left(
\frac{(x-z)}{(x-z)^2} -\frac{(y-z)}{(y-z)^2}\right)^{\mu_1...\mu_n} =
$$
$$
= \frac{\Gamma(A+n)}{\Gamma(A)} \;
\Bigl(\frac{\Gamma(A+B+n)}{\Gamma(A+B)} \; \Bigr)^{-1}
\left. \partial_t^n \frac{1}{(y-z)^{2B} (x-z -t u )^{2(A+B)}
(y-z -t u )^{2(1-A-B-n)}} \right|_{t=0}
$$
where $B$ is an arbitrary parameter.
We depict this identity as \\

\unitlength=7mm
\begin{picture}(25,4)(1,0)

\put(3,3){\line(1,0){3}}
\put(3,1){\line(1,0){3}}
\put(3,1){\line(3,2){3}}
\put(3.5,3.2){\scriptsize $1-A-n$}
\put(4.5,0.5){\scriptsize $B$}
\put(4.7,1.8){\scriptsize $A$}
\put(5.9,3.1){\footnotesize $z+tu$}
\put(6.2,0.8){\footnotesize $z$}
\put(2.4,0.8){\footnotesize $x$}
\put(2.4,3){\footnotesize $y$}

\put(7.6,2.8){$\leftarrow
\partial^n_{t}$}

%%%%%%%%%%%%%%%%%%%%%%%%%%

 \put(10,2){$= \;\;
 \frac{\Gamma(A+n)\; \Gamma(A+B)}{\Gamma(A)\; \Gamma(A+B+n)} \;
 \cdot$}

 \put(16,3){\line(1,0){3}}
\put(16,1){\line(1,0){3}}
\put(16,3){\line(3,-2){3}}
\put(17.5,3.2){\scriptsize $B$}
\put(16.5,0.5){\scriptsize $A+B$}
\put(17.6,2){\scriptsize $1-A-B-n$}
\put(19.1,3){\footnotesize $z$}
\put(18.8,0.6){\footnotesize $z+tu$}
\put(15.4,0.8){\footnotesize $x$}
\put(15.4,3){\footnotesize $y$}

\put(20.8,0.8){$\leftarrow
\partial^n_{t}$}

\end{picture}

\begin{itemize}
\item
\end{itemize}

Now we are going to derive the following tensor generalization of the
star-triangle identity:
\begin{align*}
\int \frac{d^D z  \;\;\; \bigl( \frac{y-z}{(y-z)^2} -
\frac{y-x_1}{(y-x_1)^2} \bigr)^{\mu_1...\mu_n}}{(z-x_2)^{2\beta'}
(z-x_1)^{2(\alpha+\beta)}(z-y)^{2\alpha'}} =
\tau(\alpha,\beta,n)
\frac{ \bigl(
 \frac{y-x_2}{(y-x_2)^2} -
\frac{y-x_1}{(y-x_1)^2}
\bigr)^{\mu_1...\mu_n}}{(y-x_2)^{2(\alpha'-\beta)}
(x_2-x_1)^{2\alpha}(x_1-y)^{2\beta}}\,.
\end{align*}
After contractions with the auxiliary vector $u^{\mu}$
one obtains an equivalent form of this relation
\begin{align*}
\int \frac{d^D z  \;\;\; \bigl( \frac{(u,y-z)}{(y-z)^2} -
\frac{(u,y-x_1)}{(y-x_1)^2} \bigr)^{n}}{(z-x_2)^{2\beta'}
(z-x_1)^{2(\alpha+\beta)}(z-y)^{2\alpha'}} =
\tau(\alpha,\beta,n)
\frac{ \bigl(
 \frac{(u,y-x_2)}{(y-x_2)^2} -
\frac{(u,y-x_1)}{(y-x_1)^2} \bigr)^{n}}{(y-x_2)^{2(\alpha'-\beta)}
(x_2-x_1)^{2\alpha}(x_1-y)^{2\beta}}\, ,
\end{align*}
This statement follows from the chain
of equalities: first one uses \eqref{magic02}, then
the star-triangle relation \eqref{startr2}
and again \eqref{magic02}
$$
 \begin{array}{c}
\int d^D z \;
\frac{\bigl( \frac{(u,y-z)}{(y-z)^2} -
\frac{(u,y-x_1)}{(y-x_1)^2} \bigr)^{n}}
{(x_2-z)^{2\beta'} (z-x_1)^{2(\alpha+\beta)}(y-z)^{2\alpha'}}
 \stackrel{(\ref{magic02})}{=}
\\ [0.3cm]
= \frac{\Gamma(\alpha')}{2^n \Gamma(\alpha'+n)}
\int  \;
\frac{d^D z\, \;}{(x_2-z)^{2\beta'} (z-x_1)^{2(\alpha+\beta)}}
\partial^n_t \Bigl( \frac{1}{
(y-z-t u)^{2\alpha'}(y-x_1-tu)^{2(1-\alpha'-n)}} \Bigr)
\frac{1}{(y-x_1)^{2(\alpha'+n-1)}}
\stackrel{(\ref{startr2}) \to z}{=}  \\ [0.3cm]
= \frac{\Gamma(\alpha')}{2^n \Gamma(\alpha'+n)}
\frac{\pi^{D/2} \Gamma(\alpha) \Gamma(\beta) \Gamma(\alpha'-\beta)}{
\Gamma(\alpha') \Gamma(\beta') \Gamma(\alpha+\beta)}
\frac{1}{(x_1-x_2)^{2\alpha}
(y-x_1)^{2(\alpha'+n-1)}}
  \partial^n_t  \frac{1}{
(y-x_1-t u)^{2(\beta+1-\alpha'-n)}(y-x_2-t u)^{2(\alpha'-\beta)}}
\stackrel{(\ref{magic02})}{=}
\\ [0.3cm]
=
\frac{\pi^{D/2} \Gamma(\beta) \Gamma(\alpha)
\Gamma(\alpha'-\beta)}{2^n  \Gamma(\beta') \Gamma(\alpha'+n)
\Gamma(\alpha+\beta)}
\frac{2^n \Gamma(\alpha'-\beta+n)}{\Gamma(\alpha'-\beta)}
\frac{1}{(x_1-x_2)^{2\alpha}
(y-x_1)^{2(\alpha'+n-1)}}
  \frac{\bigl( \frac{(u|y-x_2)}{(y-x_2)^2} -
\frac{(u|y-x_1)}{(y-x_1)^2} \bigr)^{n}}{(y-x_2)^{2(\alpha'-\beta)}
(y-x_1)^{2(\beta+1-\alpha'-n)}}  =
\end{array}
$$
$$
\begin{array}{c}
\frac{\pi^{D/2} \Gamma(\beta) \Gamma(\alpha) \Gamma(\alpha'-\beta+n)}{
\Gamma(\beta') \Gamma(\alpha'+n)
\Gamma(\alpha+\beta)}\frac{1}{(x_1-x_2)^{2\alpha}}
  \frac{\bigl( \frac{(u|y-x_2)}{(y-x_2)^2} -
\frac{(u|y-x_1)}{(y-x_1)^2} \bigr)^{n}}{(y-x_2)^{2(\alpha'-\beta)}
(y-x_1)^{2\beta}}
\end{array}
$$

\begin{itemize}
\item
\end{itemize}

Let us prove the following generalization of the chain relation
\begin{multline*}
\int d^D x \frac{(x-z)^{\mu_1\ldots\mu_n} (x-y)^{\mu_1\ldots\mu_n}
}{(x-z)^{2\alpha} (x-y)^{2\beta}} = \frac{\pi^{D/2} \Gamma(D/2
-\alpha+n)
 \Gamma(D/2 -\beta+n)\Gamma(\alpha+\beta-D/2-n)}{
 \Gamma(\alpha)\Gamma(\beta)\Gamma(D -\alpha-\beta+n)} \\
 \frac{\Gamma(n+D-2)\Gamma(D/2-1)}{2^{n} \Gamma(n+D/2-1)\Gamma(D-2)}
\cdot \frac{1}{(y-z)^{2(\alpha+\beta-D/2-n)}}
\end{multline*}
To prove it, we start with the relation
\be
\lb{zig01}
\frac{\Gamma(\alpha-n)\Gamma(\beta-n)}{
2^{2n}\Gamma(\alpha)\Gamma(\beta)}
\partial_z^{\mu_1\ldots\mu_n} \partial_y^{\mu_1\ldots\mu_n}
\frac{1}{(x-z)^{2(\alpha-n)} (x-y)^{2(\beta-n)}} =
\frac{(x-z)^{\mu_1\ldots\mu_n} (x-y)^{\mu_1\ldots\mu_n}
}{(x-z)^{2\alpha} (x-y)^{2\beta}}.
\ee
Then we integrate both parts of (\ref{zig01}) over $x$
and apply the chain relation (\ref{chain})
\begin{multline}
\lb{zig02}
\int d^D x \frac{(x-z)^{\mu_1\ldots\mu_n} (x-y)^{\mu_1\ldots\mu_n}
}{(x-z)^{2\alpha} (x-y)^{2\beta}} =
\frac{\Gamma(\alpha-n)\Gamma(\beta-n)}{
2^{2n}\Gamma(\alpha)\Gamma(\beta)}
 \; \cdot\\
\cdot \; \frac{\tilde{a}(D-\alpha-\beta+2n)}{\tilde{a}(D/2-\alpha+n)
\tilde{a}(D/2 -\beta+n)} \;
\partial_z^{\mu_1\ldots\mu_n} \partial_y^{\mu_1\ldots\mu_n}
\frac{1}{(y-z)^{2(\alpha+\beta-2n-D/2)}} .
\end{multline}
Finally, in the right-hand side of (\ref{zig02}),
we substitute the identity
\be
\lb{zig03b}
\begin{array}{c}
\partial_z^{\mu_1\ldots\mu_n} \partial_y^{\mu_1\ldots\mu_n}
\frac{1}{(y-z)^{2\gamma}} =
\partial_z^{\mu_1\ldots\mu_n} \partial_y^{\mu_1\ldots\mu_n}
\frac{1}{\tilde{a}(\gamma')} \int \frac{d^D k}{(2\pi)^D}
\frac{e^{ik(y-z)}}{k^{2\gamma'}} = \\ [0.3cm]
 =
\frac{1}{\tilde{a}(\gamma')} \int \frac{d^D k}{(2\pi)^D}
\frac{k^{\mu_1\ldots\mu_n}
k^{\mu_1\ldots\mu_n}e^{ik(y-z)}}{k^{2\gamma'}}
 \stackrel{(\ref{conv})}{=}
\frac{1}{\tilde{a}(\gamma')}
\frac{\Gamma(n+D-2)\Gamma(D/2-1)}{2^n\Gamma(n+D/2-1)\Gamma(D-2)}
\int \frac{d^D k}{(2\pi)^D}
\frac{e^{ik(y-z)}}{k^{2(\gamma'-n)}} = \\ [0.3cm]
= \frac{\tilde{a}(\gamma'-n)}{
 \tilde{a}(\gamma')} \;
 \frac{\Gamma(n+D-2)\Gamma(D/2-1)}{2^n \Gamma(n+D/2-1)\Gamma(D-2)}
\frac{1}{(y-z)^{2(\gamma+n)}}
 \end{array}
\ee
where we used (\ref{gr3}) and (\ref{conv}).
As a result we have
\be
\lb{zig04}
\begin{array}{c}
\int d^D x \frac{(x-z)^{\mu_1\ldots\mu_n} (x-y)^{\mu_1\ldots\mu_n}
}{(x-z)^{2\alpha} (x-y)^{2\beta}} =
\frac{\Gamma(\alpha-n)\Gamma(\beta-n)}{
2^{2n}\Gamma(\alpha)\Gamma(\beta)}
 \; \cdot \;
 \frac{\tilde{a}(D-\alpha-\beta+2n)}{\tilde{a}(D/2-\alpha+n)
\tilde{a}(D/2 -\beta+n)} \; \cdot \\ [0.3cm]
 \cdot \; \frac{\tilde{a}(D-\alpha-\beta+n)}{
 \tilde{a}(D-\alpha-\beta+2n)} \;
 \frac{\Gamma(n+D-2)\Gamma(D/2-1)}{2^n \Gamma(n+D/2-1)\Gamma(D-2)}
\frac{1}{(y-z)^{2(\alpha+\beta-D/2-n)}} =  \\ [0.3cm]
 = \frac{\pi^{D/2} \Gamma(\frac{D}{2} -\alpha+n)
 \Gamma(\frac{D}{2} -\beta+n)\Gamma(\alpha+\beta-D/2-n)}{
 \Gamma(\alpha)\Gamma(\beta)\Gamma(D -\alpha-\beta+n)} \cdot
 \frac{\Gamma(n+D-2)\Gamma(D/2-1)}{2^{n} \Gamma(n+D/2-1)\Gamma(D-2)}
\cdot \frac{1}{(y-z)^{2(\alpha+\beta-D/2-n)}}
 \end{array}
\ee

\begin{itemize}
\item
\end{itemize}
Now we turn to the derivation of the formula for the convolution of two
conformal propagators.
The following integral can be calculated by the conditions $A+B=D$ and
$\vec{u}\cdot\vec{u}=0 = \vec{v}\cdot\vec{v}=0$ in the standard
way using the Feynman-Schwinger $\alpha$-representation and then
gaussian integration
\begin{align*}
\int \mathrm{d}^D\, \vec{z}\,
\mathrm{d}^D\, \vec{x}\,
\frac{\left(\vec{u}-2\frac{(uz)\vec{z}}{z^2}
\left.\right|
\vec{v}-2\frac{(v|z-x)(\vec{z}-\vec{x})}{(z-x)^2}\right)^n\,e^{i\vec{p}
\vec{x}}}
{z^{2A}(z-x)^{2B}} = \\
(uv)^n\,
\frac{\pi^{D}\Gamma(\frac{D}{2}+n-1)\Gamma(\frac{D}{2}-B)\Gamma(\frac{D}{2}-A)}
{\Gamma(A+n)\Gamma(B+n)\Gamma(A-1)\Gamma(B-1)}\,
\frac{\Gamma(A+n-1)\Gamma(B+n-1)}
{\Gamma(\frac{D}{2}+n-1)}
\end{align*}
We rewrite this formula in an evident way without
Fourier transformation and in a tensor form
\begin{multline*}
\int \mathrm{d}^D\, z\,
\frac{S^{\mu_1 ... \mu_n}_{\alpha_1 ... \alpha_n}(x,z)
S^{\alpha_1 ... \alpha_n}_{\nu_1 ... \nu_n}(z,y)}
{(x-z)^{2A}(z-y)^{2B}} = \\
\frac{\pi^{D}\Gamma(\frac{D}{2}-B)\Gamma(\frac{D}{2}-A)}
{(A+n-1)(B+n-1)\Gamma(A-1)\Gamma(B-1)}\,
P^{\mu_1 ... \mu_n}_{\nu_1 ... \nu_n}\,\delta^D(x-y)
\end{multline*}
and in the special case $A=\frac{D}{2}+2i\nu$ and
 $B=\frac{D}{2}-2i\nu$ we obtain
\begin{multline}\label{convolution1}
\int \mathrm{d}^D\, z\,
\frac{S^{\mu_1 ... \mu_n}_{\alpha_1 ... \alpha_n}(x,z)
S^{\alpha_1 ... \alpha_n}_{\nu_1 ... \nu_n}(z,y)}
{(x-z)^{2\left(\frac{D}{2}+2i\nu\right)}
(z-y)^{2\left(\frac{D}{2}-2i\nu\right)}} = \\
\frac{\pi^{D}\Gamma(2i\nu)\Gamma(-2i\nu)}
{\left(\left(\frac{D}{2}+n-1\right)^2+4\nu^2\right)
\Gamma\left(\frac{D}{2}+2i\nu-1\right)\Gamma\left(\frac{D}{2}-2i\nu-1\right)}
\,P^{\mu_1 ... \mu_n}_{\nu_1 ... \nu_n}\,\delta^D(x-y)
\end{multline}

\begin{itemize}
\item
\end{itemize}

The last needed formula is the formula for the
convolution of a conformal triangle and a conformal propagator.
The following integral can be calculated by conditions $A+B+C=D-n$ and
$\vec{u}\cdot\vec{u}=0$ using \eqref{magic02} and then
the Feynman-Schwinger $\alpha$-representation
and gaussian integration
\begin{multline*}
\int \mathrm{d}^D\, \vec{z}\,
\frac{\left(\vec{u}-2\frac{(uz)}{z^2}
\vec{z}\left.\right|\frac{\vec{x}-\vec{z}}{(x-z)^2}-
\frac{\vec{y}-\vec{z}}{(y-z)^2}\right)^n}
{z^{2A}(x-z)^{2B}(y-z)^{2C}} =
\frac{\left(\frac{(uy)}{y^2}-\frac{(ux)}{x^2}\right)^n}
{(y-x)^{2(B+C+n-\frac{D}{2})}x^{2(\frac{D}{2}-C-n)}y^{2(\frac{D}{2}-B-n)}}
\\
\frac{\pi^{\frac{D}{2}}\,\Gamma(B+C-\frac{D}{2}+n)}
{\Gamma(B+n)\Gamma(C+n)}
\frac{\Gamma(B+C+2n-1)}{\Gamma(B+C+n-1)}
\frac{\Gamma(\frac{D}{2}-B)\Gamma(\frac{D}{2}-C)}{\Gamma(D-B-C)}
\end{multline*}
In a tensor form we have
\begin{multline*}
\int \mathrm{d}^D\, z\,
\frac{S^{\mu_1\cdots\mu_n}_{\nu_1\cdots\nu_n}(x,z)
\left(\frac{z-x_1}{(z-x_1)^2}-
\frac{z-x_2}{(z-x_2)^2}\right)^{\nu_1\cdots\nu_n}}
{(x-z)^{2A}(z-x_1)^{2B}(z-x_2)^{2C}} =
\frac{\left(\frac{x-x_1}{(x-x_1)^2}-
\frac{x-x_2}{(x-x_2)^2}\right)^{\mu_1\cdots\mu_n}}
{(x_1-x_2)^{2(B+C+n-\frac{D}{2})}(x_1-x)^{2(\frac{D}{2}-C-n)}(x_2-x)^{2(\frac{D}{2}-B-n)}}
\\
(-1)^n\frac{\pi^{\frac{D}{2}}\,\Gamma(B+C-\frac{D}{2}+n)}
{\Gamma(B+n)\Gamma(C+n)}
\frac{\Gamma(B+C+2n-1)}{\Gamma(B+C+n-1)}
\frac{\Gamma(\frac{D}{2}-B)\Gamma(\frac{D}{2}-C)}{\Gamma(D-B-C)}
\end{multline*}
Using parametrization $\Delta = \frac{D}{2}+2i\nu$
\begin{align*}
B = \frac{D}{4}+\frac{\Delta_1-\Delta_2}{2}-\frac{n}{2}+i\nu \ \ ;\ \
C = \frac{D}{4}-\frac{\Delta_1-\Delta_2}{2}-\frac{n}{2}+i\nu \ \ ;\ \
A = D-B-C-n = \frac{D}{2}-2i\nu \, ,
\end{align*}
and after multiplying on
$\frac{1}{(x_1-x_2)^{\Delta_1+\Delta_2-\frac{D}{2}-2i\nu +n}}$,
we obtain
\begin{multline*}
\int \mathrm{d}^D\, z\,
\frac{S^{\mu_1\cdots\mu_n}_{\nu_1\cdots\nu_n}(x,z)
\left(\frac{z-x_1}{(z-x_1)^2}-
\frac{z-x_2}{(z-x_2)^2}\right)^{\nu_1\cdots\nu_n}}
{(x-z)^{2\left(\frac{D}{2}-2i\nu\right)}
(z-x_1)^{\Delta_1-\Delta_2+\frac{D}{2}+2i\nu - n}
(z-x_2)^{\Delta_2-\Delta_1+\frac{D}{2}+2i\nu - n}
(x_1-x_2)^{\Delta_1+\Delta_2-\frac{D}{2}-2i\nu +n}} = \\
\frac{\left(\frac{x-x_1}{(x-x_1)^2}-
\frac{x-x_2}{(x-x_2)^2}\right)^{\mu_1\cdots\mu_n}}
{(x_1-x_2)^{\Delta_1+\Delta_2-\frac{D}{2}+2i\nu+n}
(x_1-x)^{\Delta_1-\Delta_2+\frac{D}{2}-2i\nu - n}
(x_2-x)^{\Delta_2-\Delta_1+\frac{D}{2}-2i\nu - n}}
\\
(-1)^n\pi^{\frac{D}{2}}\,C_{\Delta_1\Delta_2}(n\,,\nu)\,
\frac{\Gamma(2i\nu)\Gamma\left(\frac{D}{2}+2i\nu+n-1\right)}
{\Gamma\left(\frac{D}{2}+2i\nu-1\right)
\Gamma\left(\frac{D}{2}-2i\nu+n\right)}
\end{multline*}
or in compact notation
\begin{multline}\label{amp1}
\int \mathrm{d}^D\, z\,
\frac{S^{\mu_1\cdots\mu_n}_{\nu_1\cdots\nu_n}(x,z)}
{(x-z)^{2\left(\frac{D}{2}-2i\nu\right)}}
\Psi^{\nu_1\cdots\nu_n}_{n,\nu,z}(x_1\,,x_2) = \\
\Psi^{\mu_1\cdots\mu_n}_{n,-\nu,x}(x_1\,,x_2)\,
(-1)^n\pi^{\frac{D}{2}}\,
C_{\Delta_1\Delta_2}(n\,,\nu)\,
\frac{\Gamma(2i\nu)\Gamma\left(\frac{D}{2}+2i\nu+n-1\right)}
{\Gamma\left(\frac{D}{2}+2i\nu-1\right)\Gamma\left(\frac{D}{2}-2i\nu+n\right)}
\end{multline}
where
\begin{align*}
C_{\Delta_1\Delta_2}(n\,,\nu) =
\frac{\Gamma\left(\frac{D}{4}-\frac{\Delta_1-\Delta_2}{2}+
\frac{n}{2}-i\nu\right)}
{\Gamma\left(\frac{D}{4}-\frac{\Delta_1-\Delta_2}{2}+
\frac{n}{2}+i\nu\right)}\,
\frac{\Gamma\left(\frac{D}{4}+\frac{\Delta_1-\Delta_2}{2}+
\frac{n}{2}-i\nu\right)}
{\Gamma\left(\frac{D}{4}+\frac{\Delta_1-\Delta_2}{2}+
\frac{n}{2}+i\nu\right)}\,
\end{align*}

\section{Appendix D}
\setcounter{equation}0

In this Appendix we are going to derive
a formula for the scalar product of two conformal triangles and an
 universal relation between the coefficients $C_1$ and $C_2$.

\subsection{Orthogonality relation}

For simplicity we shall use the following notation:
\begin{align}
&A_1 = \frac{D}{4}+\frac{\Delta_1-\Delta_2}{2} - \frac{n}{2} + i\nu
\ \ \ ; \ \ \
A_2 = \frac{D}{4}-\frac{\Delta_1-\Delta_2}{2} - \frac{m}{2} - i\lambda
\\
&B_1 = \frac{D}{4}-\frac{\Delta_1-\Delta_2}{2} - \frac{n}{2} + i\nu
\ \ \ ; \ \ \
B_2 = \frac{D}{4}+\frac{\Delta_1-\Delta_2}{2} - \frac{m}{2} - i\lambda
\\
&A = \frac{D}{2} + \frac{n+m}{2} - i(\nu-\lambda)
\ \ \ ;\ \ \
A_1+A_2 = B_1+B_2 = \frac{D}{2} - \frac{n+m}{2} + i(\nu-\lambda)
\end{align}
and introduce appropriate regularization
\begin{multline}
\langle \overline{\Psi^{(m,v)}_\lambda(y)} | \Psi^{(n,
u)}_{\nu}(x)\rangle   = \lim_{\varepsilon \to 0}\,
\langle \overline{\Psi^{(m,v)}_\lambda(y)} | \Psi^{(n,
u)}_{\nu}(x)\rangle_{\varepsilon} =
\\
\lim_{\varepsilon \to 0}\,\int \, \frac{d^D x_1 \, d^D x_2
\left(\frac{(v\,,y-x_2)}{(y-x_2)^2}-\frac{(v\,,y-x_1)}{(y-x_1)^2}\right)^{m}\,
\left(\frac{(u\,,x-x_1)}{(x-x_1)^2}-\frac{(u\,,x-x_2)}{(x-x_2)^2}\right)^{n}}
{(y-x_2)^{2(B_2-\varepsilon)}
(y-x_1)^{2(A_2+\varepsilon)}
(x_1-x_2)^{2(A-2\varepsilon)}
(x-x_1)^{2(A_1+\varepsilon)}
(x-x_2)^{2(B_1-\varepsilon)}}
\end{multline}
where $\varepsilon > 0$ is the real parameter of regularization.
At the first step we use identity \eqref{magic}
(see Appendix {\bf C} for details) twice to mimic
 the tensor structure
\begin{multline*}
\frac{\Gamma(A_1+\varepsilon)}{2^n\Gamma(A_1+\varepsilon+n)}\,
\frac{(-1)^m\Gamma(A_2+\varepsilon)}{2^m\Gamma(A_2+\varepsilon+m)}\,
\partial_t^n\,\partial_s^m \\
\int \, \frac{d^D x_2}
{(y-x_2)^{2(A_2+B_2+m-1)}
(y-x_2-s v)^{2(1-A_2-\varepsilon-m)}
(x-x_2)^{2(A_1+B_1+n-1)}
(x-x_2-t u)^{2(1-A_1-\varepsilon-n)}} \\
\int \, \frac{d^D x_1}
{(y-x_1-s v)^{2(A_2+\varepsilon)}
(x-x_1-t u)^{2(A_1+\varepsilon)}
(x_1-x_2)^{2(A-2\varepsilon)}}\bigg|_{s = t = 0}
\end{multline*}
Note that $A_1+A_2+A = D$ so that integral over $x_1$ can
be calculated using the star-triangle relation \eqref{startr2}
\begin{multline*}
\frac{\Gamma(A_1+\varepsilon)}{2^n\Gamma(A_1+\varepsilon+n)}\,
\frac{(-1)^m\Gamma(A_2+\varepsilon)}{2^m\Gamma(A_2+\varepsilon+m)}\,
\pi^{\frac{D}{2}}
\frac{\Gamma\left(\frac{D}{2}-A_1-\varepsilon\right)
\Gamma\left(\frac{D}{2}-A_2-\varepsilon\right)\Gamma\left(\frac{D}{2}-A+2\varepsilon\right)}
{\Gamma(A_1+\varepsilon)\Gamma(A_2+\varepsilon)\Gamma(A-2\varepsilon)}\\
\partial_t^n\,\partial_s^m  \frac{1}
{(y-x-s v +t u)^{2(\frac{D}{2}-A+2\varepsilon)}} \\
\int \, \frac{d^D x_2}
{(y-x_2)^{2(A_2+B_2+m-1)}
(y-x_2-s v)^{2(C-m)}
(x-x_2)^{2(A_1+B_1+n-1)}
(x-x_2-t u)^{2(C-n)}}\bigg|_{s = t = 0}
\end{multline*}
where $C = \frac{D}{2}+1-A_1-A_2-2\varepsilon$.

Next we use series expansion in the form
\begin{align}
\frac{1}{(y-x-t u)^{2\alpha}} =
\sum_{k=0}^{\infty}
\frac{t^k 2^k \Gamma(\alpha+k)}{k!\Gamma(\alpha)}
\frac{(u\,,y-x)^k}{(y-x)^{2(\alpha+k)}}
\end{align}
twice and reduce the $x_2$-integral to calculable form
\begin{multline*}
\frac{\Gamma(A_1+\varepsilon)}{2^n\Gamma(A_1+\varepsilon+n)}\,
\frac{(-1)^m\Gamma(A_2+\varepsilon)}{2^m\Gamma(A_2+\varepsilon+m)}\,
\pi^{\frac{D}{2}}
\frac{\Gamma\left(\frac{D}{2}-A_1-\varepsilon\right)
\Gamma\left(\frac{D}{2}-A_2-\varepsilon\right)\Gamma\left(\frac{D}{2}-A+2\varepsilon\right)}
{\Gamma(A_1+\varepsilon)\Gamma(A_2+\varepsilon)\Gamma(A-2\varepsilon)}\\
\partial_t^n\,\partial_s^m  \frac{1}
{(y-x-s v +t u)^{2(\frac{D}{2}-A+2\varepsilon)}}
\sum_{k,p=0}^{\infty}
\frac{t^k s^p 2^k 2^p
\Gamma(C-n+k)\Gamma(C-m+p)}{k!p!\Gamma(C-n)\Gamma(C-m)}\\
\int \, \frac{d^D x_2\, (v\,,y-x_2)^p (u\,,x-x_2)^k}
{(y-x_2)^{2(\frac{D}{2}+B_2-A_1-2\varepsilon+p)}
(x-x_2)^{2(\frac{D}{2}+B_1-A_2-2\varepsilon+k)}}\bigg|_{s = t = 0}
\end{multline*}
so that
\begin{multline*}
\int \, \frac{d^D x_2\, (v\,,y-x_2)^p (u\,,x-x_2)^k}
{(y-x_2)^{2(\frac{D}{2}+B_2-A_1-2\varepsilon+p)}
(x-x_2)^{2(\frac{D}{2}+B_1-A_2-2\varepsilon+k)}} = \\
\frac{\Gamma(\frac{D}{2}+B_2-A_1-2\varepsilon)
\Gamma(\frac{D}{2}+B_1-A_2-2\varepsilon)}
{2^p2^k\Gamma(\frac{D}{2}+B_2-A_1-2\varepsilon+p)
\Gamma(\frac{D}{2}+B_1-A_2-2\varepsilon+k)}
\partial^p_{\alpha}\partial^k_{\beta} \\
\int \, \frac{d^D x_2}
{(y-x_2-\alpha v)^{2(\frac{D}{2}+B_2-A_1-2\varepsilon)}
(x-x_2-\beta u)^{2(\frac{D}{2}+B_1-A_2-2\varepsilon)}}\bigg|_{\alpha =
\beta = 0} = \\
\frac{\pi^{\frac{D}{2}}\,\Gamma(A_1-B_2+2\varepsilon)
\Gamma(A_2-B_1+2\varepsilon)\Gamma(\frac{D}{2}-4\varepsilon)}
{2^p2^k\Gamma(\frac{D}{2}+B_2-A_1-2\varepsilon+p)
\Gamma(\frac{D}{2}+B_1-A_2-2\varepsilon+k)\Gamma(4\varepsilon)}
\partial^p_{\alpha}\partial^k_{\beta}
\frac{1}{(y-x+\beta u-\alpha
v)^{2(\frac{D}{2}-4\varepsilon)}}\bigg|_{\alpha = \beta = 0}
\end{multline*}
and finally after returning to the initial variables we obtain
\begin{multline*}
\langle \overline{\Psi^{(m,v)}_\lambda(y)} | \Psi^{(n,
u)}_{\nu}(x)\rangle_{\varepsilon} = \\
\pi^{D}\,\Gamma\left(\frac{D}{2}-4\varepsilon\right)\,
\frac{\Gamma\left(\frac{D}{4}-\frac{\Delta_1-\Delta_2}{2}+
\frac{n}{2}-i\nu -\varepsilon\right)}
{2^n \Gamma\left(\frac{D}{4}+\frac{\Delta_1-\Delta_2}{2}+
\frac{n}{2}+i\nu +\varepsilon\right)}\,
\frac{(-1)^m \Gamma\left(\frac{D}{4}+\frac{\Delta_1-\Delta_2}{2}+
\frac{m}{2}+i\lambda -\varepsilon\right)}
{2^m \Gamma\left(\frac{D}{4}-\frac{\Delta_1-\Delta_2}{2}+
\frac{m}{2}-i\lambda +\varepsilon\right)}\,\\
\frac{\Gamma\left(\frac{n-m}{2}-i(\nu+\lambda)+2\varepsilon\right)
\Gamma\left(\frac{m-n}{2}+i(\nu+\lambda)+2\varepsilon\right)
\Gamma\left(-\frac{n+m}{2}+i(\nu-\lambda)+2\varepsilon\right)}
{\Gamma\left(1+\frac{m-n}{2}+i(\lambda-\nu)-2\varepsilon\right)
\Gamma\left(1+\frac{n-m}{2}+i(\lambda-\nu)-2\varepsilon\right)
\Gamma\left(\frac{D}{2}+\frac{n+m}{2}-i(\nu-\lambda)-2\varepsilon\right)\Gamma(4\varepsilon)}\\
\partial_t^n\,\partial_s^m
\sum_{k,p=0}^{\infty}
\frac{t^k s^p
\Gamma\left(1+\frac{m-n}{2}+i(\lambda-\nu)-2\varepsilon+k\right)
\Gamma\left(1+\frac{n-m}{2}+i(\lambda-\nu)-2\varepsilon+p\right)}
{k!p!\Gamma\left(\frac{D}{2}+\frac{m-n}{2}+i(\lambda+\nu)-2\varepsilon+k\right)
\Gamma\left(\frac{D}{2}+\frac{n-m}{2}-i(\lambda+\nu)-2\varepsilon+p\right)}\\
\frac{1}
{(y-x-s v +t u)^{2(i(\nu-\lambda)-\frac{n+m}{2}+2\varepsilon)}}\,
\partial^p_{\alpha}\partial^k_{\beta}
\frac{1}{(y-x+\beta u-\alpha v)^{2(\frac{D}{2}-4\varepsilon)}}\bigg|_{s
= t = \alpha = \beta = 0}
\end{multline*}
The presence of the $\delta_{nm}$ in the scalar product can be proved
in a standard way by analogy with the proof of
the orthogonality of the eigenvectors of a self-adjoint
operator with different eigenvalues so that for simplicity we put $n =
m$
\begin{multline*}
\langle \overline{\Psi^{(m,v)}_\lambda(y)} | \Psi^{(n,
u)}_{\nu}(x)\rangle_{\varepsilon}  = \\
\pi^{D}\,\Gamma\left(\frac{D}{2}-4\varepsilon\right)\,
\frac{\Gamma\left(\frac{D}{4}-\frac{\Delta_1-\Delta_2}{2}+
\frac{n}{2}-i\nu -\varepsilon\right)}
{2^n \Gamma\left(\frac{D}{4}+\frac{\Delta_1-\Delta_2}{2}+
\frac{n}{2}+i\nu +\varepsilon\right)}\,
\frac{(-1)^n \Gamma\left(\frac{D}{4}+\frac{\Delta_1-\Delta_2}{2}+
\frac{n}{2}+i\lambda -\varepsilon\right)}
{2^n \Gamma\left(\frac{D}{4}-\frac{\Delta_1-\Delta_2}{2}+
\frac{n}{2}-i\lambda +\varepsilon\right)}\,\\
\frac{\Gamma\left(-i(\nu+\lambda)+2\varepsilon\right)
\Gamma\left(i(\nu+\lambda)+2\varepsilon\right)
\Gamma\left(-n+i(\nu-\lambda)+2\varepsilon\right)}
{\Gamma\left(1+i(\lambda-\nu)-2\varepsilon\right)
\Gamma\left(1+i(\lambda-\nu)-2\varepsilon\right)
\Gamma\left(\frac{D}{2}+n-i(\nu-\lambda)-2\varepsilon\right)\Gamma(4\varepsilon)}\\
\partial_t^n\,\partial_s^n
\sum_{k,p=0}^{\infty}
\frac{t^k s^p \Gamma\left(1+i(\lambda-\nu)-2\varepsilon+k\right)
\Gamma\left(1+i(\lambda-\nu)-2\varepsilon+p\right)}
{k!p!\Gamma\left(\frac{D}{2}+i(\lambda+\nu)-2\varepsilon+k\right)
\Gamma\left(\frac{D}{2}-i(\lambda+\nu)-2\varepsilon+p\right)}\\
\frac{1}
{(y-x-s v +t u)^{2(i(\nu-\lambda)-n+2\varepsilon)}}\,
\partial^p_{\alpha}\partial^k_{\beta}
\frac{1}{(y-x+\beta u-\alpha v)^{2(\frac{D}{2}-4\varepsilon)}}\bigg|_{s
= t = \alpha = \beta = 0}
\end{multline*}
Note that due to the factor $\Gamma(4\varepsilon)$ in the denominator
this expression is non-zero in the limit $\varepsilon \to 0$ only
in two cases $\nu = +\lambda$ and
$\nu = -\lambda$ so that in the sense of distribution we obtain
a distribution with support at the points $\nu = +\lambda$ and
$\nu = -\lambda$. Let us start from the first point $\nu = -\lambda$
and use the formula
\begin{align*}
\lim_{\varepsilon \to 0}
\frac{\Gamma\left(-i(\nu+\lambda)+2\varepsilon\right)
\Gamma\left(i(\nu+\lambda)+2\varepsilon\right)}{\Gamma(4\epsilon)} = 2
\pi \delta(\nu+\lambda)
\end{align*}
so that
\begin{multline*}
\langle \overline{\Psi^{(m,v)}_\lambda(y)} | \Psi^{(n,
u)}_{\nu}(x)\rangle_{\varepsilon}  \to \\
2 \pi \delta(\nu+\lambda)
\pi^{D}\,\Gamma\left(\frac{D}{2}\right)\,
\frac{\Gamma\left(\frac{D}{4}-\frac{\Delta_1-\Delta_2}{2}+
\frac{n}{2}-i\nu\right)}
{2^n \Gamma\left(\frac{D}{4}+\frac{\Delta_1-\Delta_2}{2}+
\frac{n}{2}+i\nu\right)}\,
\frac{(-1)^n \Gamma\left(\frac{D}{4}+\frac{\Delta_1-\Delta_2}{2}+
\frac{n}{2}-i\nu\right)}
{2^n \Gamma\left(\frac{D}{4}-\frac{\Delta_1-\Delta_2}{2}+
\frac{n}{2}+i\nu\right)}\,\\
\frac{\Gamma\left(-n+2i\nu\right)}
{\Gamma\left(1-2i\nu\right)
\Gamma\left(1-2i\nu\right)
\Gamma\left(\frac{D}{2}+n-2i\nu\right)}\,
\partial_t^n\,\partial_s^n
\sum_{k,p=0}^{\infty}
\frac{t^k s^p \Gamma\left(1-2i\nu+k\right)
\Gamma\left(1-2i\nu+p\right)}
{k!p!\Gamma\left(\frac{D}{2}+k\right)
\Gamma\left(\frac{D}{2}+p\right)}\\
\frac{1}
{(y-x-s v +t u)^{2(2i\nu-n)}}\,
\partial^p_{\alpha}\partial^k_{\beta}
\frac{1}{(y-x+\beta u-\alpha v)^{D}}\bigg|_{s = t = \alpha = \beta =
0}
\end{multline*}
Now it is possible to calculate the sum over $p$
\begin{multline*}
\sum_{p=0}^{\infty}
\frac{s^p \Gamma\left(1-2i\nu+p\right)}
{p!\Gamma\left(1-2i\nu\right)}
\frac{\Gamma\left(\frac{D}{2}\right)}
{\Gamma\left(\frac{D}{2}+p\right)}
\partial^p_{\alpha}
\frac{1}{(y-x+\beta u-\alpha v)^{D}}\bigg|_{\alpha = 0} = \\
=
\frac{1}{(y-x+\beta u)^{2(\frac{D}{2}+2i\nu-1 )}}
\frac{1}
{(y-x+\beta u -s v)^{2(1-2i\nu)}}
\end{multline*}
then derivative with respect to $s$
\begin{multline*}
\partial_s^n \frac{1}
{(y-x+t u - s v)^{2(2i\nu-n)}(y-x+\beta u -s v)^{2(1-2i\nu)}}\bigg|_{s
= 0} = \\
2^n\,\frac{\Gamma(1-2i\nu+n)}{\Gamma(1-2i\nu)}
\frac{(t-\beta)^n\bigl((v\,,u)(y-x)^2 - 2(v\,,y-x)(u\,,y-x)\bigr)^n}
{(y-x+t u)^{2(2i\nu)}(y-x+\beta u)^{2(1-2i\nu+n)}}
\end{multline*}
and after all these steps one obtains the
following intermediate result for the
contribution at the first point $\nu = -\lambda$
\begin{multline*}
\langle \overline{\Psi^{(m,v)}_\lambda(y)} | \Psi^{(n,
u)}_{\nu}(x)\rangle_{\varepsilon}  \to \\
2 \pi\, \delta(\nu+\lambda)\,
\pi^{D}\,\frac{(-1)^n}{2^n}\,
\frac{\Gamma\left(\frac{D}{4}-\frac{\Delta_1-\Delta_2}{2}+
\frac{n}{2}-i\nu\right)}
{\Gamma\left(\frac{D}{4}+\frac{\Delta_1-\Delta_2}{2}+
\frac{n}{2}+i\nu\right)}\,
\frac{\Gamma\left(\frac{D}{4}+\frac{\Delta_1-\Delta_2}{2}+
\frac{n}{2}-i\nu\right)}
{\Gamma\left(\frac{D}{4}-\frac{\Delta_1-\Delta_2}{2}+
\frac{n}{2}+i\nu\right)}\,\\
\left((v\,,u)(y-x)^2 - 2(v\,,y-x)(u\,,y-x)\right)^n\,
\frac{\Gamma\left(-n+2i\nu\right)\Gamma(1-2i\nu+n)}
{\Gamma\left(1-2i\nu\right)
\Gamma\left(1-2i\nu\right)
\Gamma\left(\frac{D}{2}+n-2i\nu\right)}\\
\partial_t^n\,
\sum_{k=0}^{\infty}
\frac{t^k \Gamma\left(1-2i\nu+k\right)}
{k!\Gamma\left(\frac{D}{2}+k\right)}
\partial^k_{\beta}
\frac{(t-\beta)^n}
{(y-x+t u)^{2(2i\nu)}(y-x+\beta
u)^{2(1-2i\nu+n+\frac{D}{2}+2i\nu-1)}}\bigg|_{t = \beta = 0}
\end{multline*}
The differentiation is at the point $\beta=t=0$ so that
it is possible to change the variable $\beta \to t\beta $ and
calculate everything in a closed form using the Gauss summation
formula
\begin{align*}
\sum_k\binom{n}{k}(-1)^{k}
\frac{\Gamma(A+k)}{\Gamma(B+k)} =
\frac{\Gamma(A)\Gamma(B-A+n)}{\Gamma(B-A)\Gamma(B+n)}\,.
\end{align*}
Indeed we have
\begin{multline*}
\left.\partial_t^n\, t^n
\sum_{k=0}^{\infty}
\frac{\Gamma\left(1-2i\nu+k\right)}
{k!\Gamma\left(\frac{D}{2}+k\right)}
\partial^k_{\beta}
\frac{(1-\beta)^n}
{(y-x+t u)^{2(2i\nu)}(y-x+t \beta
u)^{2(\frac{D}{2}+n)}}\right|_{t=\beta=0} = \\
n!\,\sum_{k=0}^{\infty}
\frac{\Gamma\left(1-2i\nu+k\right)}
{k!\Gamma\left(\frac{D}{2}+k\right)}
\frac{\left.\partial^k_{\beta}(1-\beta)^n\right|_{\beta=0}}
{(y-x)^{2(2i\nu+\frac{D}{2}+n)}} =
\frac{\Gamma(1-2i\nu)\Gamma\left(\frac{D}{2}+2i\nu-1+n\right)}
{\Gamma\left(\frac{D}{2}+2i\nu-1\right)\Gamma\left(\frac{D}{2}+n\right)}\,
\frac{n!}{(y-x)^{2(\frac{D}{2}+2i\nu+n)}}
\end{multline*}
so that
\begin{multline*}
\langle \overline{\Psi^{(m,v)}_\lambda(y)} | \Psi^{(n,
u)}_{\nu}(x)\rangle_{\varepsilon} \to \\
2 \pi \delta(\nu+\lambda)
\pi^{D}\,\frac{(-1)^n n!}{2^n}\,
\frac{\Gamma\left(\frac{D}{4}-\frac{\Delta_1-\Delta_2}{2}+
\frac{n}{2}-i\nu\right)}
{\Gamma\left(\frac{D}{4}+\frac{\Delta_1-\Delta_2}{2}+
\frac{n}{2}+i\nu\right)}\,
\frac{\Gamma\left(\frac{D}{4}+\frac{\Delta_1-\Delta_2}{2}+
\frac{n}{2}-i\nu\right)}
{\Gamma\left(\frac{D}{4}-\frac{\Delta_1-\Delta_2}{2}+
\frac{n}{2}+i\nu\right)}\,\\
\frac{\Gamma\left(-n+2i\nu\right)\Gamma(1-2i\nu+n)}
{\Gamma\left(1-2i\nu\right)
\Gamma\left(\frac{D}{2}+n-2i\nu\right)}
\frac{\Gamma\left(\frac{D}{2}+2i\nu-1+n\right)}
{\Gamma\left(\frac{D}{2}+2i\nu-1\right)\Gamma\left(\frac{D}{2}+n\right)}\,
\frac{\left((v\,,u)- 2\frac{(v\,,y-x)(u\,,y-x)}{(y-x)^2}\right)^n}
{(y-x)^{2(\frac{D}{2}+2i\nu)}}\,.
\end{multline*}
This expression can be rewritten in an equivalent form \eqref{C2}
\begin{multline*}
\langle \overline{\Psi^{(m,v)}_\lambda(y)} | \Psi^{(n,
u)}_{\nu}(x)\rangle_{\varepsilon} \to 2 \pi \delta(\nu+\lambda)
\pi^{D}\,\frac{n!}{2^n}\,
\frac{\Gamma\left(\frac{D}{4}-\frac{\Delta_1-\Delta_2}{2}+
\frac{n}{2}-i\nu\right)}
{\Gamma\left(\frac{D}{4}+\frac{\Delta_1-\Delta_2}{2}+
\frac{n}{2}+i\nu\right)}\,
\frac{\Gamma\left(\frac{D}{4}+\frac{\Delta_1-\Delta_2}{2}+
\frac{n}{2}-i\nu\right)}
{\Gamma\left(\frac{D}{4}-\frac{\Delta_1-\Delta_2}{2}+
\frac{n}{2}+i\nu\right)}\,\\
\frac{\Gamma\left(2i\nu\right)\Gamma\left(\frac{D}{2}+2i\nu-1+n\right)}
{\Gamma\left(\frac{D}{2}+n-2i\nu\right)
\Gamma\left(\frac{D}{2}+2i\nu-1\right)\Gamma\left(\frac{D}{2}+n\right)}\,
\frac{\left((v\,,u)-
2\frac{(v\,,y-x)(u\,,y-x)}{(y-x)^2}\right)^n}{(y-x)^{2(\frac{D}{2}+2i\nu)}}
\end{multline*}
using the reflection relation for the $\Gamma$-function
\begin{equation*}
    \Gamma(\alpha-n)\,\Gamma(1-\alpha+n) =
    (-1)^n\,\Gamma(\alpha)\,\Gamma(1-\alpha)\,.
\end{equation*}

Now we turn to the contribution at the second point $\nu = \lambda$
and from the very beginning we simplify everything as much as possible
by
extracting the contribution which leads to the delta-function
$\delta(\nu-\lambda)$
\begin{multline*}
\langle \overline{\Psi^{(m,v)}_\lambda(y)} | \Psi^{(n,
u)}_{\nu}(x)\rangle_{\varepsilon} \to
\pi^{D}\,\Gamma\left(\frac{D}{2}-4\varepsilon\right)\,\frac{(-1)^n}{4^n}
\frac{\Gamma\left(-2i\nu\right)\Gamma\left(2i\nu\right)
\Gamma\left(-n+i(\nu-\lambda)+2\varepsilon\right)}
{\Gamma\left(\frac{D}{2}+n-i(\nu-\lambda)-2\varepsilon\right)\Gamma(4\varepsilon)}\,
\\
\partial_t^n\,\partial_s^n
\sum_{k,p=0}^{\infty}
\frac{t^k s^p}
{\Gamma\left(\frac{D}{2}+2i\nu+k\right)
\Gamma\left(\frac{D}{2}-2i\nu+p\right)}\\
\frac{1}
{(y-x-s v +t u)^{2(i(\nu-\lambda)-n+2\varepsilon)}}\,
\partial^p_{\alpha}\partial^k_{\beta}
\frac{1}{(y-x+\beta u-\alpha v)^{2(\frac{D}{2}-4\varepsilon)}}\bigg|_{t
= s = \alpha = \beta = 0}
\end{multline*}
and using the standard formula for the Fourier transformation
\begin{multline*}
\langle \overline{\Psi^{(m,v)}_\lambda(y)} | \Psi^{(n,
u)}_{\nu}(x)\rangle_{\varepsilon} \to
\frac{(-1)^n}{4^n}\,
2^{2\left(-i(\nu-\lambda)+n+2\varepsilon-\frac{D}{2}\right)}
\Gamma\left(-2i\nu\right)\Gamma\left(2i\nu\right) \\
\partial_t^n\,\partial_s^n
\sum_{k,p=0}^{\infty}
\frac{t^k s^p\,\partial^p_{\alpha}\partial^k_{\beta}}
{\Gamma\left(\frac{D}{2}+2i\nu+k\right)
\Gamma\left(\frac{D}{2}-2i\nu+p\right)}\,
\int d^D p \frac{e^{i(p\,,y-x-s v +t u)}}
{p^{2\left(\frac{D}{2}-i(\nu-\lambda)+n-2\varepsilon\right)}}\,
\int d^D k \frac{e^{i(k\,,y-x-\alpha v + \beta u)}}
{k^{2\left(4\varepsilon\right)}}
\end{multline*}
All derivatives have to be calculated at the point
$s = t = \alpha = \beta = 0$ and for simplicity of notation we
do not show this explicitly. The
next step is the shift $k \to k-p $ and scaling $\alpha \to s\alpha$
and $\beta \to t\beta$
\begin{multline*}
\langle \overline{\Psi^{(m,v)}_\lambda(y)} | \Psi^{(n,
u)}_{\nu}(x)\rangle_{\varepsilon} \to
\frac{(-1)^n}{4^n}\,
2^{2\left(-i(\nu-\lambda)+n+2\varepsilon-\frac{D}{2}\right)}
\Gamma\left(-2i\nu\right)\Gamma\left(2i\nu\right) \\
\partial_t^n\,\partial_s^n
\sum_{k,p=0}^{\infty}
\frac{\partial^p_{\alpha}\partial^k_{\beta}}
{\Gamma\left(\frac{D}{2}+2i\nu+k\right)
\Gamma\left(\frac{D}{2}-2i\nu+p\right)}\,
\int d^D p \frac{e^{i(p\,,s(\alpha-1) v + t(1-\beta) u)}}
{p^{2\left(\frac{D}{2}-i(\nu-\lambda)+n-2\varepsilon\right)}}\,
\int d^D k \frac{e^{i(k\,,y-x-s\alpha v + t\beta u)}}
{(k-p)^{2\left(4\varepsilon\right)}}
\end{multline*}
and now it is possible to put $\varepsilon = 0$ because there
is no singularities at $\varepsilon \to 0$
\begin{multline*}
\langle \overline{\Psi^{(m,v)}_\lambda(y)} | \Psi^{(n,
u)}_{\nu}(x)\rangle_{\varepsilon} \to
\frac{(-1)^n}{4^n}\,
2^{2\left(-i(\nu-\lambda)+n-\frac{D}{2}\right)}
\Gamma\left(-2i\nu\right)\Gamma\left(2i\nu\right) \\
\partial_t^n\,\partial_s^n
\sum_{k,p=0}^{\infty}
\frac{\partial^p_{\alpha}\partial^k_{\beta}}
{\Gamma\left(\frac{D}{2}+2i\nu+k\right)
\Gamma\left(\frac{D}{2}-2i\nu+p\right)}\,
\int d^D p \frac{e^{i(p\,,s(\alpha-1) v + t(1-\beta) u)}}
{p^{2\left(\frac{D}{2}-i(\nu-\lambda)+n\right)}}\,
\int d^D k e^{i(k\,,y-x-s\alpha v + t\beta u)}\, .
\end{multline*}
Due to the orthogonality relation \eqref{orthog}
\begin{align*}
\int d^D p \dfrac{(v\,,p)^{n}(u\,,p)^{m}}{p^{2\left(\frac{D}{2} + n
+i(\nu-\lambda)\right)}} =
\dfrac{\pi^{D/2 + 1}n!}{2^{n - 1}\Gamma\left(\frac{D}{2} + n\right)}
\delta_{nm}\delta\left(\nu - \lambda\right)\,(v\,,u)^n
\end{align*}
the nonzero contribution appears when the $s$- and $t$-derivatives
act on the first exponent only
\begin{multline*}
\sum_{k,p=0}^{\infty}
\frac{\partial^p_{\alpha}\,\partial^k_{\beta}\,(\alpha-1)^n\,(\beta-1)^n}
{\Gamma\left(\frac{D}{2}+2i\nu+k\right)
\Gamma\left(\frac{D}{2}-2i\nu+p\right)}\,
\int d^D p \frac{(p\,,v)^n\,(p\,,u)^n}
{p^{2\left(\frac{D}{2}-i(\nu-\lambda)+n\right)}}\,
\int d^D k e^{i(k\,,y-x)} = \\
\dfrac{\pi^{D/2 + 1}n!}{2^{n - 1}\Gamma\left(\frac{D}{2} + n\right)}
\delta\left(\nu - \lambda\right)\,(v\,,u)^n\, (2\pi)^D \delta^D(y-x)\,
\sum_{k,p=0}^{\infty}
\frac{\partial^p_{\alpha}\,\partial^k_{\beta}\,(\alpha-1)^n\,(\beta-1)^n}
{\Gamma\left(\frac{D}{2}+2i\nu+k\right)
\Gamma\left(\frac{D}{2}-2i\nu+p\right)}
\end{multline*}
and it remains to calculate the sums over $k$ and $p$ using
the Gauss summation formula. For the sum over $k$ one obtains
\begin{align*}
\sum_{k=0}^{n}
\frac{\left.\partial^k_{\beta}\,(\beta-1)^n\right|_{\beta=0}}
{\Gamma\left(\frac{D}{2}+2i\nu+k\right)} =
(-1)^n\,\sum_{k=0}^{n}\binom{n}{k}\,(-1)^k\,
\frac{\Gamma(k+1)}
{\Gamma\left(\frac{D}{2}+2i\nu+k\right)} = \\
(-1)^n\,\frac{\Gamma\left(\frac{D}{2}+2i\nu-1+n\right)}
{\Gamma\left(\frac{D}{2}+2i\nu-1\right)\Gamma\left(\frac{D}{2}+2i\nu+n\right)}
=
\frac{(-1)^n}
{\left(\frac{D}{2}+2i\nu+n-1\right)\,
\Gamma\left(\frac{D}{2}+2i\nu-1\right)}
\end{align*}
and the sum over $p$ is obtained by the substitution $\nu \to -\nu$.
Collecting everything together we obtain the contribution \eqref{C1} at
the
second point $\nu = \lambda$
\begin{multline*}
\langle \overline{\Psi^{(m,v)}_\lambda(y)} | \Psi^{(n,
u)}_{\nu}(x)\rangle_{\varepsilon} \to \\
\frac{(-1)^n}{4^n}\,
2^{2\left(n-\frac{D}{2}\right)}
\Gamma\left(-2i\nu\right)\Gamma\left(2i\nu\right)
\dfrac{\pi^{D/2 + 1}n!}{2^{n - 1}\Gamma\left(\frac{D}{2} + n\right)}
\delta\left(\nu - \lambda\right)\,(v\,,u)^n\, (2\pi)^D
\delta^D(y-x)\,\\
\frac{1}{\left(\left(\frac{D}{2}+n-1\right)^2 + 4\nu^2\right)\,
\Gamma\left(\frac{D}{2}+2i\nu-1\right)\,
\Gamma\left(\frac{D}{2}-2i\nu-1\right)}\,.
\end{multline*}

\subsection{Universal relation between coefficients $C_1$ and $C_2$}

Let us start with the orthogonality relation
\begin{multline}
\langle\Psi^{\nu_1\cdots\nu_m}_{m,\lambda,y}|
\Psi^{\mu_1\cdots\mu_n}_{n,\nu,x}\rangle =
C_1(n\,,\nu)\,\delta_{n m}\,\delta(\nu -\lambda)\,\delta^D(x-y)\,
P^{\mu_1 ... \mu_n}_{\nu_1 ... \nu_n} + \\
C_2(n\,,\nu)\,\delta_{n m}\,\delta(\nu +\lambda)\,
\frac{S^{\mu_1 ... \mu_n}_{\nu_1 ... \nu_n}(x,y)}
{(x-y)^{2\left(\frac{D}{2}+2i\nu\right)}}
\end{multline}
We use two formulae from the previous
Appendix {\bf C}: convolution of two conformal propagators
\begin{multline}\label{convolution1}
\int \mathrm{d}^D\, z\,
\frac{S^{\mu_1 ... \mu_n}_{\alpha_1 ... \alpha_n}(x,z)
S^{\alpha_1 ... \alpha_n}_{\nu_1 ... \nu_n}(z,y)}
{(x-z)^{2\left(\frac{D}{2}+2i\nu\right)}(z-y)^{
2\left(\frac{D}{2}-2i\nu\right)}} = \\
\frac{\pi^{D}\Gamma(2i\nu)\Gamma(-2i\nu)}
{\left(\left(\frac{D}{2}+n-1\right)^2+4\nu^2\right)
\Gamma\left(\frac{D}{2}+2i\nu-1\right)\Gamma\left(\frac{D}{2}-2i\nu-1\right)}
\,P^{\mu_1 ... \mu_n}_{\nu_1 ... \nu_n}\,\delta^D(x-y)
\end{multline}
and convolution of the conformal propagator with conformal triangle
\begin{multline}\label{amp1}
\int \mathrm{d}^D\, z\,
\frac{S^{\mu_1\cdots\mu_n}_{\nu_1\cdots\nu_n}(x,z)}
{(x-z)^{2\left(\frac{D}{2}-2i\nu\right)}}
\Psi^{\nu_1\cdots\nu_n}_{n,\nu,z}(x_1\,,x_2) = \\
\Psi^{\mu_1\cdots\mu_n}_{n,-\nu,x}(x_1\,,x_2)\,
(-1)^n\pi^{\frac{D}{2}}\,C_{\Delta_1\Delta_2}(n\,,\nu)\,
\frac{\Gamma(2i\nu)\Gamma\left(\frac{D}{2}+2i\nu+n-1\right)}
{\Gamma\left(\frac{D}{2}+2i\nu-1\right)\Gamma\left(\frac{D}{2}-2i\nu+n\right)}
\end{multline}
Now we are going to derive certain universal relation between
the coefficients
$C_1$ and $C_2$ using \eqref{convolution1} and \eqref{amp1}.
Let us start with the orthogonality relation in the form
\begin{multline}
\langle\Psi^{\nu_1\cdots\nu_m}_{m,\lambda,y}|
\Psi^{\mu_1\cdots\mu_n}_{n,\nu,z}\rangle =
C_1(n\,,\nu)\,\delta_{n m}\,\delta(\nu -\lambda)\,\delta^D(z-y)\,
P^{\mu_1 ... \mu_n}_{\nu_1 ... \nu_n} + \\
C_2(n\,,\nu)\,\delta_{n m}\,\delta(\nu +\lambda)\,
\frac{S^{\mu_1 ... \mu_n}_{\nu_1 ... \nu_n}(z,y)}
{(z-y)^{2\left(\frac{D}{2}+2i\nu\right)}}
\end{multline}
multiply it by $S^{\mu_1\cdots\mu_n}_{\nu_1\cdots\nu_n}(x,z)$,
integrate over $z$ and use the convolution formula \eqref{amp1}
\begin{multline*}
(-1)^n\pi^{\frac{D}{2}}\,C_{\Delta_1\Delta_2}(n\,,\nu)\,
\frac{\Gamma(2i\nu)\Gamma\left(\frac{D}{2}+2i\nu+n-1\right)}
{\Gamma\left(\frac{D}{2}+2i\nu-1\right)
\Gamma\left(\frac{D}{2}-2i\nu+n\right)}
\langle\Psi^{\nu_1\cdots\nu_m}_{m,\lambda,y}|
\Psi^{\mu_1\cdots\mu_n}_{n,-\nu,x}\rangle = \\
C_1(n\,,\nu)\,\delta_{n m}\,\delta(\nu -\lambda)\,
\frac{S^{\mu_1\cdots\mu_n}_{\nu_1\cdots\nu_n}(x,y)}
{(x-y)^{2\left(\frac{D}{2}-2i\nu\right)}} + \\
C_2(n\,,\nu)\,\delta_{n m}\,\delta(\nu +\lambda)\,
\int \mathrm{d}^D\, z\,
\frac{S^{\mu_1\cdots\mu_n}_{\alpha_1\cdots\alpha_n}(x,z)}
{(x-z)^{2\left(\frac{D}{2}-2i\nu\right)}}
\frac{S^{\alpha_1 ... \alpha_n}_{\nu_1 ... \nu_n}(z,y)}
{(z-y)^{2\left(\frac{D}{2}+2i\nu\right)}}
\end{multline*}
Next we use again the formula for the scalar product with $\nu \to
-\nu$
and formula \eqref{convolution1}
\begin{multline}
(-1)^n\pi^{\frac{D}{2}}\,
C_{\Delta_1\Delta_2}(n\,,\nu)\,
\frac{\Gamma(2i\nu)\Gamma\left(\frac{D}{2}+2i\nu+n-1\right)}
{\Gamma\left(\frac{D}{2}+2i\nu-1\right)\Gamma\left(\frac{D}{2}-2i\nu+n\right)}
\times \\
\left( C_1(n\,,-\nu)\,\delta_{n m}\,\delta(\nu
+\lambda)\,\delta^D(x-y)\,
P^{\mu_1 ... \mu_n}_{\nu_1 ... \nu_n} +
C_2(n\,,-\nu)\,\delta_{n m}\,\delta(\nu -\lambda)\,
\frac{S^{\mu_1 ... \mu_n}_{\nu_1 ... \nu_n}(x,y)}
{(x-y)^{2\left(\frac{D}{2}-2i\nu\right)}} \right) = \\
C_1(n\,,\nu)\,\delta_{n m}\,\delta(\nu -\lambda)\,
\frac{S^{\mu_1\cdots\mu_n}_{\nu_1\cdots\nu_n}(x,y)}
{(x-y)^{2\left(\frac{D}{2}-2i\nu\right)}} + \\
C_2(n\,,\nu)\,\delta_{n m}\,\delta(\nu +\lambda)\,
\frac{\pi^{D}\Gamma(2i\nu)\Gamma(-2i\nu)}
{\left(\left(\frac{D}{2}+n-1\right)^2+4\nu^2\right)
\Gamma\left(\frac{D}{2}+2i\nu-1\right)\Gamma\left(\frac{D}{2}-2i\nu-1\right)}
\,P^{\mu_1 ... \mu_n}_{\nu_1 ... \nu_n}\,\delta^D(x-y)
\end{multline}
The coefficients in front of different structures should coincide,
 and there are
two consistency relations
\begin{multline}\label{1}
(-1)^n\pi^{\frac{D}{2}}\,C_{\Delta_1\Delta_2}(n\,,\nu)\,
\frac{\Gamma\left(\frac{D}{2}+2i\nu+n-1\right)}
{\Gamma\left(\frac{D}{2}-2i\nu+n\right)}\,C_1(n\,,-\nu) = \\
\frac{\pi^{D}\Gamma(-2i\nu)}
{\left(\left(\frac{D}{2}+n-1\right)^2+4\nu^2\right)
\Gamma\left(\frac{D}{2}-2i\nu-1\right)}\,C_2(n\,,\nu)
\end{multline}
and
\begin{align}\label{2}
(-1)^n\pi^{\frac{D}{2}}\,C_{\Delta_1\Delta_2}(n\,,\nu)\,
\frac{\Gamma(2i\nu)\Gamma\left(\frac{D}{2}+2i\nu+n-1\right)}
{\Gamma\left(\frac{D}{2}+2i\nu-1\right)\Gamma\left(\frac{D}{2}-2i\nu+n\right)}
C_2(n\,,-\nu) = C_1(n\,,\nu)
\end{align}
It is possible to check that it is in fact only one relation and
everything works for the coefficients $C_1$ and $C_2$ from
\eqref{C1},\eqref{C2}.

\section{Appendix E\label{FF}}
\setcounter{equation}0

In the
 integral kernel representation
the proof of equation \eqref{confR}
consists in a direct check.
We represent $Q_{_{\Delta_1\Delta_2}}(u)$
as an integral operator acting on the function of two variables
\begin{align}
\left[Q_{_{\Delta_1\Delta_2}}(u)\Phi\right](x_1\,,x_2) =
{\sf C} \cdot (x_1-x_2)^{2(\frac{D}{2}-\Delta_1)}\,
\int d^D y_1
\frac{(y_1-x_2)^{2\left(u+\frac{\Delta_1+\Delta_2-D}{2}\right)}}
{(x_1-y_1)^{2\left(\frac{D}{2}
+u+\frac{\Delta_2-\Delta_1}{2}\right)}}\, \Phi(y_1\,,x_2)\,,
\end{align}
where the constant ${\sf C}$ arises after Fourier transformation and
its exact form is not important at this stage.
Next we have ($\frac{1}{x} : = x^\mu/x^2$)
\begin{align}
\left[Q_{_{\Delta_1\Delta_2}}(u)\,
{\cal I}^{(1)}_{\Delta_1}\,
{\cal I}^{(2)}_{\Delta_2}\Phi\right](x_1\,,x_2) =
 {\sf C} \cdot (x_1-x_2)^{2(\frac{D}{2}-\Delta_1)}\,
\int d^D y_1
\frac{(y_1-x_2)^{2\left(u+\frac{\Delta_1+\Delta_2-D}{2}\right)}}
{(x_1-y_1)^{2\left(\frac{D}{2}
+u+\frac{\Delta_2-\Delta_1}{2}\right)}}\,
\frac{\Phi\left(\frac{1}{y_1}\,,\frac{1}{x_2}\right)}
{y_1^{2\Delta_1}x_2^{2\Delta_2}}\,,
\lb{QII}
\end{align}
and
\begin{align}
\left[{\cal I}^{(1)}_{\frac{\Delta_1+\Delta_2}{2}+u}\,
{\cal I}^{(2)}_{\frac{\Delta_1+\Delta_2}{2}
-u}\,Q_{_{\Delta_1\Delta_2}}(u)\Phi\right](x_1\,,x_2) = \\
{\sf C} \cdot
\frac{\left(\frac{1}{x_1}-\frac{1}{x_2}\right)^{2(\frac{D}{2}-\Delta_1)}}
{x_1^{2\left(\frac{\Delta_1+\Delta_2}{2}+u\right)}
x_2^{2\left(\frac{\Delta_1+\Delta_2}{2}-u\right)}}\,
\int d^D y_1
\frac{\left(y_1-\frac{1}{x_2}\right)^{2\left(u+
\frac{\Delta_1+\Delta_2-D}{2}\right)}}
{\left(\frac{1}{x_1}-y_1\right)^{2\left(\frac{D}{2}
+u+\frac{\Delta_2-\Delta_1}{2}\right)}}\,
\Phi\left(y_1\,,\frac{1}{x_2}\right) =  \\
 {\sf C} \cdot (x_1-x_2)^{2(\frac{D}{2}-\Delta_1)}\,
\int d^D y_1
\frac{(y_1-x_2)^{2\left(u+\frac{\Delta_1+\Delta_2-D}{2}\right)}}
{(x_1-y_1)^{2\left(\frac{D}{2}
+u+\frac{\Delta_2-\Delta_1}{2}\right)}}\,
\frac{\Phi\left(\frac{1}{y_1}\,,\frac{1}{x_2}\right)}
{y_1^{2\Delta_1}x_2^{2\Delta_2}}
\lb{IIQ}
\end{align}
where at the last step we change the variables
$y_1 \to \frac{1}{y_1}$ so that $d^Dy_1 \to \frac{d^Dy_1}{y_1^{2D}}$
and use the formula
$\left(\frac{1}{x}-\frac{1}{y}\right)^{2} = \frac{(x-y)^2}{x^2y^2}$.
Expressions (\ref{QII})
and (\ref{IIQ}) are equal, which proves (\ref{confR}).

\end{document}